\documentclass[sigconf]{acmart}
\renewcommand\footnotetextcopyrightpermission[1]{} 
\setcopyright{none}

\usepackage{nopageno}

\usepackage{booktabs} 

\newif\ifconf
\conffalse

\usepackage{etex}

\usepackage{caption}
\usepackage{subcaption}
\usepackage[sc]{mathpazo}

\usepackage{dblfloatfix}

\usepackage{graphicx}
\usepackage{booktabs}
\usepackage{epstopdf}
\usepackage{url}
\usepackage{placeins}
\usepackage{amsmath,amsfonts}
\usepackage{mathtools,mathrsfs}
\usepackage{multirow}
\usepackage{rotating}
\usepackage{pbox}
\usepackage[normalem]{ulem}
\usepackage{listings}

\usepackage{cleveref}
\usepackage[utf8]{inputenc}
\crefname{section}{§}{§§}
\Crefname{section}{§}{§§}

\usepackage[10pt]{moresize}

\usepackage{xcolor}
\definecolor{darkgrey}{RGB}{70,70,70}
\definecolor{lightgrey}{RGB}{200,200,200}

\usepackage{enumitem}

\usepackage[linesnumbered,ruled,vlined]{algorithm2e}
\SetAlFnt{\scriptsize\tt}
\SetAlCapFnt{\scriptsize\sf}
\SetAlCapNameFnt{\scriptsize\sf}

\newcommand{\dimitri}[1]{\textcolor{green}{[Dimitri: #1]}}

\newcommand{\goal}[1]{\noindent\textcolor{red}{[Goal: #1]}\par}

\newcommand{\macb}[1]{\textbf{\textsf{#1}}}

\usepackage[hang,flushmargin]{footmisc}

\usepackage{bm}

\usepackage{scalerel,stackengine}
\stackMath
\newcommand\rwh[1]{%
\savestack{\tmpbox}{\stretchto{%
  \scaleto{%
      \scalerel*[\widthof{\ensuremath{#1}}]{\kern-.6pt\bigwedge\kern-.6pt}%
          {\rule[-\textheight/2]{1ex}{\textheight}}
            }{\textheight}%
}{0.5ex}}%
\stackon[1pt]{#1}{\tmpbox}%
}

\def\HiLiGA{\leavevmode\rlap{\hbox to \hsize{\color{black!10}\leaders\hrule height 1\baselineskip depth 1ex\hfill}}}
\def\HiLiGB{\leavevmode\rlap{\hbox to \hsize{\color{black!25}\leaders\hrule height 1\baselineskip depth 1ex\hfill}}}
\def\HiLiGC{\leavevmode\rlap{\hbox to \hsize{\color{black!40}\leaders\hrule height 1\baselineskip depth 1ex\hfill}}}
\def\HiLiGD{\leavevmode\rlap{\hbox to \hsize{\color{black!55}\leaders\hrule height 1\baselineskip depth 1ex\hfill}}}
\def\HiLiGE{\leavevmode\rlap{\hbox to \hsize{\color{black!70}\leaders\hrule height 1\baselineskip depth 1ex\hfill}}}
\def\HiLiGF{\leavevmode\rlap{\hbox to \hsize{\color{black!85}\leaders\hrule height 1\baselineskip depth 1ex\hfill}}}

\usepackage{algpseudocode}
\usepackage{tikz}
\usetikzlibrary{calc}
\usepackage{xcolor}
\makeatletter

\makeatletter 
\renewcommand{\@seccntformat}[1]{\csname the#1\endcsname\ \ }
\makeatother

\renewcommand{\goal}[1]{}
\renewcommand{\dimitri}[1]{}

\usepackage{makecell}

\usepackage{tikz}
\usetikzlibrary{tikzmark}

\tikzstyle{comment} = [draw, fill=blue!70, text=white, text width=3cm, minimum height=1cm, rounded corners, align=left, font=\scriptsize]
\tikzstyle{background_alg} = [draw, fill=blue!20, opacity=0.4, inner sep=4pt, rounded corners=2pt]

\usetikzlibrary{shapes}
\usetikzlibrary{plotmarks}
\usetikzlibrary{calc,fit}

 \newcommand\encircle[1]{%
\tikz[baseline=(X.base)]
\node (X) [draw, shape=circle, inner sep=0, fill=black, text=white, inner sep=-0.5pt] {\strut #1};}

\newcommand*{\affmark}[1][*]{\textsuperscript{#1}}

\renewcommand{\textrightarrow}{$\rightarrow$}

\begin{document}
\thispagestyle{empty}

\title{Log(Graph): A Near-Optimal\\ High-Performance Graph Representation}

\author{
  {\large\mbox{Maciej Besta\affmark[\ensuremath\dagger], Dimitri Stanojevic, Tijana Zivic, Jagpreet Singh, Maurice Hoerold, Torsten Hoefler\affmark[\ensuremath\dagger]}}\\
  {\large Department of Computer Science, ETH Zurich}\\
  {\large \affmark[\ensuremath\dagger]Corresponding authors (maciej.besta@inf.ethz.ch, htor@inf.ethz.ch)}
}

\begin{abstract}
%
%
Today's graphs used in domains such as machine learning or social network
analysis may contain hundreds of billions of edges. Yet, they are not
necessarily stored efficiently, and standard graph representations such as
adjacency lists waste a significant number of bits while graph compression
schemes such as WebGraph often require time-consuming decompression.
To address this, we propose Log(Graph): a graph representation that combines
high compression ratios with very low-overhead decompression to enable cheaper
and faster graph processing.
%
%
The key idea is to encode a graph so that the parts of the representation
approach or match the respective storage lower bounds. We call our approach
``graph logarithmization'' because these bounds are usually logarithmic.
%
%
Our high-performance Log(Graph) implementation based on modern bitwise
operations and state-of-the-art succinct data structures achieves high
compression ratios as well as performance. For example, compared to the tuned
Graph Algorithm Processing Benchmark Suite (GAPBS), it reduces 
graph sizes by 20-35\% while matching GAPBS' performance or even delivering speedups
due to reducing amounts of transferred data. It approaches the
compression ratio of the established WebGraph compression library
while enabling speedups of up to more than 2$\times$.
Log(Graph) can improve the design of various graph processing engines or
libraries on single NUMA nodes as well as distributed-memory systems. 
\end{abstract}




\maketitle

{\vspace{-0.5em}\noindent \textbf{This is a full version of a paper published at\\ PACT'18 under the same title}}

{\vspace{1em}\noindent\textbf{Code:}\\\url{http://spcl.inf.ethz.ch/Research/Performance/LogGraph}}


\section{Introduction}
\label{sec:intro}



Large graphs form the basis of many problems in machine learning, social
network analysis, and computational
sciences~\cite{DBLP:journals/ppl/LumsdaineGHB07}. For example, graph clustering
is important in discovering relationships in graph data. The sheer size of such
graphs, up to hundreds of billions of edges, exacerbates the number of
needed memory banks, increases the amount of data transferred between CPUs and
memory, and may lead to I/O accesses while processing graphs.  Thus, reducing
the size of such graphs is becoming increasingly important. 

However, state-of-the-art graph representations and compression schemes, for
example the well-known WebGraph~\cite{boldi2004webgraph}, use techniques such
as reference encoding or interval encoding that may require costly
decompression. For example, consider two vertices, $v_1$ and $v_2$, and assume
that some of the neighbors of $v_1$ and $v_2$ are identical.  In reference
encoding, these shared neighbors are stored only once, in the adjacency array
of either $v_1$ or $v_2$.  The other adjacency array contains a pointer to the
location of these neighbors in the first array. Such encoding may be nested
arbitrarily deeply, leading to pointer chasing. Such schemes may 
degrade performance of \emph{graph accesses} (e.g., verifying if an edge
exists) that are performance critical operations in various graph
algorithms such as triangle counting.
An ideal graph representation should not only provide \emph{high compression
ratios} but also \emph{reduce or eliminate decompression overheads when
accessing a graph}.


In this work, we propose Log(Graph): a representation that achieves the above
goals. 
The key idea is to \emph{encode different graph elements
using the associated storage lower bounds}. We apply this idea to
the popular adjacency array (AA) graph representation and its elements, including
vertex IDs, edge weights, offsets, and the whole arrays with offsets and with adjacency data.
We call this approach ``graph logarithmization'' as most considered storage lower bounds
are \emph{logarithmic} (one needs at least
$\left\lceil \log |S| \right\rceil$ bits to store an object from a set
$S$).
%
%
We illustrate that the main advantage of this approach is its \emph{very low
overhead of decompression} combined with \emph{high compression ratios}.
For example, the compression ratio of Log(Graph)
is often negligibly lower than that of WebGraph. Yet, processing
these graphs with algorithms such as BFS or PageRank is faster (up to $>$2$\times$) when
using the Log(Graph) schemes.
%


\begin{figure}[b]
\centering
\includegraphics[width=0.44\textwidth]{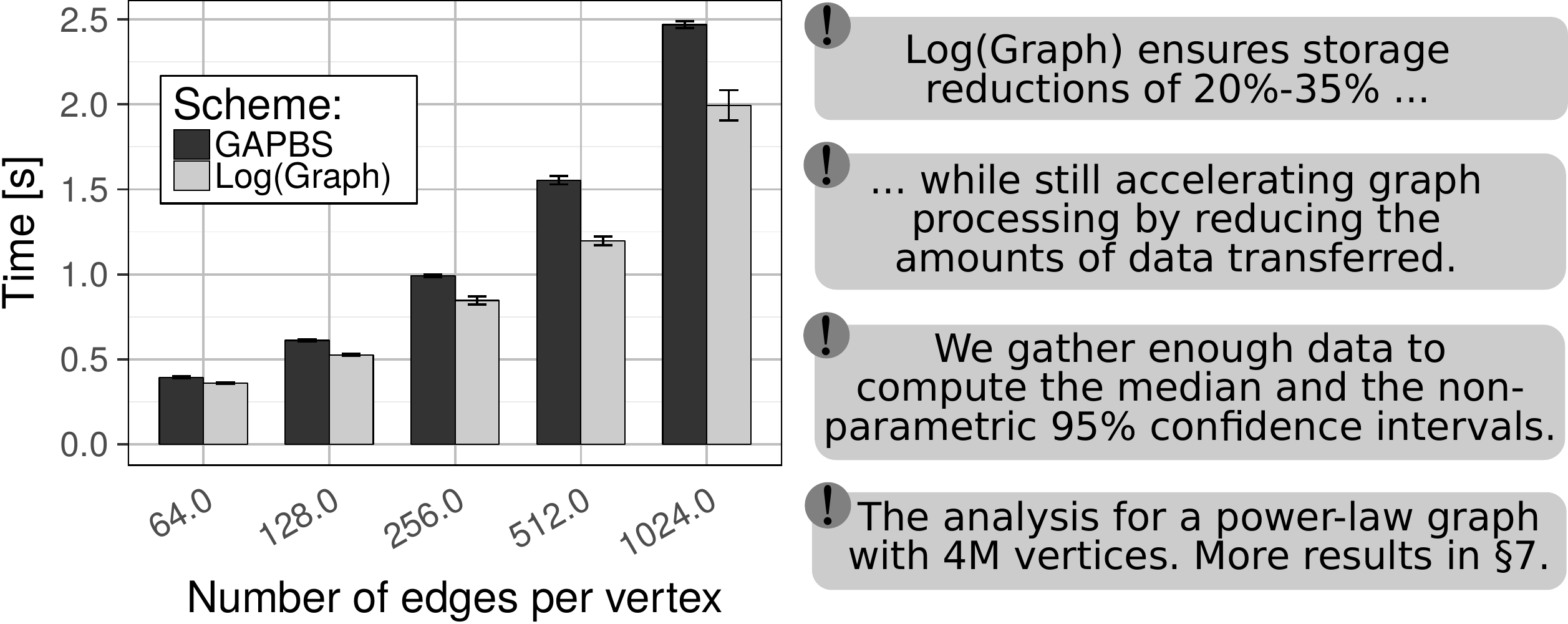}
%
\caption{Log(Graph) performance with SSSP 
when compressing vertex IDs (the \emph{local} approach~\cref{sec:local_approach_log}), compared to the tuned GAPBS~\cite{beamer2015gap}.}
%
%
%
\label{fig:motivation}
%
%
\end{figure}

\begin{figure*}
\centering
\includegraphics[width=0.98\textwidth]{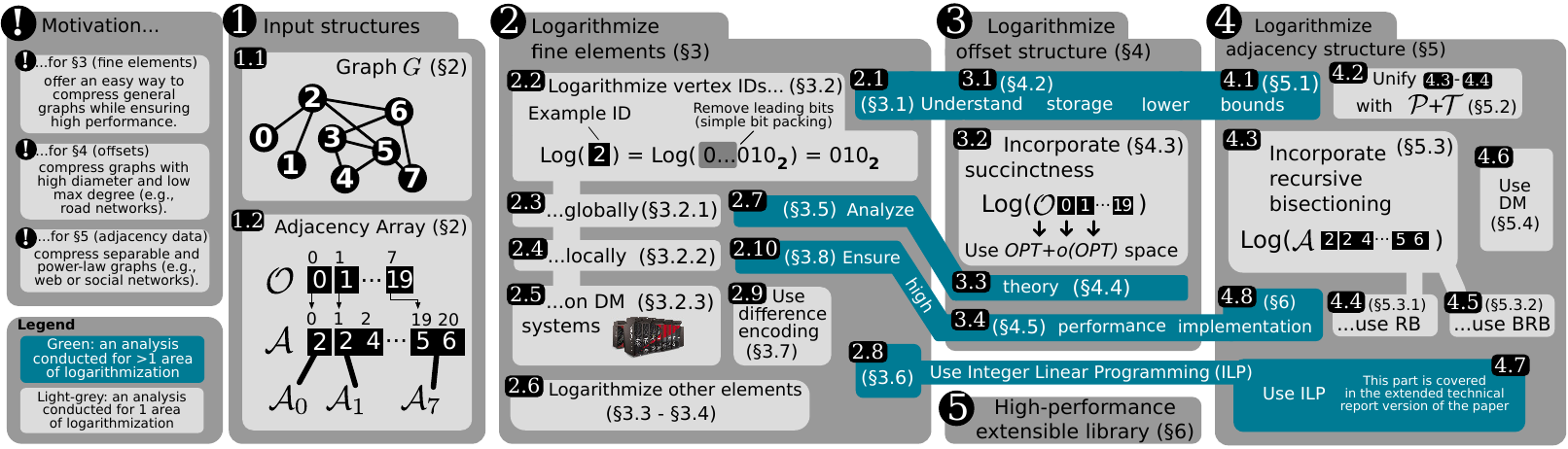}
%
\caption{(\cref{sec:bck}) The roadmap of incorporated schemes.
The green areas indicate analyzes and themes shared by multiple
logarithmization areas.}
%
\label{fig:roadmap}
\end{figure*}

Simultaneously, we illustrate that one must be careful when selecting an
element of AA to logarithmize. For example, a straightforward bit packing
scheme~\cite{suel2001compressing, adler2001towards}, in
which one uses $\left\lceil \log n \right\rceil$ bits to store a vertex ID from
the set of all vertices $V = \{1, ..., n\}$ in a given graph, brings only
modest storage reductions, as shown later (\cref{sec:evaluation}). In this
particular case, we \emph{separately consider the neighborhood $N_v$} of each
vertex $v$, and encode any vertex $w \in N_v$ using $\left\lceil \log
|N_v|\right\rceil$ bits.

We motivate Log(Graph) in Figure~\ref{fig:motivation} with example results for
the Single Source Shortest Path (SSSP) algorithm.  
Log(Graph) outperforms a
tuned SSSP code from the GAP Benchmark Suite~\cite{beamer2015gap}, a
state-of-the-art graph benchmarking platform, by 20\%, while reducing the
required storage by 20-35\%.
Thus, Log(Graph) does \emph{not only compress graphs with negligible
preprocessing costs}, but it even \emph{delivers speedups} by combining storage
reductions with fast decompression.
%

To further accelerate Log(Graph) and enhance its compression ratios, we use
\emph{modern bitwise operations} to efficiently extract data from its
logarithmic encoding. 
Moreover, we develop an Integer Linear Programming (ILP) heuristic that
reorders vertex IDs to save space.
Third, to compress offsets into graph adjacency data, we use \emph{succinct
data structures}~\cite{Jacobson:1988:SSD:915547} that approach theoretical
storage lower bounds while enabling constant-time data accesses.  We show that
they asymptotically reduce the usual $O(n \log n)$ bits used in traditional
offset arrays.  In addition, we provide \emph{the first performance analysis of
succinct data structures in a parallel setting} and we conclude that, in the
context of graph accesses, succinct data structures deliver nearly identical
performance to that of offset arrays.

%
%


We target shared- and distributed-memory settings, and a wide selection of
graph algorithms: BFS, PageRank (PR), Connected Components (CC), Betweenness
Centrality (BC), Triangle Counting (TC), and SSSP.  We conclude that
\emph{Log(Graph) reduces storage required for graphs while enabling
low-overhead decompression, matching the performance of tuned graph processing
codes, and outperforming established compression systems}. 

\section{BACKGROUND AND NOTATION}
\label{sec:bck}

We first describe the used concepts and notation; 
see Table~\ref{tab:symbols} for summarized symbols.
In Log(Graph), we use multiple techniques and we
postpone describing some of them to their related sections
for better readability.


\begin{table*}[b]
\centering
 \setlength{\tabcolsep}{2pt}
\renewcommand{\arraystretch}{0.9}
\sf
\begin{tabular}{@{}l|ll@{}}
\toprule
\multirow{6}{*}{\begin{turn}{90}\shortstack{Graph model}\end{turn}} 
      & $G$&A graph $G=(V,E)$; $V$ and $E$ are sets of vertices and edges.\\
      & $n,m$&Numbers of vertices and edges in $G$; $|V| = n, |E| = m$.\\
      & $\mathcal{W}_{(v,w)}$& The weight of an edge $(v,w)$.\\ 
      & $d_v, N_v, N_{i,v}$ & Degree, neighbors, and $i$th neighbor of a vertex $v$; $N_{0,v} \equiv v$. \\
      & $\overline{x}, \rwh{x}$ & The average and the maximum value in a set or sequence $x$.\\
      & $\alpha, \beta; p$ & Parameters of a power-law graph and an Erdős-Rényi graph.\\
      \midrule
\multirow{5}{*}{\begin{turn}{90}\shortstack{Adjacency\\array}\end{turn}} 
      & $\mathcal{A}, \mathcal{A}_v$ & The adjacency array of a given graph and a given vertex.\\
      & $\mathcal{O}, \mathcal{O}_v$ & The offset structure of a given graph and an offset to $\mathcal{A}_v$.\\
      & $|\mathcal{A}|, |\mathcal{O}|$ & The sizes of $\mathcal{A}$ and $\mathcal{O}$.\\
      & $\mathcal{L}[\mathcal{A}], \mathcal{L}[\mathcal{O}]$ & Logarithmization schemes acting upon $\mathcal{A}$ and $\mathcal{O}$.\\
      & $B,L,W$ & Various parameters of $\mathcal{A}$ and $\mathcal{O}$; see~\cref{sec:O_LB}--\cref{sec:offset_arrays_types} for details.\\
\midrule
\multirow{5}{*}{\begin{turn}{90}\shortstack{Machine\\model}\end{turn}}
      & $N$&The number of levels in a hierarchical machine.\\
      & $H_i, \mathcal{H}$ & The number of elements at level $i$, the number of compute nodes.\\
      & $T,P$ & The number of threads/processes.\\
      & $W$ & The memory word size [bits].\\
      & $t_x$ & Time to do a given operation $x$.\\
      \midrule
\multirow{3}{*}{\begin{turn}{90}\shortstack{Others}\end{turn}} 
      & $\mathcal{P}$ & Permuter: function that relabels vertices.\\
      & $\mathcal{T}_x, \mathcal{T}$ & Transformers: functions that arbitrarily modify $\mathcal{A}$.\\
      & $G_x$ & Subgraphs of $G$ constructed in recursive bisectioning. \\
%
\bottomrule
\end{tabular}
\caption{Symbols used in the paper gathered for the reader's convenience.}
\label{tab:symbols}
\end{table*}

\subsection{Used Models}
\label{sec:models}

We first describe models used in this work.

\subsubsection{Graph Model}

%
We model an undirected graph $G$ as a tuple $(V,E)$; $V$ is a set of vertices
and $E \subseteq V \times V$ is a set of edges; $|V|=n$, $|E|=m$.
Vertices are identified by contiguous IDs ($\equiv$ labels) from $\{1, ..., n\}$.
$N_v$ and $d_v$ denote the neighbors and the degree of a vertex $v$.
$N_{i,v}$ is $v$'s $i$th neighbor.
{$\rwh{X}$} indicates the maximum value in a given set or sequence $X$,
for example {$\rwh{N}_{v}$} is the maximum ID among $v$'s neighbors.
{$\rwh{\mathcal{W}}$} is the maximal edge weight in a given graph~$G$. 

\subsubsection{Machine and System Model}

For more storage reductions on today's hardware, we consider arbitrary
hierarchical machines where, for example, cores reside on a socket, sockets
constitute a node, and nodes form a rack. $N$ is the
number of hierarchy levels and $H_i$ is the total number of elements from
level~$i$.  The first level corresponds to the whole machine; thus $H_1 = 1$.
We also refer specifically to the number of compute nodes as $\mathcal{H}$.
Finally, the numbers of used threads per node and processes are $T$ and $P$.
The memory word size is $W$.


\subsection{Used Concepts}

We next explain concepts related to the structures used in Log(Graph) and their
size; see also Figure~\ref{fig:roadmap} for an overview.

First, we discuss \textbf{succinctness}.
Assume $OPT$ is the optimal number of bits to store some data. A representation
of this data is \emph{compact} if it uses $\mathcal{O}(OPT)$ bits and \emph{succinct} if it
uses $OPT + o(OPT)$ bits.
%
%
They all should support \emph{a reasonable set of
queries in (ideally) $\mathcal{O}(1)$ time}~\cite{Blandford:2003:CRS:644108.644219}.
Now, they differ from \emph{compression} mechanisms such as
zlib as they
do not entail expensive decompression.
To avoid confusion, we use the term \emph{condensing} to refer in general to
reducing the size of some data.

Finally, we describe \textbf{graph separability}. Intuitively, $G$ is vertex
(or edge) \emph{separable} (i.e., has \emph{good separators}) if we can divide
$V$ into two subsets of vertices of approximately the same size so that the
size of a vertex (or edge) cut between these two subsets is much smaller than
$|V|$.

\subsection{Used Data Structures}

\subsubsection{Adjacency Array Data Structures}

Log(Graph) builds upon the traditional \textbf{adjacency array} (AA)
representation. One part of AA is an array (denoted as $\mathcal{A}$) with adjacency data.
$\mathcal{A}$ consists of $n$ subarrays $\mathcal{A}_i, i \in V$.
Subarray $\mathcal{A}_v$
contains neighbors of
vertex~$v$. We have 
$\mathcal{A} = \{\mathcal{A}_1, ..., \mathcal{A}_v\}$. 
%
%
Each $\mathcal{A}_v$ is sorted by vertex IDs.
The second part of AA is an array $\mathcal{O}$ with offsets (or pointers) to
each $\mathcal{A}_v$. $\mathcal{O}_v$ denotes the offset
to $\mathcal{A}_v$. 
%
%
%
%
%
%
%
$|\mathcal{O}|$ and $|\mathcal{A}|$ denote the sizes of $\mathcal{O}$ and
$\mathcal{A}$.

\subsubsection{Bit Vectors}

Next, we use simple \textbf{bit vectors} to enhance $\mathcal{O}$
(\cref{sec:O_BV}, \cref{sec:offset_arrays_types}).  A bit vector $S$ of length
$L$ takes only $L$ bits, but uses $\mathcal{O}(L)$ time to answer two important
queries: $rank_S(x)$ and $select_S(x)$.  For a given $S$, $rank_S(x)$ returns
the number of ones in $S$ up to and including the $x$th bit.  Conversely,
$select_S(x)$ returns the position of the $x$th one in $S$.
These queries are widely used for implementing structures such as binary trees
and sets~\cite{succ-category, DBLP:journals/corr/abs-0705-0552}. For example,
if $S$ represents a set with an order imposed on its elements, $rank_S(x)$
  gives the number of elements in the set up to $x$. Thus, many designs with
  $rank$ and $select$ answering in $o(L)$ time have been proposed.  An
  important family are \emph{succinct bit vectors}.
\subsection{Roadmap of Schemes}
To enhance readability, we
summarize Log(Graph) in Figure~\ref{fig:roadmap}. 
%
%
We divide the Log(Graph) schemes into three categories, based on what they
compress: fine graph elements (\cref{sec:log_fines}), offset structure
$\mathcal{O}$ (\cref{sec:compress_offsets}), and adjacency data $\mathcal{A}$
(\cref{sec:compress_adj_lists}).
Each of these sections is structured similarly. We first describe lower bounds
associated with a compressed object (e.g.,~\cref{sec:F_LB}) and the actual
compression schemes (e.g.,~\cref{sec:log_v_ids}--\cref{sec:fine_offsets}), then
conduct a theoretical analysis (e.g.,~\cref{sec:theory_fine}), describe further
enhancements such as an ILP heuristic (e.g.,~\cref{sec:relabeling_scheme}) or
gap-encoding (\cref{sec:fine_gap_encoding}), and the implementation
(e.g.,~\cref{sec:bitwise_ops}).



As we show empirically in~\cref{sec:evaluation}, each of the three classes of
logarithmization schemes has slightly different characteristics and thus
application domains.
First, compressing fine graph elements (\cref{sec:log_fines}) brings
\emph{storage reductions of 20-35\%} compared to the traditional AA while
delivering performance close to or matching or \emph{even exceeding} that of
tuned graph processing codes.
Second, compressing offset structures $\mathcal{O}$
(\cref{sec:compress_offsets}) can enhance \emph{any} parallel graph processing
computation because it does \emph{not} impact performance in parallel settings
while it \emph{does} reduce storage required for offsets $\mathcal{O}$ even by
$>$90\%.
Finally,  adjacency data $\mathcal{A}$
(\cref{sec:compress_adj_lists}) is the most aggressive in reducing storage, in
some cases by \emph{up to $\approx$80\%} compared to the adjacency array,
approaching the compression ratios of modern graph compression schemes and
simultaneously offering speedups of more than 2$\times$.

\section{LOGARITHMIZING FINE ELEMENTS}
\label{sec:log_fines}

We first logarithmize fine elements of the adjacency array: 
vertex IDs, vertex offsets, and edge weights; see Figure~\ref{fig:roadmap} (\encircle{2}).
%
%

\subsection{Understanding Storage Lower-Bounds}
\label{sec:F_LB}

A simple storage lower bound is the logarithm of the number of possible
instances of a given entity, which corresponds to the number of bits required
to distinguish between these instances.
Now, bounds derived for fine-grained graph elements are illustrated in
Table~\ref{tab:bounds}.
First, a storage lower bound for a single vertex ID is $\left\lceil \log n
\right\rceil$ bits as there are $n$ possible numbers to be used for a single
vertex ID.
Second, a corresponding bound to store an offset into the neighborhood of a
single vertex is $\lceil \log 2m \rceil$; this is because in an undirected
graph with $m$ edges there are $2m$ cells.
Third, a storage lower bound of an edge weight from a discrete set
{\footnotesize$\left\{0, ..., \rwh{\mathcal{W}}\right\}$} is
{\footnotesize$\left\lceil \log \rwh{\mathcal{W}} \right\rceil$} (for
continuous weights, we first scale them appropriately to become elements of a
discreet set).


\begin{table}
\centering
\small
\sf
\begin{tabular}{@{}l|l|ll@{}}
\toprule
& \textbf{Entity} & \textbf{Bound} & \textbf{Assumptions, remarks} \\
\midrule
\multirow{4}{*}{\begin{turn}{90}{\scriptsize\shortstack{Fine elements\\(\cref{sec:F_LB})}}\end{turn}} 
& Vertex ID   & $\log n$ & In the global approach. \\
& Vertex ID   & $\log \rwh{N}_v$ & In the local approach. \\
& Offset      & $\log 2m$ & For unweighted graphs. \\
& Edge weight & $\log \rwh{\mathcal{W}}$ & - \\
\midrule
\multirow{1}{*}{\scriptsize\shortstack{$\mathcal{O}$\\(\cref{sec:O_LB})}} 
%
& Bit vector & $\log \dbinom{\frac{2Wm}{B}}{n}$ & $n$ set bits among $\frac{2Wm}{B}$ bits. \\
\midrule
\multirow{1}{*}{\scriptsize\shortstack{$\mathcal{A}$\\(\cref{sec:A_LB})}}
& Graph & $\log {\dbinom {\binom {n} {2}}{m}}$ & The graph is undirected. \\
\bottomrule
\end{tabular}
\caption{Storage lower bounds for various parts of an AA.
The ceiling function $\lceil \cdot \rceil$ surrounding each $\log \cdot$ was
omitted for aesthetic purposes.} \label{tab:bounds}
\end{table}

\subsection{Logarithmization of Vertex IDs}
\label{sec:log_v_ids}


We first logarithmize vertex IDs.

\subsubsection{Vertex IDs: The Global Approach}

The first and simplest step in Log(Graph) is to use $\lceil \log n \rceil$ bits
to store a vertex ID in a graph with $n$ vertices. In this case, the total size
of the adjacency data $|\mathcal{A}|$ is $2m \lceil \log n \rceil$.
This simple bit packing was mentioned in past work~\cite{suel2001compressing,
adler2001towards} and it modestly reduces $|\mathcal{A}|$ (as we show in
\cref{sec:evaluation}). We now \emph{extend it with schemes that deliver
more storage reductions and performance}
(\cref{sec:local_approach_log}--\cref{sec:log_fine_dm}).
%


\subsubsection{Vertex IDs: The Local Approach}
\label{sec:local_approach_log}

Even if the above global approach uses an optimum number or bits to store a
vertex ID, it may be far from optimum when considering \emph{subsets} of these
vertices.
For example, consider a vertex $v$ with very few neighbors ($d_v \ll n$) that
all have small IDs ($\rwh{N}_v \ll n$).  Here, the optimum number of bits for a
vertex ID in $\mathcal{A}_v$ is {$\left\lceil \log \rwh{N}_v \right\rceil$},
\emph{which may be much lower than $\lceil \log n \rceil$}.
However, one must also keep the information on the number of bits required for
each $N_v$. We use a fixed number of bits; it is lower bounded by {$\left\lceil
\log \log \rwh{N}_v \right\rceil$} bits. The size of $\mathcal{A}$ is in this
case

\begin{gather}
%
%
%
\nonumber|\mathcal{A}| = \sum_{v \in V} \left(d_v \left\lceil \log \rwh{N}_v \right\rceil + \left\lceil \log \log \rwh{N}_v \right\rceil\right)
%
%
\end{gather}
\normalsize

%


\subsubsection{Vertex IDs in Distributed-Memories}
\label{sec:log_fine_dm}

We now extend vertex logarithmization to the distributed-memory setting.
We divide a vertex ID into an \emph{intra} part that ensures the uniqueness of
IDs within a given machine element (e.g., a compute
node), and an \emph{inter} part that encodes the position of a vertex in the
distributed-memory structure.
%
%
The intra part can be encoded with either the local or the global approach.

We first only consider the level of compute nodes; each node constitutes a
cache-coherent domain and they are connected with non-coherent network.
The number of vertices in one node is $\frac{n}{\mathcal{H}}$. The intra ID part
takes {$\left\lceil \log \frac{n}{\mathcal{H}} \right\rceil$}; the inter one takes
$\left\lceil \log \mathcal{H} \right\rceil$. As the inter part is unique for a
given node, it is stored once per node. Thus

\begin{gather}
\nonumber|\mathcal{A}| = n \left\lceil \log \frac{n}{\mathcal{H}} \right\rceil + \mathcal{H} \left\lceil \log \mathcal{H} \right\rceil
\end{gather}
\normalsize


Next, we consider
the arbitrary number of memory hierarchy levels.
Here, the number of vertices in one element from the bottom of the hierarchy
(e.g., a die) is $\frac{n}{H_{N}}$. Thus, the intra ID part requires
{$\left\lceil \log \frac{n}{H_{N}} \right\rceil$} bits.
The inter part needs {$\sum_{j \in \{2..N-1\}} \left\lceil \log H_{j}
\right\rceil$} bits and has to be stored once per each machine element, thus

\begin{gather}
\nonumber|\mathcal{A}| = n \left\lceil \log \frac{n}{H_{N}} \right\rceil + \sum_{j = 2}^{N-1} H_{j} \left\lceil \log H_{j} \right\rceil
\end{gather}
\normalsize
\subsection{Logarithmization of Edge Weights}

We similarly condense edge weights.  The storage lower bound for storing a
maximal edge weight is {$\left\lceil \log \rwh{\mathcal{W}} \right\rceil$} bits.
Thus, if $G$ is weighted, we respectively have (for the global and local
approach applied to the weights)

\small
\begin{gather}
\nonumber|\mathcal{A}| = 2m \left(\left\lceil \log n \right\rceil + \left\lceil \log \rwh{\mathcal{W}} \right\rceil\right)
\end{gather}

\begin{gather}
%
%
%
%
\nonumber|\mathcal{A}| = \sum_{v \in V} \left( d_v \left(\left\lceil \log \rwh{N}_v \right\rceil + \left\lceil \log \rwh{\mathcal{W}} \right\rceil \right) + \left\lceil \log \log \rwh{N}_v \right\rceil + \left\lceil \log \log \rwh{\mathcal{W}} \right\rceil \right)
\end{gather}
\normalsize

\subsection{Logarithmization of Single Offsets}
\label{sec:fine_offsets}

Finally, one can also ``logarithmize'' other AA elements, including offsets.
Each offset must be able to address any position in $\mathcal{A}$ that may
reach $2m$.  Thus, the related lower bound is $\left\lceil \log 2m
\right\rceil$, giving $|\mathcal{O}| = n \left\lceil \log{2m} \right\rceil$.
\subsection{Theoretical Storage Analysis}
\label{sec:theory_fine}

Next, we show how the above schemes reduce the size of graphs generated using
two synthetic graph models (random uniform and power-law) for the global
approach.
%

\subsubsection{Erdős-Rényi (Uniform) Graphs}

We start with Erdős-Rényi random uniform graphs. Here, every edge is present
with probability $p$. The expected degree of any vertex is $pn$, thus

\begin{gather}
\nonumber E[|\mathcal{A}|] = \left(\left\lceil \log n \right\rceil + \left\lceil \log \rwh{\mathcal{W}} \right\rceil \right) p n^2 \\
\nonumber E[|\mathcal{O}|] = n \left\lceil \log \left(2pn^2\right)  \right\rceil = n \left\lceil \log2p + 2\log n \right\rceil
\end{gather}
\normalsize

\subsubsection{Power-Law Graphs}

We next analyze power-law graphs; the derivation is in the Appendix. 
Here, the probability that a vertex has
degree $d$ is $f(d) = \alpha d^{-\beta}$, and

\begin{gather}
\nonumber E[|\mathcal{A}|] \approx \frac{\alpha}{2-\beta} \left(\left(\frac{\alpha n \log n}{\beta - 1}\right)^{\frac{2-\beta}{\beta - 1}} - 1\right) \left(\left\lceil \log n \right\rceil + \left\lceil \log \rwh{\mathcal{W}} \right\rceil\right) 
\end{gather}
\normalsize

\begin{figure}
  \centering
  \begin{subfigure}[t]{0.21 \textwidth}
    \centering
    \includegraphics[width=\textwidth]{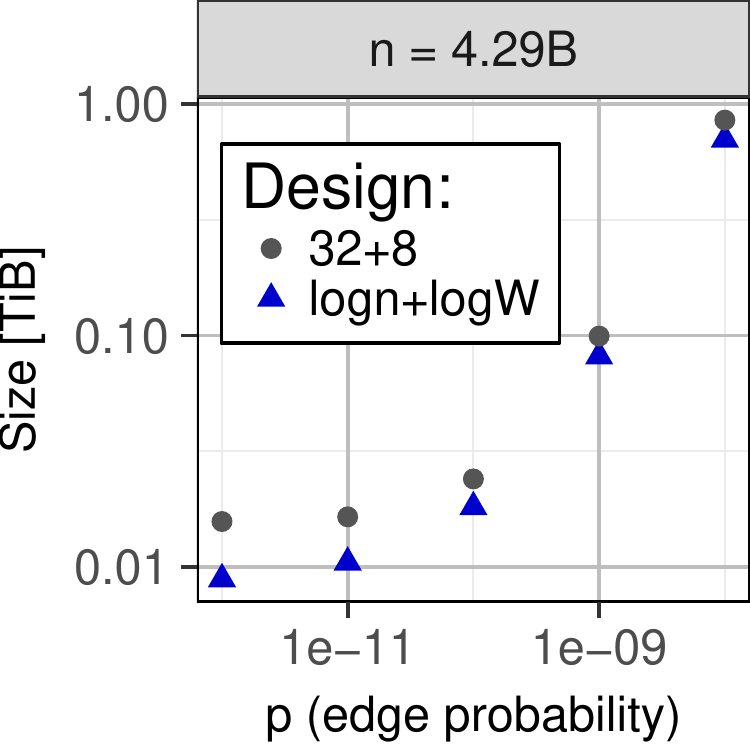}
    \caption{Random-uniform.}
    \label{fig:er}
  \end{subfigure}
  \quad
  \begin{subfigure}[t]{0.21 \textwidth}
    \centering
    \includegraphics[width=\textwidth]{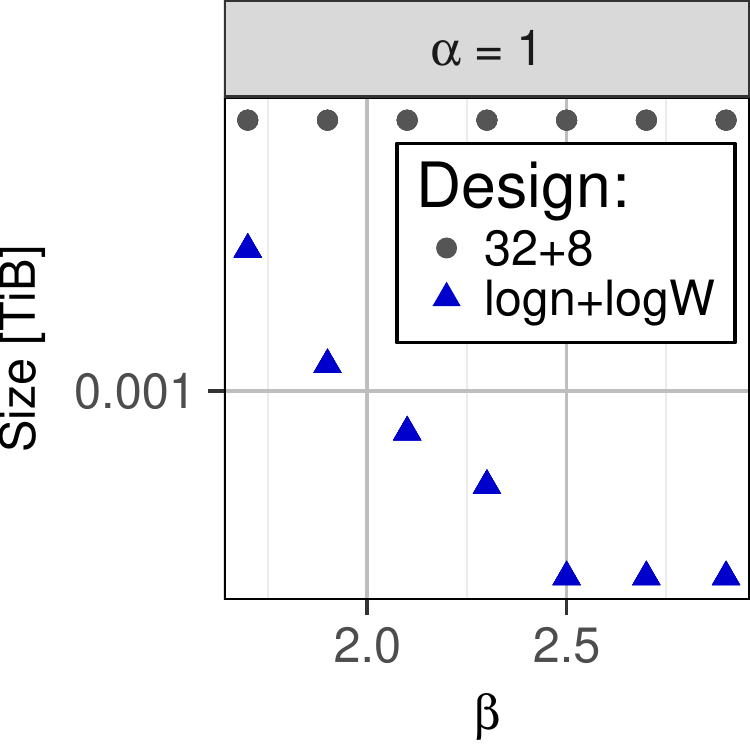}
    \caption{Power-law.}
    \label{fig:pl}
  \end{subfigure}
  \caption{(\cref{sec:theory_fine}) The analysis of the size of random-uniform and power-law graphs with Log(Graph) and traditional adjacency array.}
  \label{fig:theory-analysis}
\end{figure}

The results are in Figure~\ref{fig:theory-analysis}.  ``32+8'' indicates an AA
with 32 bits for a vertex ID and 8 bits for a weight; the other target is
Log(Graph).
Compressing fine-grained elements consistently reduces storage. Yet, it may
offer suboptimal space and performance results as it ignores the
\emph{structure of the graph} and \emph{the structure of the memory with
fixed-size words}.
We now address these issues with ILP and gap encoding (for less storage) and
efficient design (for more performance).

\subsection{Adding Integer Linear Programming}
\label{sec:relabeling_scheme}


%
We now enhance the local logarithmization
(\cref{sec:log_v_ids}) to further reduce $|\mathcal{A}|$.
Consider any $\mathcal{A}_v$. We observe that a single neighbor in
$\mathcal{A}_v$ with a large ID may vastly increase $|\mathcal{A}_v|$:
we use {\small$\left\lceil \rwh{N}_v \right\rceil$} bits to
store each neighbor in $\mathcal{A}_v$ but other neighbors may have much
lower IDs.
Thus, we permute vertex IDs to 
\emph{reduce maximal IDs in as many neighborhoods as possible} without modifying the graph structure. We illustrate an ILP
formulation and then propose a heuristic.
%
%
%

The new objective function is shown in Eq.~(\ref{eq:obj_fun_2}).  It
minimizes the weighted sum of $\rwh{N}_v, v \in V$.
Each maximal ID is given a positive weight which is
the inverse of the neighborhood size; this intuitively decreases
$\rwh{N}_v$ in smaller $\mathcal{A}_v$.

\begin{alignat}{2}
& \min \ \ \sum_{v \in V} \rwh{N}_v \cdot \frac{1}{d_v}  \label{eq:obj_fun_2}
\end{alignat}
\normalsize




%


In Constraint~(\ref{eq:max}), we set $\rwh{N}_v$ to be the maximum of the new IDs
assigned to the neighbors of $v$. $\mathcal{N}(v)$ is the new ID of $v$.

\begin{alignat}{2} 
\forall_{v,u \in V}  \left(u\in N_v \right)\ \Rightarrow \left[\mathcal{N}(u) \leq \rwh{N}_v \right] \label{eq:max}
\end{alignat}
\normalsize

Listing~\ref{lst:relabeling} describes a greedy polynomial time heuristic for
changing IDs. We sort vertices in the increasing order of their degrees
(Line~\ref{algo:sort}). Next, we traverse vertices in the sorted order,
beginning with the smallest $|\mathcal{A}_v|$, and assign a new smallest ID
possible (Line~\ref{algo:relabel}). The remaining vertices that are not renamed
by now are relabeled in Line~\ref{algo:left}. This scheme acts similarly to the
proposed ILP. 
%

\begin{lstlisting}[aboveskip=0em,belowskip=0em,float=h!,label=lst:relabeling,caption=(\cref{sec:relabeling_scheme}) The
greedy heuristic for vertex relabeling.]
/* Input: graph $G$, Output: a new relabeling $\mathcal{N}(v), \forall v \in V$. */
void relabel($G$) {
 $ID[0..n-1]$ = $[0..n-1]$; //An array with vertex IDs.
 $D[0..n-1]$ = $[d_0..d_{n-1}]$; //An array with degrees of vertices.
 //An auxiliary array for determining if a vertex was relabeled:
 $visit[0..n-1]$ = $[false..false]$; 
 $nl$ = 1; //An auxiliary variable ``new label''.
 sort($ID$); sort($D$); |\label{algo:sort}|
 for(int $i$ = 1; $i$ < $n$; ++$i$) //For each vertex... |\label{algo:relabel}|
  for(int $j$ = 0; $j$ < $D[i]$; ++$j$) { //For each neighbor...
   int $id$ = $N_{j,ID[i]}$; //$N_{j,ID[i]}$ is $j$th neighbor of vertex with ID $ID[i]$
   if($visit[id]$ == $false$) { 
    $\mathcal{N}(id)$ = $nl$++; 
    $visit[id]$ = $true$; 
 }}
 for(int $i$ = 1; $i$ < $n$; ++$i$) |\label{algo:left}| 
  if($visit[i]$ == $false$) 
   $\mathcal{N}(id)$ = $nl$++; 
}
\end{lstlisting}

\subsection{Adding Fixed-Size Gap Encoding}
\label{sec:fine_gap_encoding}

We next use gap encoding to further reduce
$|\mathcal{A}|$.
Traditionally, in gap encoding one calculates differences between
all consecutive neighbors in each $\mathcal{A}_v$.  
These differences are then encoded with a variable-length code
  such as Varint~\cite{Blandford:2003:CRS:644108.644219}.
Now, this may entail significant decoding overheads. We alleviate
these overheads with \emph{fixed-size gap encoding} where the maximum
difference within a given neighborhood determines the number of bits used to encode
other differences in this neighborhood.
%
%
In the local approach (\cref{sec:local_approach_log}), the maximum difference in each $\mathcal{A}_v$
determines the number of bits to encode any difference in the same
$\mathcal{A}_v$.

\subsection{High-Performance Implementation}
\label{sec:bitwise_ops}

We finally describe the high-performance implementation.
We focus on the global approach due to space
constraints, the local approach entails an almost identical design.

\subsubsection{Bitwise Operations}

We analyzed Intel bitwise operations to ensure the fastest implementation.
Table~\ref{tab:ops} presents the used operations together with the number of
CPU cycles that each operation requires~\cite{granlund2012instruction}.

\begin{table}[h]
\centering
\sf
\footnotesize
\begin{tabular}{@{}l|l|ll@{}}
\toprule
\textbf{Name} & \textbf{C++ syntax} & \textbf{Description} & \textbf{Cycles} \\ \midrule
\texttt{BEXTR} & \texttt{\_bextr\_u64} & Extracts a contiguous number of bits. & 2 \\
\texttt{SHR} & \texttt{>>} & Shifts the bits in the value to the right. & 1 \\
\texttt{AND} & \texttt{\&} & Performs a bitwise AND operation.  & 1 \\
\texttt{ADD} & \texttt{+} & Performs an addition between two values.  & 2 \\
\bottomrule
\end{tabular}
\caption{(\cref{sec:bitwise_ops}) The utilized Intel bitwise operations.}
\label{tab:ops}
\end{table}

\subsubsection{Accessing An Edge ($N_i(v)$)}

We first describe how to access a given edge in $\mathcal{A}$ that corresponds
to a given neighbor of $v$ ($N_{i,v}$); see
Listing~\ref{lst:accessing-edge} for details.
The main issue is to access an $s$-bit value from a byte-addressable memory
with $s = \lceil \log n \rceil$.
In short, we fetch a $64$-bit word that contains the required $s$-bit edge.
In more detail, we first load the offset $\mathcal{O}[v]$ of $v$'s neighbors'
array (Line~3).
%
%
%
This usually involves a cache miss, taking $t_{cm}$.
Second, we derive $o = s \cdot \mathcal{O}[v]$ (the exact bit position of
$N_{i,v}$); it takes $t_{mul}$ (Line~3).
Third, we find the closest byte alignment before $o$ by
right-shifting $o$ by $3$ bits, taking $t_{shf}$ (Line~4).
Instead of byte alignment we also considered any other alignment but it
entailed negligible ($<$1\%) performance differences.
%
%
Next, we derive the distance $d$ from this alignment with a bitwise
\texttt{and} acting on $o$ and binary $111$, taking $t_{and}$.
We can then access the derived $64$-bit value; this involves another cache miss
($t_{cm}$). If we shift this value by $d$ bits and mask it, we obtain
$N_i(v)$. Here, we use the x86 \texttt{bextr} instruction that combines these
two operations and takes $t_{bxr}$.
In the local approach, we also maintain the bit length for each neighborhood.
It is stored next to the associated offset to avoid another cache miss.

\begin{lstlisting}[aboveskip=0em,belowskip=0em,float=h!,label=lst:accessing-edge,caption=(\cref{sec:bitwise_ops}) Accessing an edge in Log(Graph) ($N_i(v)$).]
/* v_ID is an opaque type for IDs of vertices. */
v_ID $N_{i,v}$(v_ID $v$, int32_t $i$, int64_t* $\mathcal{O}$, int64_t* $\mathcal{A}$, int8_t $s$){
  int64_t exactBitOffset = $s$ * ($\mathcal{O}$[v] + $i$);|\label{ln:1}|
  int8_t* address = (int8_t*) $\mathcal{A}$ + (exactBitOffset >> 3);|\label{ln:2}|
  int64_t distance = exactBitOffset & 7;|\label{ln:3}|
  int64_t value = ((int64_t*) (address))[0];|\label{ln:4}|
  return _bextr_u64(value, distance, $s$);|\label{ln:5}| }
\end{lstlisting}

\subsubsection{Accessing Neighbors ($N_v$)}

%
Once we have calculated the exact bit position of the first neighbor as
described in Listing~\ref{lst:accessing-edge}, we simply add $s = \lceil \log n
\rceil$ to obtain the bit position of the next neighbor. Thus, the
multiplication that is used to get the exact bit position is only needed for
the first neighbour, while others are obtained with additions instead.

\subsubsection{Accessing a Degree ($d_v$)}

$d_v$ is simply calculated as the difference
between two offsets: $\mathcal{O}[v+1] - \mathcal{O}[v]$.

\subsubsection{Accessing an Edge Weight}

Finally, we describe how Log(Graph) handles edge weights.
%
%
We store the weight of each edge directly after the corresponding vertex
ID in $\mathcal{A}$.
The downside is that every weight is thus stored twice for undirected graphs.
However, it enables accessing the weight and the vertex ID together. We model
fetching the ID and the weight as a single cache line miss overhead
$t_{cm}$.

\subsubsection{Performance Model}

We finally present the performance model that we use to understand better the
behavior of Log(Graph).  First, we model accessing an edge
($N_{i,v}$) as

\begin{gather}
t_{edge} = 2 t_{cm} + t_{mul} + t_{shf} + t_{and} + t_{bxr} \nonumber
\end{gather}
\normalsize

The model for accessing $N_v$ looks similar with the
difference that only one multiplication is used:

\begin{gather}
%
t_{neigh}(v) = t_{edge} + (d_v-1)(t_{add} + t_{shf} + t_{and} + t_{bxr}) \nonumber
\end{gather}
\normalsize

As $\mathcal{A}$ is contiguous we assume there are no more
cache misses from prefetching. Now, we model the latency of $d_v$ as

\begin{gather}
t_{degree} = 2t_{cm} + t_{sub} \nonumber
\end{gather}
\normalsize

\section{LOGARITHMIZING OFFSETS}
\label{sec:compress_offsets}

\goal{Introduce and motivate the section}
We now logarithmize \emph{the whole offset structure $\mathcal{O}$}, treating
it as a single entity with its own associated storage lower bound.
%
%
Our main technique for combining storage reductions and low-overhead
decompression is to \emph{store the offsets $\mathcal{O}$ as a bitvector} and
then \emph{encode it as a succinct bit vector}~\cite{succinct}
that approaches the storage lower bound while providing
fast accesses to its contents.

\subsection{Arrays of Offsets vs. Bit Vectors for $\mathcal{O}$}
\label{sec:O_BV}

Usually, $\mathcal{O}$ is an array of $n$ offsets and the size of
$\mathcal{O}$, $|\mathcal{O}|$, is much smaller than $|\mathcal{A}|$. Still, in
sparse graphs with low maximal degree {\small$\rwh{d}$}, $|\mathcal{O}|
\approx |\mathcal{A}|$ or even $|\mathcal{O}| > |\mathcal{A}|$.
For example, for the USA road network, if $\mathcal{O}$ contains 32-bit
offsets, $|\mathcal{O}| \approx 0.83 |\mathcal{A}|$.
%
%
To reduce $|\mathcal{O}|$ in such cases, one can use a bit vector instead of an
array of offsets.
For this, $\mathcal{A}$ is divided into blocks (e.g., bytes or words) of a size
$B$ [bits].
Then, if the $i$th bit of $\mathcal{O}$ is set (i.e., $\mathcal{O}[i] = 1$) and if
this is the $j$th set bit in $\mathcal{O}$, then $\mathcal{A}_j$ starts at the $i$th
block.
The key insight is that getting the position of $j$th set bit and thus the
offset of the array $\mathcal{A}_j$ is equivalent to performing a certain
operation called $select_{\mathcal{O}}(j)$ that returns the position of the
$j$th one in $\mathcal{O}$.
%
%
%
%
%
%
Yet, $select$ on a raw bit vector of size $L$ takes $O(L)$ time.
Thus, we first incorporate two designs that enhance $select$: \emph{Plain}
(bvPL) and \emph{Interleaved} (bvIL) bit vectors~\cite{gbmp2014sea}. They both
trade some space for a faster $select$.  bvPL uses up to $0.2|\mathcal{O}|$
additional bits in an auxiliary data structure to enable $select$ in $O(1)$
time. In bvIL, the original bit vector data is interleaved (every $L$ bits)
with 64-bit cumulative sums of set bits up to given positions; $select$ has
$O(\log |\mathcal{O}|)$ time~\cite{gbmp2014sea}. Neither bvPL nor bvIL are
succinct; we use them (1) as reference points and (2) because they also enable
a smaller yet simple offset structure $\mathcal{O}$.


\subsection{Understanding Storage Lower Bounds}
\label{sec:O_LB}

A bit vector that serves as an $\mathcal{O}$ and corresponds to an offset array
of a $G$ takes $\frac{2Wm}{B}$ bits. This is because it must be able to address
up to $2m \cdot W$ bits (there are $2m$ edges in $\mathcal{A}$, each stored
using $W$ bits in a memory word) grouped in blocks of size $B$ bits.
There are exactly {\small$\mathcal{Q} = \dbinom{\frac{2Wm}{B}}{n}$} bit vectors of
length {$\frac{2Wm}{B}$} with $n$ ones and the storage lower bound
is {$\left\lceil \mathcal{Q} \right\rceil$}
bits.

\begin{table*}[t]
\centering
\sf
\begin{tabular}{l|llllll}
\toprule
$\mathcal{O}$ & {\textbf{ID}} & {\textbf{Asymptotic size}} [bits] & {\textbf{Exact size}} [bits] & $rank$ & $select$ & Deriving $\mathcal{O}_v$ \\ \midrule
Pointer array & ptr$W$ & $O(Wn)$ & $W(n+1)$ & - & - & $O(1)$  \\
Plain~\cite{gbmp2014sea} & bvPL & $O\left(\frac{Wm}{B}\right)$ & $\frac{2Wm}{B}$ & $O\left(\frac{Wm}{B}\right)$ & $O(1)$ & $O(1)$ \\
Interleaved~\cite{gbmp2014sea} & bvIL & $O\left(\frac{Wm}{B} + \frac{Wm}{L}\right)$ & ${2Wm}\left(\frac{1}{B} + \frac{64}{L}\right)$ & $O(1)$ & $O\left(\log{\frac{Wm}{B}}\right)$ & $O\left(\log{\frac{Wm}{B}}\right)$ \\
%
%
Entropy based~\cite{DBLP:journals/corr/abs-0705-0552, Claude08practicalrankselect} & bvEN & \makecell[l]{$O\left(\frac{Wm}{B} \log{\frac{Wm}{B}}\right)$} & {\ssmall$\approx \log \dbinom{\frac{2Wm}{B}}{n}$} & $O(1)$ & $O\left(\log{\frac{Wm}{B}}\right)$ & $O\left(\log{\frac{Wm}{B}}\right)$ \\
Sparse~\cite{Okanohara07practicalentropycompressed} & bvSD & $O\left(n + n \log {\frac{Wm}{Bn}} \right)$ & {\ssmall$\approx n \left( 2 + \log \frac{2 W m}{B n} \right)$} & $O\left(\log{\frac{Wm}{Bn}}\right)$ & $O(1)$ & $O(1)$ \\
B-tree based~\cite{dynamic} & bvBT & $O\left(\frac{Wm}{B}\right)$ & $\approx 1.1 \cdot \frac{2Wm}{B}$ & $O(\log{n})$ & $O(\log{n})$ & $O(\log{n})$ \\
Gap-compressed~\cite{dynamic} & bvGC & $O\left( \frac{Wm}{B} \log{\frac{Wm}{Bn}} \right)$  & $\approx 1.3 \cdot \frac{2Wm}{B} \log{\frac{2Wm}{Bn}}$ & $O(\log{n})$ & $O(\log{n})$ & $O(\log{n})$ \\
\bottomrule
\end{tabular}
\caption{(\cref{sec:offset_arrays_types}) Theoretical analysis of various types of $\mathcal{O}$ and time complexity of associated queries.}
\label{tab:theory_bit_vectors}
\end{table*}

\subsection{Incorporating Succinct Bit Vectors}
\label{sec:offset_arrays_types}

To reduce the size of $\mathcal{O}$ and improve the performance of its $select$
query, we use \emph{succinct bit vectors}.

\subsubsection{Succinct Data Structures}

Assume $OPT$ is the optimal number of bits to store some data. A representation
of this data is \emph{succinct} if it uses $OPT + o(OPT)$ bits \emph{and} if it
supports \emph{a reasonable set of queries in (ideally) $O(1)$
time}~\cite{Blandford:2003:CRS:644108.644219}.
Thus, succinct designs differ from {compression} mechanisms such as zlib as
they do not entail expensive decompression.

\subsubsection{Succinct Bit Vectors: Preliminaries}

Succinct bit vectors use $\lceil\mathcal{Q}\rceil + o(\mathcal{Q})$ bits
assuming a storage lower bound of $\lceil\mathcal{Q}\rceil$ and they answer the $select$ query
in $o(\mathcal{Q})$ time. Many such designs
exist~\cite{DBLP:journals/corr/abs-0705-0552} and are widely used in
space-efficient trees and other schemes such as dictionaries.
The high-level idea behind their design is to divide the bit vector to be
encoded into \emph{small} parts (i.e., contiguous bit vector chunks of equal
sizes), group these chunks in an auxiliary table, and represent them with the
indices into this table~\cite{besta2018survey}. This table should contain
\emph{all} possible chunks so that \emph{any} bit vector could be constructed
from them. These small chunks are again divided into yet smaller (\emph{tiny})
chunks, stored similarly in other auxiliary tables. Now, the size of both
small and tiny chunks is selected in such a way that the sum of the sizes of
all the indices and all the auxiliary tables is $\lceil\mathcal{Q}\rceil +
o(\mathcal{Q})$.
The central observation that enables these bounds is that the bit vector
representation consisting of small and tiny chunks can be \emph{hierarchical}:
tiny chunks only need pointers
%

%
%

\subsubsection{Succinct Bit Vectors in Log(Graph)}

First, we use the \emph{entropy based} bit vector
(bvEN)~\cite{DBLP:journals/corr/abs-0705-0552}.
The key idea behind bvEN is to use a dictionary data
structure~\cite{Brodnik:1999:MCT:347566.347587} that achieves the lower bound
for storing bit vectors of length $2Wm/B$ with $n$ ones.
%
%
%
Second, we use \emph{sparse} succinct bit vectors
(bvSD)~\cite{Okanohara07practicalentropycompressed}. Here, positions of ones
are represented as a sequence of integers, which is then encoded using the
Elias-Fano scheme for non-decreasing sequences.  As bvSD specifically targets
sparse bit vectors, we expect it to be a good match for various graphs where $m
= O(n)$. 
Third, we investigate the \emph{B-tree based} bit vector (bvBT)~\cite{dynamic}.
This data structure supports inserts, making the bit vector dynamic. It is
implemented with B-trees where leaves contain the actual bit vector data while
internal nodes contain meta data for more performance. 
Finally, the \emph{gap-compressed} (bvGC) dynamic variant is
incorporated~\cite{dynamic} that along with bvSD also compresses sequences of
zeros.

\subsection{Theoretical Storage \& Time Analysis}
\label{sec:O_ANALYSIS}

We now analyze the storage/time complexity of the described offset structures
in Table~\ref{tab:theory_bit_vectors}.  For completeness, we present the
asymptotic and the exact size as well the time to derive $\mathcal{O}_v$.
%
%
Now, \textsf{ptrW} (array of offsets) together with \textsf{bvPL} and
\textsf{bvSD} feature the fastest $\mathcal{O}_v$; we select them as the most
promising candidates for $\mathcal{O}$ in Log(Graph).


\begin{figure*}
 \centering
 \includegraphics[width=1.0\textwidth]{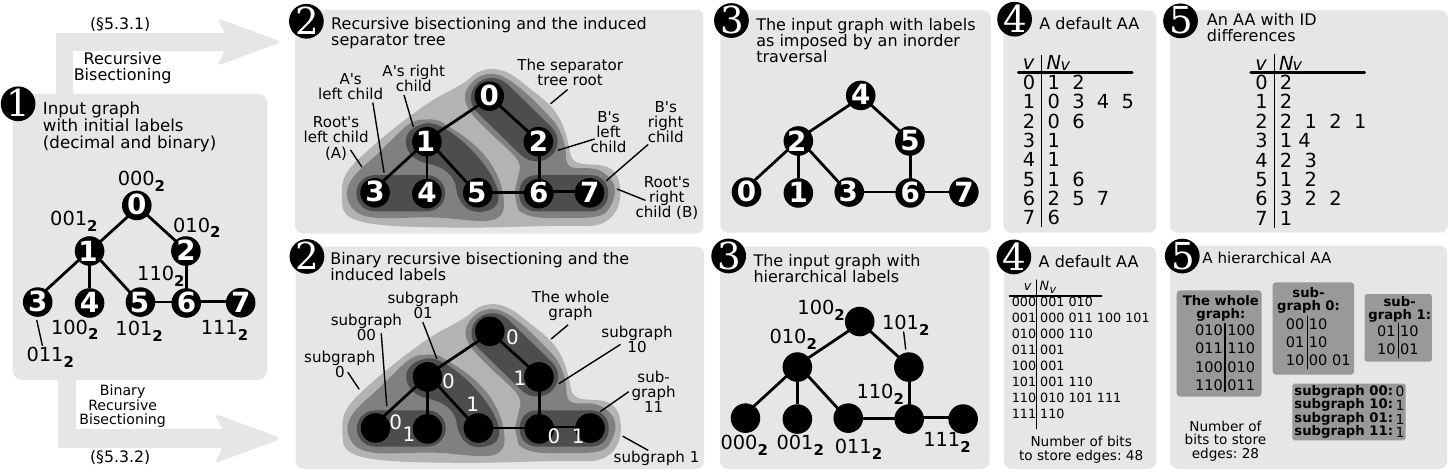}
 \caption{An example graph
 representation logarithmized with Recursive Bisectioning (RB~\cite{Blandford:2003:CRS:644108.644219},
 \cref{sec:bf_scheme}), 
and Binary Recursive Bisectioning (BRB, \cref{sec:br_scheme}).}
 \label{fig:example_blandford}
\end{figure*}


\subsection{High-Performance Implementation}

For high performance, we use the sdsl-lite library~\cite{gog2014optimized} that
provides fast codes of various succinct and compact bit vectors.  Yet, it is
fully sequential and oblivious to the utilized workload. Thus, we evaluate its
performance tradeoffs (\cref{sec:evaluation}) and identify the best designs for
respective graph families, illustrating that the empirical results follow the
theoretical analysis from~\cref{sec:O_ANALYSIS}.

\section{LOGARITHMIZING ADJACENCY DATA}
\label{sec:compress_adj_lists}

In this section, we logarithmize the adjacency data $\mathcal{A}$.
$\mathcal{A}$ is usually more complex than $\mathcal{O}$ as it encodes the whole structure of a graph.
To facilitate  
logarithmizing $\mathcal{A}$,
we first develop a formal model
and we show that various past schemes are its special cases. We illustrate that these schemes entail inherent
performance or storage issues and we then propose novel schemes to overcome
these problems.

Similarly to compressing fine elements and offsets, the schemes from 
this section provide low-overhead decompression combined with
large storage reductions, enabling high-performance graph processing running
over compressed graphs.
The difference is that \emph{the main focus is on reducing storage overheads with performance
being the secondary priority}, while compressing fine elements 
comes with \emph{reverse} objectives.
Selecting the most appropriate set of schemes depends on the specific requirements of the user of
Log(Graph). 
%
%
%

\subsection{A Model for Logarithmizing $\mathcal{A}$}
\label{sec:compression_model}

%
Log(Graph) comes with many compression schemes for
$\mathcal{A}$ that target various classes of graphs. 
We define any such scheme to
be a tuple $(\mathcal{P},\mathcal{T})$.
$\mathcal{P}$ is the \emph{permuter}: a function that relabels the vertices.
We introduce $\mathcal{P}$ to explicitly capture the notion that appropriate
labeling of vertices significantly reduces $|\mathcal{A}|$.  We have
$\mathcal{P}: V \to \mathbb{N}$ such that (the condition enforces the
uniqueness of IDs)):

\begin{alignat}{2}
\forall_{v,u \in V}\ \left(v \neq u \right) \Rightarrow \left[\mathcal{P}(v) \neq \mathcal{P}(u) \right]\label{eq:permuter_cond} 
\end{alignat}
\normalsize

Next, $\mathcal{T} = \{\mathcal{T}_{x}\ |\ x \in \mathbb{N}\}$ is a \emph{set}
of \emph{transformers}: functions that map sequences of vertex labels into
sequences of bits:

\begin{alignat}{2}
\nonumber\mathcal{T}_{x}: \overbrace{V\times ...  \times V}^{\text{$x$ times}} \to \{0,1\} \times ... \times \{0,1\} 
\end{alignat}
\normalsize

\noindent We introduce $\mathcal{T}$ to enable arbitrary operations on
sequences of relabeled vertices, for example be the Varint
encoding~\cite{dean2009challenges}.

\subsection{Understanding Storage Lower Bounds}
\label{sec:A_LB}

$\mathcal{A}$ is determined by the corresponding $G$ and thus a simple storage
lower bound is determined by the number of graphs with $n$ vertices and $m$
edges and equals {\footnotesize$\left\lceil \log \dbinom{\binom{n}{2}}{m} \right\rceil$}
(Table~\ref{tab:bounds}).
Now, today's graph codes already approach this bound. For example, the Graph500
benchmark~\cite{murphy2010introducing} requires $\approx$1,126 TB for a graph
with $2^{42}$ vertices and $2^{46}$ edges while the corresponding lower bound
is merely $\approx$350 TB.
%
%
We thus propose to assume more about $G$'s structure on top of the number of
vertices and edges. We now target \emph{separable} graphs (cf.~Section~\ref{sec:bck}).

\subsection{Incorporating Compactness}
\label{sec:adj_existing}



We use compact graph representations that take $O(n)$ bits to encode graphs.
The main technique that ensures compactness that we incorporate is
\emph{recursive bisectioning}.  We first describe an existing recursive
bisectioning scheme (\cref{sec:bf_scheme}) and then enhance it for more
performance (\cref{sec:br_scheme}).

\subsubsection{Recursive Bisectioning (RB)}
\label{sec:bf_scheme}

Here, we first illustrate a representation introduced by Blandford et
al.~\cite{Blandford:2003:CRS:644108.644219} (referred to as the \emph{RB}
scheme) that requires $O(n)$ bits to store a graph that is \emph{separable}.
Figure~\ref{fig:example_blandford} contains an example.
The basic method is to relabel vertices of a given graph $G$ to minimize
differences between the labels of consecutive neighbors of each vertex $v$ in
each adjacency list.
Then, the differences are recorded with
any variable-length gap encoding scheme 
such as Varint~\cite{dean2009challenges}.
%
%
Assuming that the new labels of $v$'s neighbors do not differ significantly,
the encoded gaps use less space than the IDs~\cite{Blandford:2003:CRS:644108.644219}.
Now, to reassign labels in such a way that the storage is reduced, the graph
(see~\encircle{1} in Figure~\ref{fig:example_blandford} for an example) is
bisected recursively until the size of a partition is one vertex (for edge
cuts) or a pair of connected vertices (for vertex cuts); in the example we
focus on edge cuts. Respective partitions form a binary \emph{separator tree}
with the leaves being single vertices~\encircle{2}. Then, the vertices are
relabeled as imposed by an inorder traversal over the \emph{leaves} of the
separator tree~\encircle{3}. The first leaf visited gets the lowest label
(e.g., 0); the label being assigned is incremented for each new visited leaf.
This minimizes the differences between the labels of the neighboring vertices
(the leaves corresponding to the neighboring vertices are close to one another
in the separator tree), reducing \textsf{AA}'s size~\encircle{4},~\encircle{5}.

\macb{Using Permuters and Transformers}
One can easily express RB using $\mathcal{P}$ and $\mathcal{T}$. First,
$\mathcal{P}$ relabels the vertices according to the order in which they appear
as leaves in the inorder traversal of the separator tree obtained after
recursive graph bipartitioning.  Here, we partition graphs to make subgraphs
[almost] equal (<0.1\% of difference) in size.
Second, each transformer $\mathcal{T} = \{\mathcal{T}_{v}(v, N_v)\}$
takes as input $v$ and $N_v$.  It then encodes the differences between
consecutive vertex labels using Varint. The respective differences are:
$|N_{1,v} - N_{0,v}|$, $N_{2,v} - N_{1,v}$, ..., and $N_{d_v,v} -
N_{d_v-1,v}$.

\goal{Say what is bad in RB and motivate our schemes}
\macb{Problems}
RB suffers from very expensive preprocessing, as we illustrate
later in~\cref{sec:evaluation} (Table~\ref{tab:rb-sucks}). Generation of
\textsf{RB} usually takes more than 20x longer than that of \textsf{AA}.



\subsubsection{Binary Recursive Bisectioning (BRB)}
\label{sec:br_scheme}

The core idea is to relabel vertices so that vertices in clusters have large
common prefixes (clusters are identified during partitioning). One prefix is
stored only once per each cluster.

\macb{What Does It Fix?}
BRB alleviates two issues inherent to RB.
First, there is no costly inorder traversal over the separator tree. 
More importantly, there is no expensive derivation of the full separator tree. 
Instead, one sets the number of partitioning levels \emph{upfront to control
the preprocessing overhead}.

\macb{Permuter (Relabeling Vertices)}
We present an example in Figure~\ref{fig:example_blandford}.
First, we recursively bipartition the input graph $G$ to identify common
prefixes and uniquely relabel the vertices.  After the first partitioning, we
label an arbitrarily selected subgraph as 0 and the other as 1, we denote these
subgraphs as $G_0$ and $G_1$, respectively.  We then apply this step
recursively to each subgraph for the specified number of steps or until the
size of each partition is one (i.e., each partition contains only one vertex).
$G_{0}$ would be partitioned into subgraphs $G_{00}, G_{01}$ with labels 00 and
01 (we refer to a partition with label $X$ as $G_{X}$). Eventually, each vertex
obtains a unique label in the form of a binary string; each bit of this label
identifies each partition that the vertex belongs to.

\macb{Transformer (Encoding Edges)}
Here, the idea is to group edges within each subgraph derived in the process of
hierarchical vertex labeling. Several leading bits are identical in each label
and are stripped off, decreasing
$|\mathcal{A}|$. 
To make such a hierarchical adjacency list decodable, we store 
(for each $v$) such labels of $v$'s neighbors from the
same subgraph contiguously in memory, together with
the common associated prefix and the neighbor count. 

\subsection{Incorporating Integer Linear Programming}

We next logarithmize $G$ with ILP to target generic graphs and not just the
ones that are separable. We first illustrate a simple existing scheme that uses
ILP for graph storage reductions (\cref{sec:odb_scheme}) and then 
accelerate it (\cref{sec:pod_scheme}). 

\subsubsection{Optimal Difference-Based (ODB)}
\label{sec:odb_scheme}

There are several variants of ILP-based schemes~\cite{diaz2002survey} where the
objective function minimizes: the sum of differences between consecutive
neighbors in adjacency lists (minimum gap arrangement (MGapA)), the sum of
logarithms of differences from MGapA (minimum logarithmic gap arrangement
(MLogGapA)), the sum of differences of each pair of neighbors (minimum linear
arrangement (MLinA)), and the sum of logarithms of differences from MLinA
(minimum logarithmic arrangement (MLogA)).

\macb{Using Permuters and Transformers}
Now, ODB's $\mathcal{T}$ is identical to that of RB as it encodes ID 
differences while $\mathcal{P}$ determines the relabeling obtained by solving
a respective ILP problem.
Consider a vector $\mathbf{v} = (\mathcal{P}(v_1), ..., \mathcal{P}(v_n))^T$
that models new vertex labels (where $v_1, ..., v_n \in V$). 
For example, the MGapA and MLogA objective functions are respectively

\begin{alignat}{2}
& \min_{\mathcal{P}(v), \forall v \in V}\ \ \sum_{v \in V} \sum_{i = 0}^{|N_v|-1} |\mathcal{P}(N_{i+1,v}) - \mathcal{P}(N_{i,v})| \label{eq:obj_fun_1} 
\end{alignat}
\normalsize
%
%

\begin{alignat}{2}
& \min_{\mathcal{P}(v), \forall v \in V}\ \ \sum_{v \in V} \sum_{u \in N_v} \log|\mathcal{P}(v) - \mathcal{P}(u)| \label{eq:obj_fun_2} 
\end{alignat}
\normalsize
Both functions use the uniqueness Constraint~(\ref{eq:permuter_cond}).
%

\macb{Problems}
All of the above schemes except MLogGapA were proved to be
NP-hard~\cite{chierichetti2009compressing} and do not scale with $n$. MLogGapA
is still an open problem.

\subsubsection{Positive Optimal Differences (POD)}
\label{sec:pod_scheme}

We now enhance the ODB MGapA
(\cref{sec:odb_scheme}) by removing the absolute value $|\cdot|$
from the objective function, which accelerates relabeling.
Yet, this requires additional constraints to enforce that the
neighbors of each vertex are sorted according to their IDs.
We present the constraints below; readers who are not interested in
the mathematical details may proceed to~\cref{sec:hybridization}.

\begin{alignat}{2}
\forall_{v \in V} \forall_{i,j \in \{1..d_v\}}  \left[\mathcal{P}\left(N^I_{j,v}\right) + (x_{vij} - 1) \cdot n  \leq N_{i,v} \right] \label{eq:pod_1}
\end{alignat}
\begin{alignat}{2}
\forall_{v \in V} \forall_{i,j \in \{1..d_v\}} \left[\mathcal{P}\left(N^I_{j,v}\right) + (1 - x_{vij}) \cdot n  \geq N_{i,v} \right] \label{eq:pod_2}
\end{alignat}

\begin{alignat}{2}
\forall_{v \in V} \forall_{i \in \{1..d_v\}} \left[\sum_{j=1}^{d_v} x_{vij} = 1 \right] \label{eq:pod_3}
\end{alignat}

\begin{alignat}{2}
\forall_{v \in V} \forall_{i \in \{1..d_v\}} \left[ N_{i,v} < N_{i+1,v} \right] \label{eq:pod_4}
\end{alignat}
\normalsize

$N^I_v$ is the initial labeling of $N_v$, 
$x_{vij}$ is a boolean variable that determines if neighbor~$j
\in N^I_v$ must be $i$th neighbor in $N_v$ according to relabeling
$\mathcal{P}$. 
If $x_{vij}=1$, constraints~(\ref{eq:pod_1}), (\ref{eq:pod_2}) use
{\footnotesize$N_{i,v} = \mathcal{P}\left(N^I_{j,v}\right)$}, otherwise they
are trivially satisfied.  Constraint~(\ref{eq:pod_3}) selects each neighbor
once. Finally, constraint~(\ref{eq:pod_4}) sorts $N_v$ in the increasing label
order.

\subsection{Combining Compactness and ILP (CMB)}
\label{sec:hybridization}

\begin{figure}[h!]
\centering
  \includegraphics[width=0.6\columnwidth]{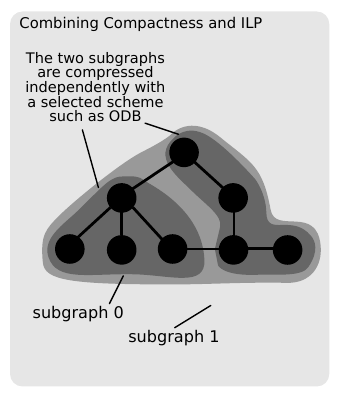}
%
\caption{An example of Hybridization (\cref{sec:hybridization}).}
%
\label{fig:example_CMB}
\end{figure}

Finally, we design combining (CMB) schemes that use the compact recursive
partitioning approach to enhance ODB and others.
The core idea is to first bisect the graph $k$ times (within the given time
constraints), and then encode independently each subgraph (cluster) with a
selected ILP scheme.
We illustrate an example of this scheme in Figure~\ref{fig:example_CMB}.

\macb{What Does It Fix?}
First, the initial partitioning does not dominate the total runtime. Second,
the NP-hardness of ODB is alleviated as it now runs on subgraphs that are $k$
times smaller than the initial graph.
Finally, it is generic and one can use an arbitrary scheme instead of ODB.

\macb{Using Permuters and Transformers}
The exact design of $\mathcal{P}$ and $\mathcal{T}$ depend on the scheme used
for condensing subgraphs. 
For example, consider ODB. The most significant $k$ bits are now
determined by $G$'s partitioning. The remaining bits are derived from
ODB independently for each subgraph. Their combination gives each final label.
$\mathcal{T}$ can be, e.g., Varint.

\subsection{Incorporating Degree-Minimizing (DM)}
\label{sec:dm_scheme}

The final step is to relabel vertices so that those with the highest degrees (and
thus occurring more often in $\mathcal{A}$) receive the smallest labels. Then,
in one scheme variant (\textsf{DMf}, proposed in the past~\cite{adler2001towards}), full labels are encoded using Varint
(``f'' stands for full). In another variant (\textsf{DMd}, offered in this work), labels are encoded
as differences (``d'' indicates differences), similarly to \textsf{RB}. Thus,
$|\mathcal{A}|$ is decreased as the edges that occur most often are stored
using fewer bits.

\macb{What Does It Fix?}
First, DM trades some space reductions for faster accesses to $\mathcal{A}$
compared to BRB (BRB's
hierarchical encoding entails expensive queries).
Second, it does not require costly recursive partitioning. 
Finally, we later (\cref{sec:eval_log_A}) show that \textsf{DMd} significantly outperforms \textsf{DMf}
and matches the compression ratios of the WebGraph library~\cite{BoVWFI}.

\macb{Permuter/Transformer}
DM's $\mathcal{T}$ is identical to that of RB. DM's $\mathcal{P}$ differs 
as the relabeling is now purely guided by vertex degrees:
higher $d_v$ enforces lower $v$'s label.

\section{HIGH-PERFORMANCE LIBRARY}
\label{sec:lib_design}

Past sections (\cref{sec:log_fines}--\cref{sec:compress_adj_lists}) illustrate
a plethora of logarithmization schemes and enhancements for various graph
families and scenarios. This large number poses design challenges. We now
present the Log(Graph) C++ library that ensures: (1) a straightforward
development, analysis, and comparison of graph representations composed of any
of the proposed schemes, and (2) high-performance. The implementation of the Log(Graph) library is
available online\footnote{\footnotesize
https://spcl.inf.ethz.ch/Research/Performance/LogGraph}.


\subsection{Modular Design and Extensibility}

Any graph compressed with Log(Graph) can be
represented as a tuple $(G,\allowbreak \mathcal{O},\allowbreak
\mathcal{A},\allowbreak \mathcal{L}[\mathcal{O}],\allowbreak \mathcal{L}[\mathcal{A}])$.
These tuple elements corresponds to \emph{modules} in the Log(Graph) library.
These modules gather variants of $\mathcal{O}$,
$\mathcal{A}$, and two logarithmization schemes, $\mathcal{L}[\mathcal{O}]$ and $\mathcal{L}[\mathcal{A}]$,
that act upon $\mathcal{O}$ and
$\mathcal{A}$.
The $\mathcal{L}[\mathcal{A}]$
module contains submodules for $\mathcal{P}$
and $\mathcal{T}$. This enables us to seamlessly implement, analyze, and
compare the described Log(Graph) variants.




\subsection{High Performance}

Combinations of the variants of $\mathcal{O}$, $\mathcal{L}[\mathcal{O}]$,
$\mathcal{A}$, and $\mathcal{L}[\mathcal{A}] = (\mathcal{P}, \mathcal{T})$ give many possible
designs. For example, $\mathcal{O}$ can be any succinct bit
vector.
Now, selecting a specific variant takes place in a
performance-critical region such as querying {\small$\rwh{d}$}.
%
%
%
We identify four C++ mechanisms for such selections: \#if pragmas, virtual
functions, runtime branches, and templates.  The first one results in
unmanageably complex code. The next two entail 
performance overheads. We thus use templates to reduce code complexity
while retaining high performance. Listing~\ref{lst:templates_construct}
illustrates: a generic \texttt{LogGraph} template class for defining different
Log(Graph) representations, the constructor of a
Log(Graph) representation, and a function $N_v$ for accessing neighbors of a given vertex~$v$. 
\texttt{LogGraph} only requires defining the following \emph{types}: the offset structure
($\mathcal{O}$) type, the offset compression structure ($\mathcal{L}[\mathcal{O}]$) type, and the
transformer ($\mathcal{T}$) type.

\begin{lstlisting}[aboveskip=0.0em, abovecaptionskip=0.0em, belowskip=0em, float=h, xleftmargin=1em,
caption=(\cref{sec:lib_design}) A graph representation from the Log(Graph)
library., label=lst:templates_construct, mathescape=true]
template<typename $\mathcal{O}$, typename $\mathcal{L}[\mathcal{O}]$, typename $\mathcal{T}$>
class LogGraph : public BaseLogGraph { //Class template.
 $\mathcal{O}$* offsets; //Opaque structure for storing offsets
 $\mathcal{L}[\mathcal{O}]$* compressor; //Logarithmization scheme for offsets
 $\mathcal{T}$* transformer; //Scheme acting upon permuted vertex IDs
};

template<typename $\mathcal{O}$, typename $\mathcal{L}[\mathcal{O}]$, typename $\mathcal{T}$> // Constructor.
LogGraph<$\mathcal{O}$, $\mathcal{L}[\mathcal{O}]$, $\mathcal{T}$>::LogGraph(Permutation $\mathcal{P}$, AA* aa) {
 //aa is an instance of an adjacency array.
 aa->permute($\mathcal{P}$); //Relabel vertices. Note that $\mathcal{P}$ isn't a type
 transformer = new $\mathcal{T}$(); //Create a new transformer object.
 transformer->transform(&aa); //Modify vertices after relabeling
 offsets = new $\mathcal{O}$(aa); //Create a new object for offsets.
 compressor = new $\mathcal{L}[\mathcal{O}]$(); 
 compressor->compress(&offsets); //Logarithmize offset structure.
}

template<typename $\mathcal{O}$, typename $\mathcal{L}[\mathcal{O}]$, typename $\mathcal{T}$> 
//v_id is an opaque type for a vertex ID.
v_id* LogGraph<$\mathcal{O}$, $\mathcal{L}[\mathcal{O}]$, $\mathcal{T}$>::getNeighbors(v_id v) {//Resolve $N_{v}$.
 v_id offset = offsets->getOffset(v);
 v_id* neighbors = tr->decodeNeighbors(v, offset);
 return neighbors; 
}
\end{lstlisting}

%
%
%

Note that the permuter $\mathcal{P}$ is an \emph{object}, not \emph{type}.  As
relabeling is executed only during preprocessing, it does not impact
time-critical functions. Thus, selecting a given permutation can be done
with simple branches based on the value of $\mathcal{P}$.

\section{EVALUATION}
\label{sec:evaluation}


We now illustrate that Log(Graph) offers the \emph{sweetspot between low-overhead decompression and
high compression ratios}, \emph{enabling high-performance graph processing
on top of compressed datasets}.


\subsection{Evaluation Methodology}

We first describe evaluation methodology.

\subsubsection{Goals}

We illustrate that the Log(Graph)
schemes offer (1) storage reductions, in many cases comparable to those of the
state-of-the-art graph compression schemes, and (2) low-overhead decompression
that enables high-performance graph processing running over compressed graph
datasets. 

\subsubsection{Considered Algorithms}

We consider the following algorithms included in the GAP Benchmark
Suite~\cite{beamer2015gap}: Breadth-First Search (BFS), PageRank (PR), Single
Source Shortest Paths (SSSP), Betweenness Centrality (BC), Connected Components
(CC), and Triangle Counting (TC).
BFS, SSSP, and BC represent various types of traversals. PR is an iterative
scheme where all the vertices are accessed in each iteration. CC represents
protocols based on pointer-chasing. Finally, TC stands for non-iterative
compute-intensive tasks.
  
\begin{itemize}[leftmargin=1.0em, noitemsep]
\item \textbf{\textsf{BFS:}} A state-of-the-art variant with
direction-optimization and other enhancements that
reduce data transfer~\cite{beamer2013direction, beamer2015gap}.
\item \textbf{\textsf{SSSP:}} An optimized $\Delta$-Stepping algorithm~\cite{meyer2003delta,
madduri2007experimental, beamer2015gap}.
\item \textbf{\textsf{CC:}} A variant of the Shiloach-Vishkin
scheme~\cite{shiloach1982logn, bader2005architectural}.
\item \textbf{\textsf{BC:}} An enhanced Brandes' scheme~\cite{brandes2001faster,
madduri2009faster, beamer2015gap}.
\item \textbf{\textsf{PR:}} A variant without atomic operations~\cite{beamer2015gap, schweizer2015evaluating, schmid2016high}.
\item \textbf{\textsf{TC:}} An optimized algorithm that reduces the computational complexity by
preprocessing the input graph~\cite{chu2011triangle}.
\end{itemize}



\begin{figure*}
\centering
  \begin{subfigure}[t]{0.2 \textwidth}
    \centering
    \includegraphics[width=\textwidth]{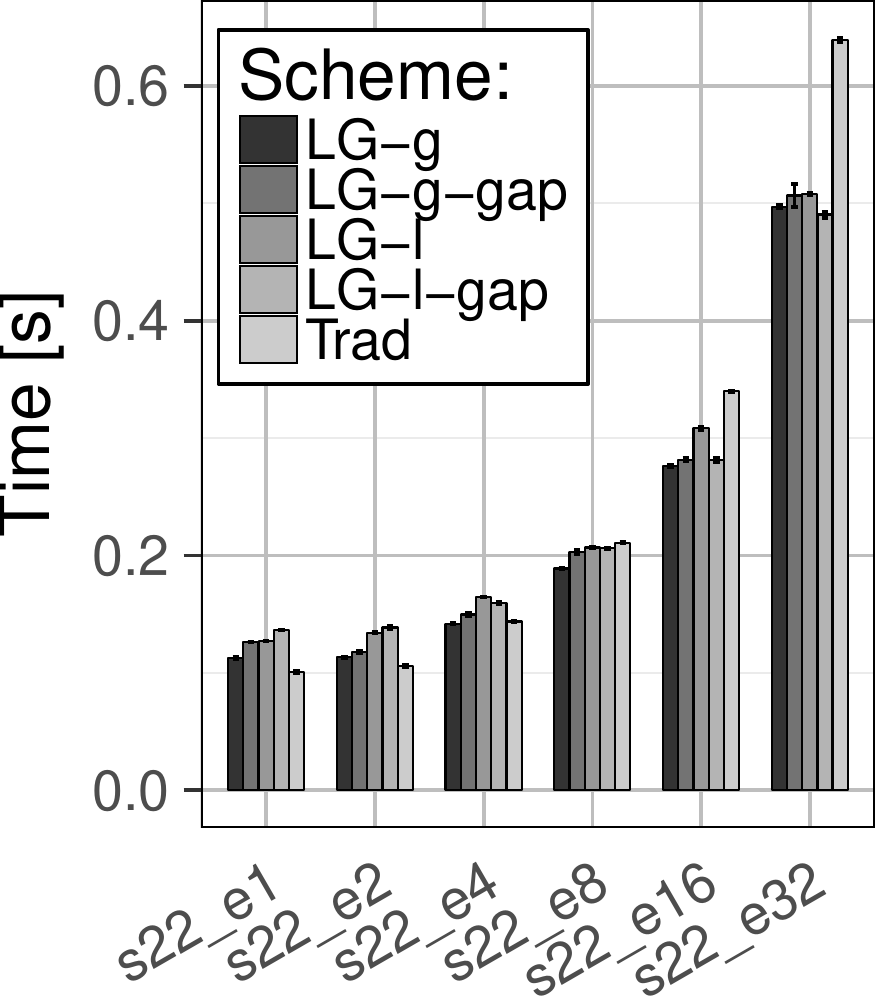}
    \caption{PR, sparse graphs.}
    \label{fig:x}
  \end{subfigure}
  \begin{subfigure}[t]{0.2 \textwidth}
    \centering
    \includegraphics[width=\textwidth]{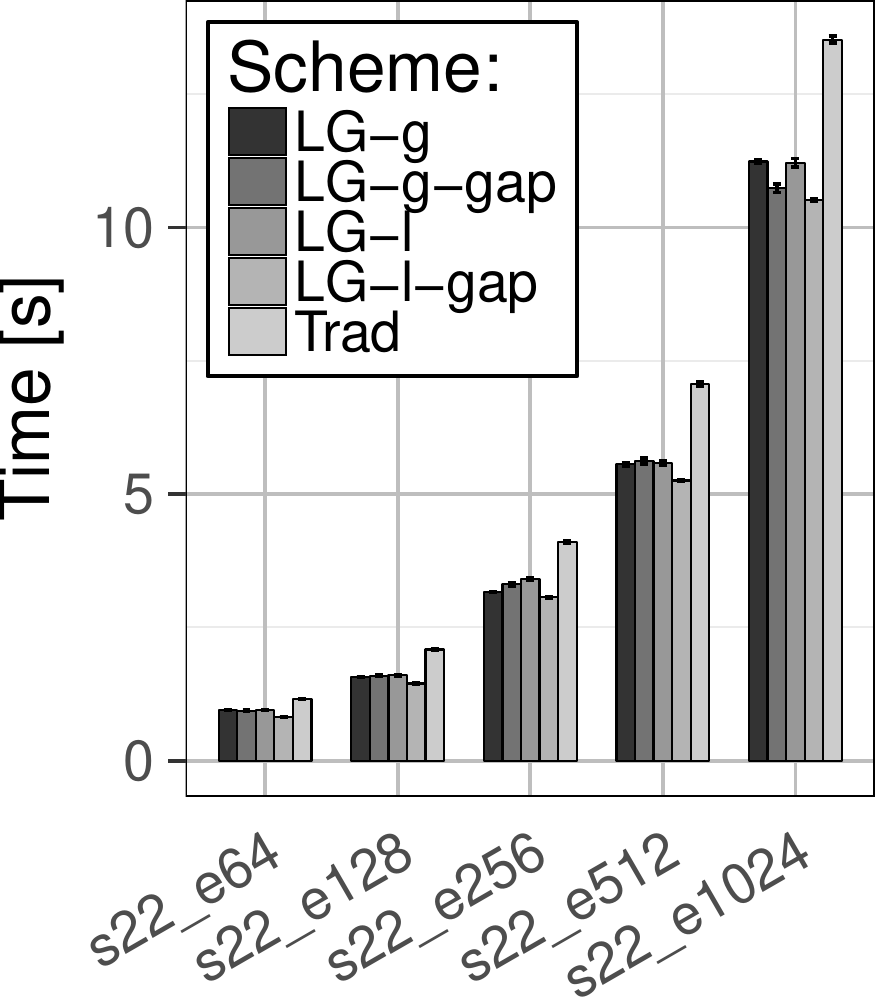}
    \caption{PR, dense graphs.}
    \label{fig:x}
  \end{subfigure}
  \begin{subfigure}[t]{0.2 \textwidth}
    \centering
    \includegraphics[width=\textwidth]{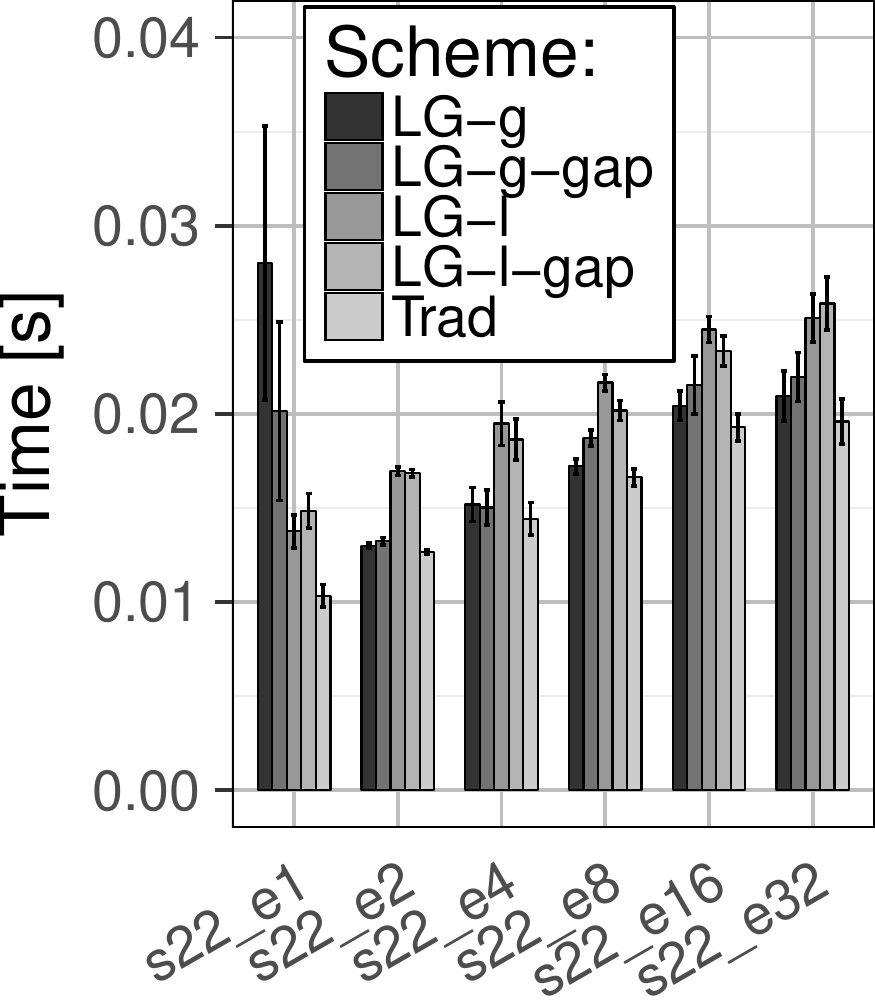}
    \caption{BFS, sparse graphs.}
    \label{fig:x}
  \end{subfigure}
\\
  \begin{subfigure}[t]{0.2 \textwidth}
    \centering
    \includegraphics[width=\textwidth]{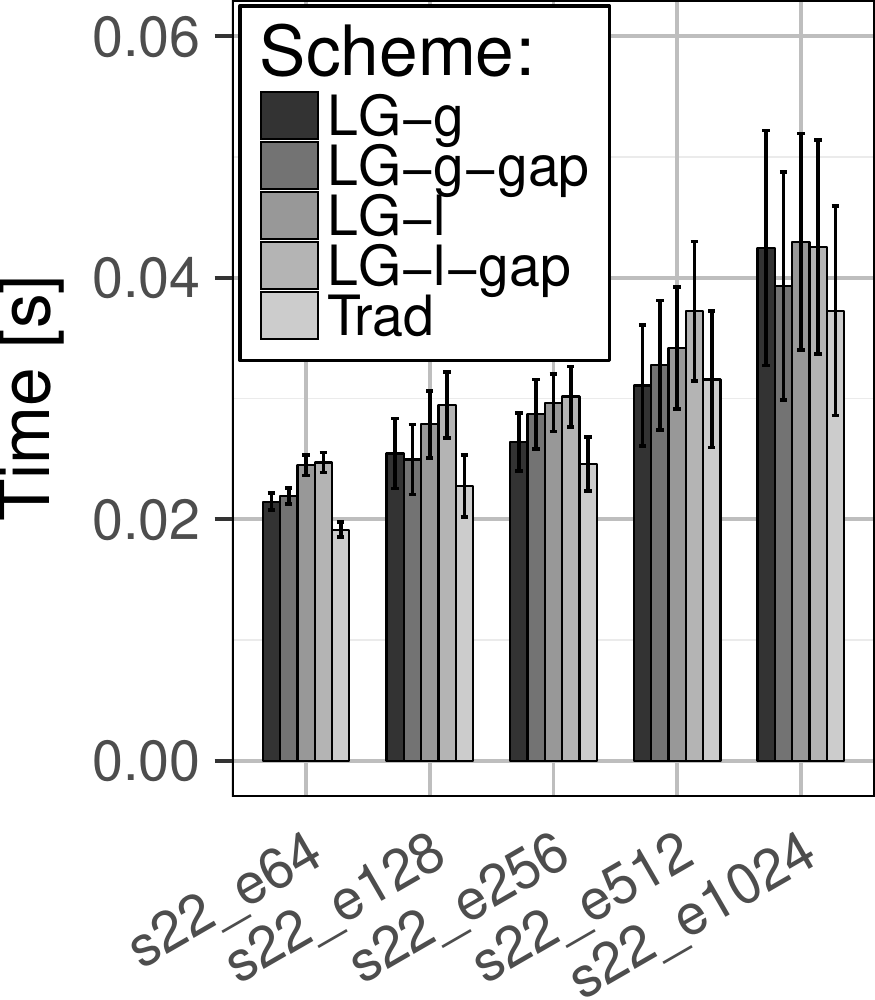}
    \caption{BFS, dense graphs.}
    \label{fig:x}
  \end{subfigure}
  \begin{subfigure}[t]{0.2 \textwidth}
    \centering
    \includegraphics[width=\textwidth]{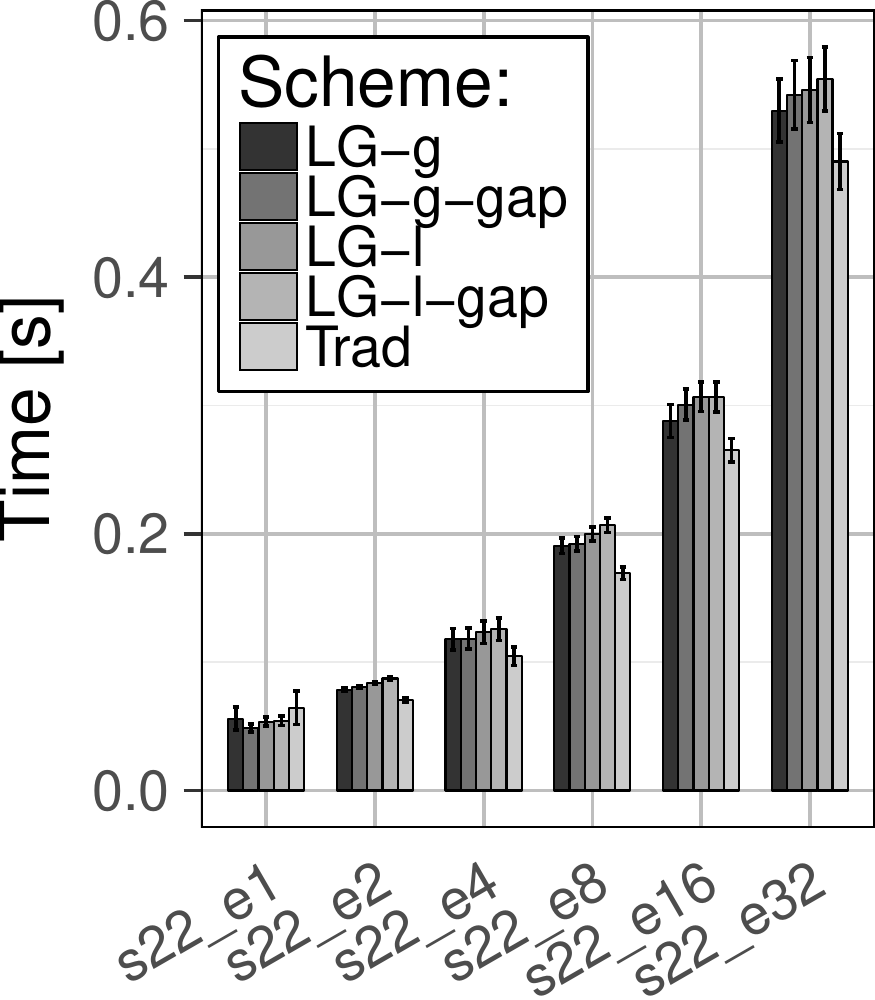}
    \caption{BC, sparse graphs.}
    \label{fig:x}
  \end{subfigure}
  \begin{subfigure}[t]{0.2 \textwidth}
    \centering
    \includegraphics[width=\textwidth]{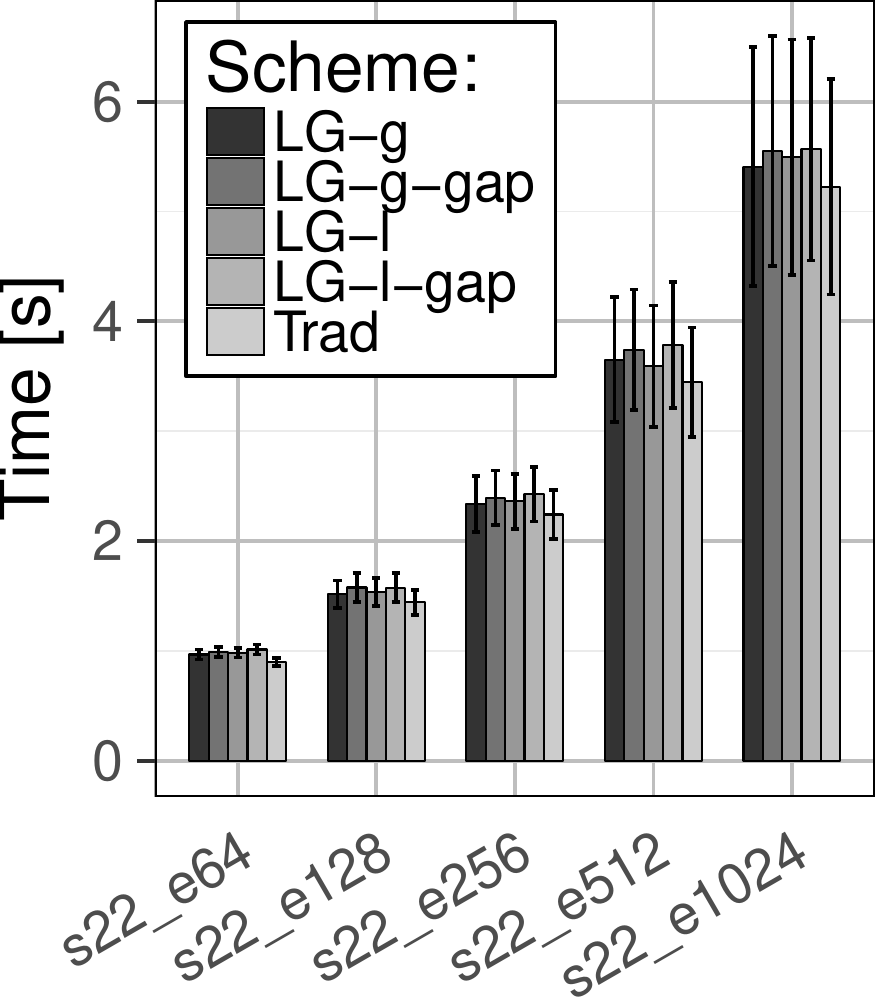}
    \caption{BC, dense graphs.}
    \label{fig:x}
  \end{subfigure}
\\
  \begin{subfigure}[t]{0.2 \textwidth}
    \centering
    \includegraphics[width=\textwidth]{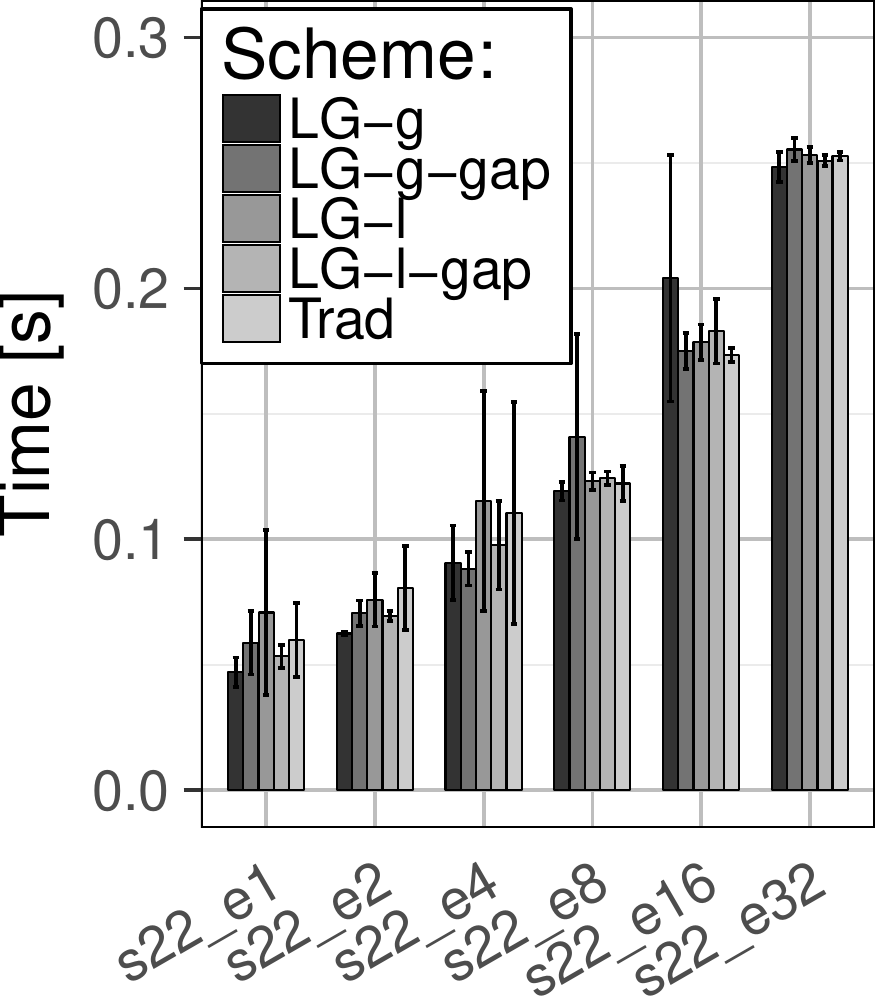}
    \caption{SSSP, sparse graphs.}
    \label{fig:x}
  \end{subfigure}
  \begin{subfigure}[t]{0.2 \textwidth}
    \centering
    \includegraphics[width=\textwidth]{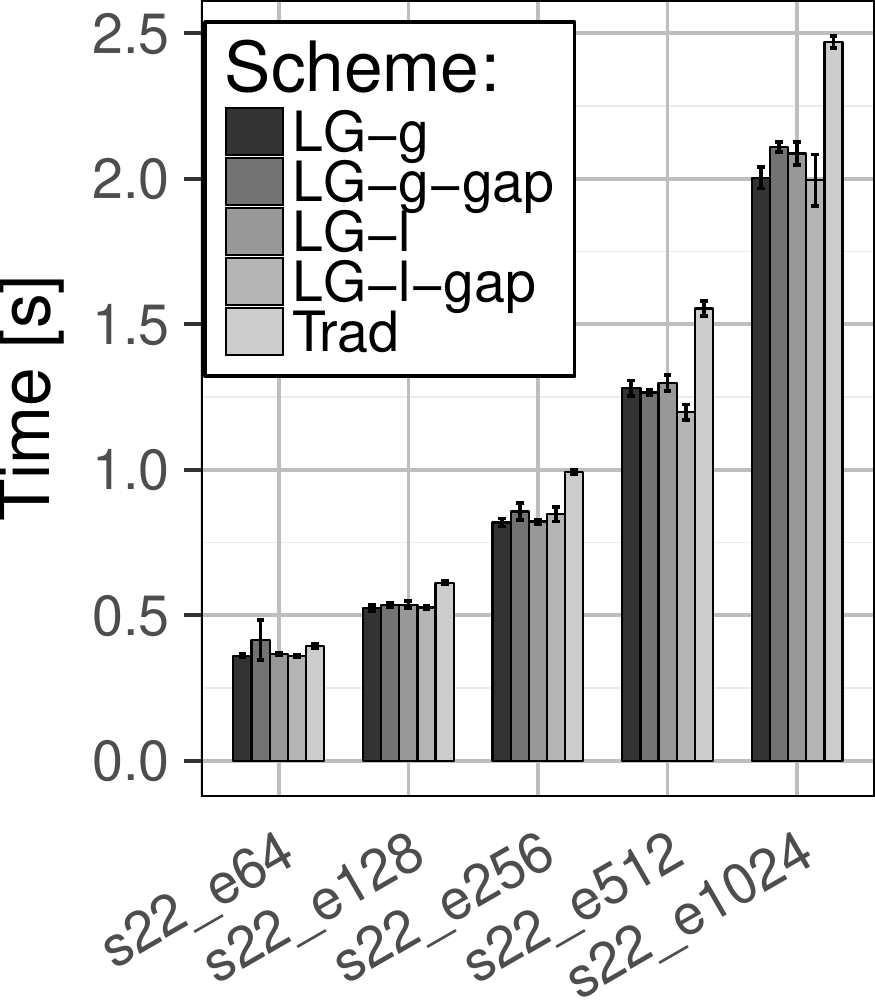}
    \caption{SSSP, dense graphs.}
    \label{fig:x}
  \end{subfigure}
  \begin{subfigure}[t]{0.2 \textwidth}
    \centering
    \includegraphics[width=\textwidth]{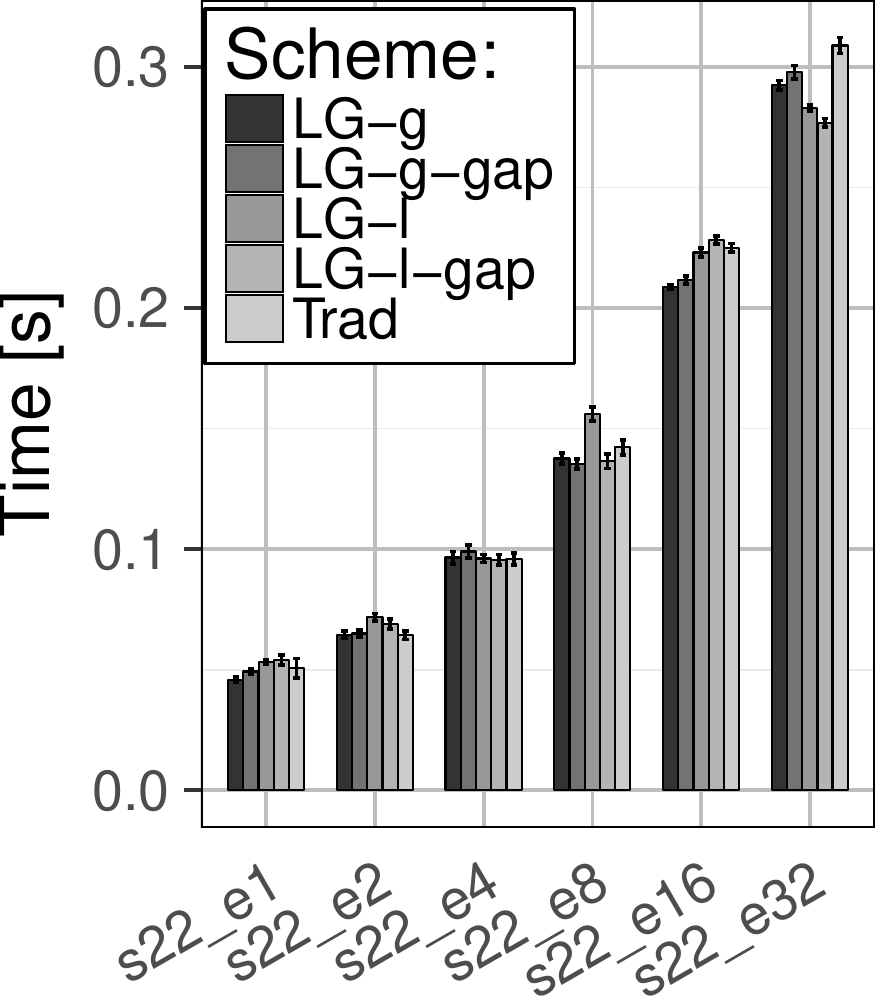}
    \caption{CC, sparse graphs.}
    \label{fig:x}
  \end{subfigure}
\\
  \begin{subfigure}[t]{0.2 \textwidth}
    \centering
    \includegraphics[width=\textwidth]{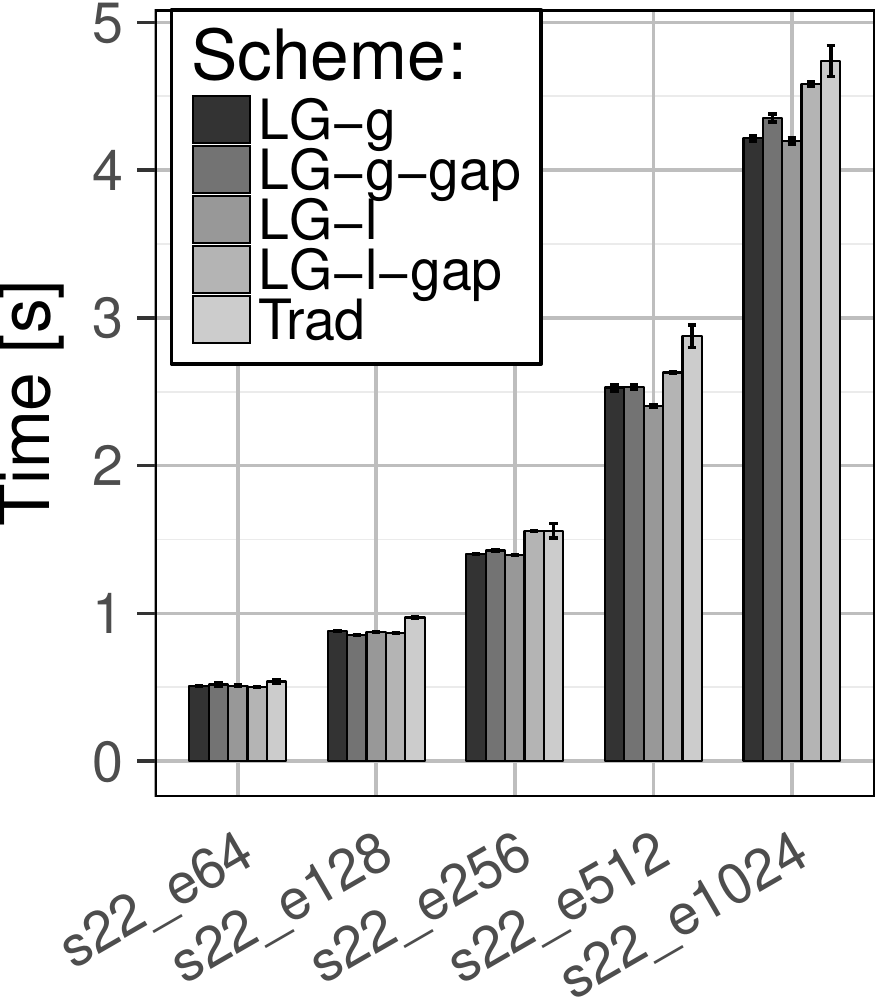}
    \caption{CC, dense graphs.}
    \label{fig:x}
  \end{subfigure}
  \begin{subfigure}[t]{0.2 \textwidth}
    \centering
    \includegraphics[width=\textwidth]{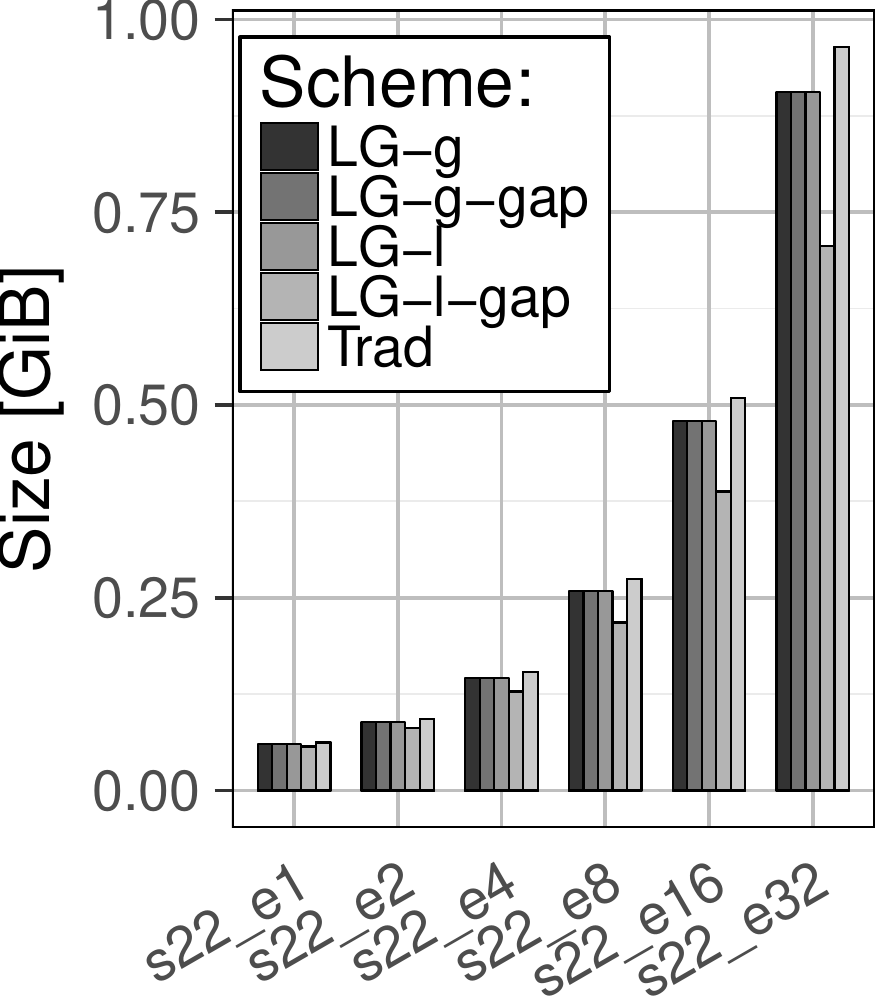}
    \caption{Size, sparse graphs.}
    \label{fig:x}
  \end{subfigure}
  \begin{subfigure}[t]{0.2 \textwidth}
    \centering
    \includegraphics[width=\textwidth]{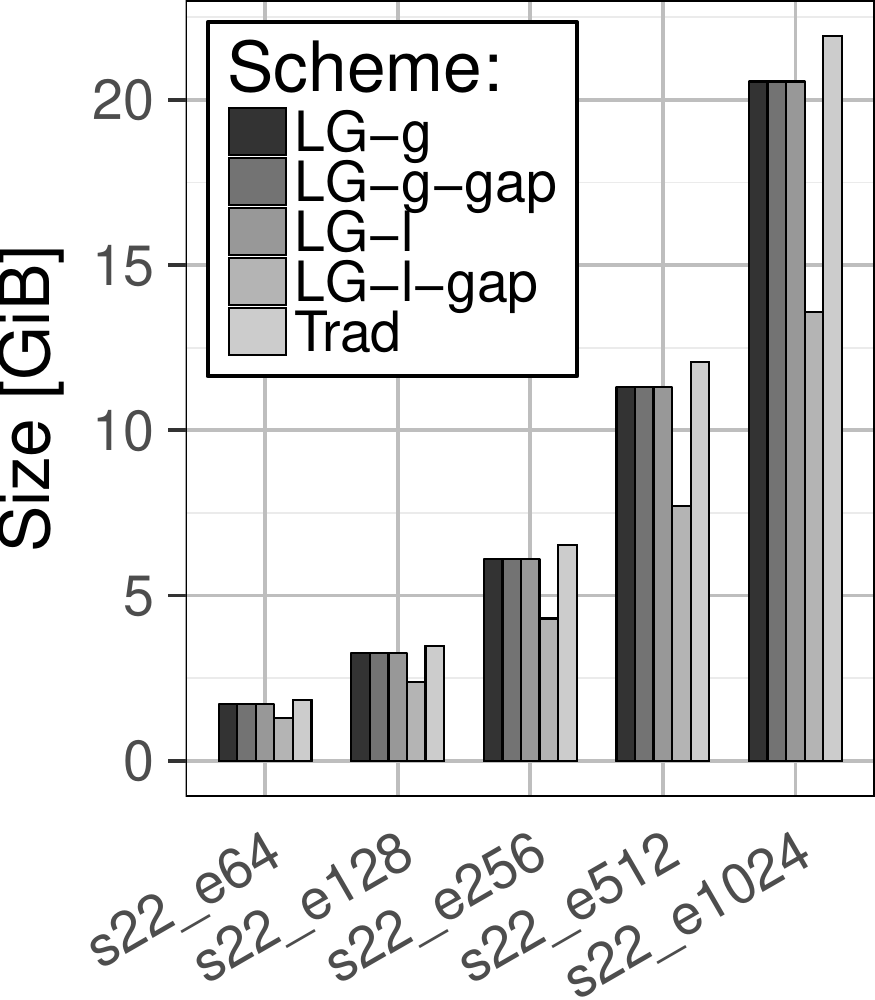}
    \caption{Size, dense graphs.}
    \label{fig:x}
  \end{subfigure}
%
\caption{(\cref{sec:eval_log_fine}) Log(Graph) performance analysis, logarithmizing fine elements, $n=2^{22}$, $T=16$ (full parallelism), Kronecker graphs.}
\label{fig:lg-fine}
\end{figure*}

\begin{figure*}[t!]
\centering
\includegraphics[width=1.0\textwidth]{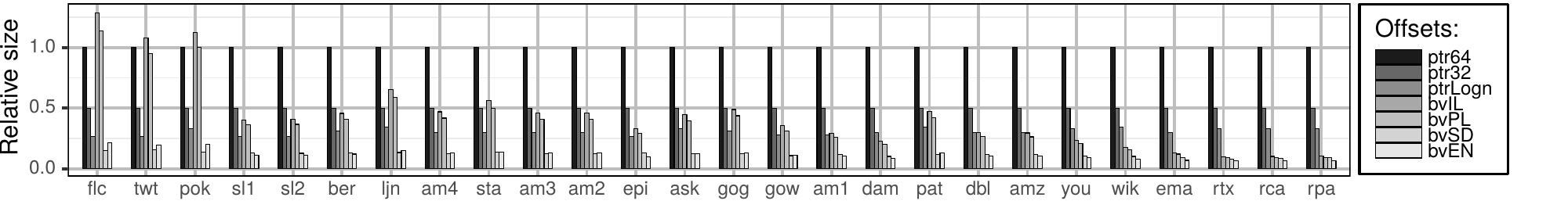}
\caption{(\cref{sec:eval_log_O}) Illustration of the size differences of various $\mathcal{O}$ (both offset arrays and bit vectors). The offset sizes are $W \in \{32, 64, \lceil \log n
\rceil\}$.} 
\label{fig:size-analysis-no-changes}
\end{figure*}

\subsubsection{Considered Graphs}

We analyze synthetic power-law (the Kronecker
model~\cite{leskovec2010kronecker}), synthetic uniform (the Erdős-Rényi
model~\cite{erdHos1976evolution}), and real-world datasets (including
SNAP~\cite{snapnets}, KONECT~\cite{kunegis2013konect},
DIMACS~\cite{demetrescu2009shortest}, and WebGraph~\cite{BoVWFI}); see
Table~\ref{tab:graphs} for details.
Now, for Kronecker graphs, we denote them with symbols \texttt{sX\_eY} where
\texttt{s} is the scale (i.e., $\log_2 n$) and \texttt{e} is the average number
of edges per vertex.
Due to a large amount of data we present and discuss in detail a comprehensive
subset; the remainder follows similar patterns.


\begin{table}[t]
\centering
\footnotesize
\sf
\begin{tabular}{@{}l|lllll@{}}
\toprule
\textbf{Type} & \multicolumn{1}{c}{\textbf{ID}} & \multicolumn{1}{c}{\textbf{Name}} & \multicolumn{1}{c}{$n$} & \multicolumn{1}{c}{$m$} & \multicolumn{1}{c}{\textbf{$\bar{d}$}} \\ \midrule
\multirow{9}{*}{\begin{tabular}[c]{@{}c@{}}Web graphs\end{tabular}}   & \textbf{uku}  & \textbf{Union of .uk domain} & 133M & 4.66B & 34.9             \\
& \textbf{uk}  & \textbf{.uk domain}   & 110M      & 3.45B           &  31.3             \\
& \textbf{sk}  & \textbf{.sk domain}   & 50.6M      & 1.81B           &  35.75             \\
& \textbf{gho}  & \textbf{Hosts of the gsh webgraph}   & 68.6M      & 1.5B           &  21.9             \\
& wb  & WebBase   & 118M      & 855M           &  7.24             \\
& tpd  & Top private domain   & 30.8M      & 490M           &  15.9             \\
& wik  & Wikipedia links   & 12.1M      & 288M           &  23.72             \\
& tra               & Trackers                      & 27.6M                           & 140M            &  5.08             \\
\cmidrule{2-6}
& \multicolumn{5}{l}{{Others:} tra, ber, gog, sta} \\\midrule
\multirow{2}{*}{\begin{tabular}[c]{@{}c@{}}Affiliation\\ graphs\end{tabular}}     & orm   & Orkut Memberships       & 8.73M            & 327M          &   37.46             \\
& ljm  & LiveJournal Memberships    & 7.48M      & 112M          & 15           \\ \midrule
\multirow{4}{*}{\begin{tabular}[c]{@{}c@{}}Social\\ networks\end{tabular}}          & \textbf{fr}   & \textbf{Friendster}      & 65.6M  & 1.8B          &   27.53             \\
& tw  & Twitter   & 49.2M      & 1.5B          & 30.5           \\
& ork  & Orkut    & 3.07M      & 117M          & 38.14           \\
\cmidrule{2-6}
& \multicolumn{5}{l}{{Others:} ljn, pok, flc, gow, sl1, sl2, epi, you, dbl, amz} \\\midrule
\multirow{2}{*}{\begin{tabular}[c]{@{}c@{}}Road networks\end{tabular}}          & \textbf{usrn}   & \textbf{USA road network}      & 23.9M  & 28.8M          &   1.2             \\
\cmidrule{2-6}
& \multicolumn{5}{l}{{Others:} rca, rtx, rpa} \\\midrule
\multirow{1}{*}{\begin{tabular}[c]{@{}c@{}}Various\end{tabular}}          &  \multicolumn{5}{l}{\makecell[l]{Purchase networks (am1--am4), communication graphs (ema, wik)} }             \\ \bottomrule
\end{tabular}
\caption{The used real-world graphs (sorted by ${m}$). The details are provided for $n>10$M or $m>100$M. The largest ones are bolded.}
%
\label{tab:graphs}
\end{table}

\subsubsection{Experimental Setup and Architectures}

We use the following systems to cover various types of machines:

\begin{itemize}[leftmargin=1em,noitemsep]
\item \macb{CSCS Piz Daint} is a Cray with various XC* nodes. Each XC50 compute
node contains a 12-core HT-enabled Intel Xeon E5-2690 CPU with 64 GiB RAM.
Each XC40 node contains two 18-core HT-enabled Intel Xeons E5-2695 CPUs with 64
GiB RAM.  The interconnection is based on Cray's Aries and it implements the
Dragonfly topology~\cite{dally08, DBLP:conf/sc/FaanesBRCFAJKHR12}. The batch system is slurm 14.03.7.
This machine represents massively parallel HPC machines.
\item \textbf{\textsf{Monte Leone}} is an HP DL 360 Gen 9 system. One node has:
two Intel E5-2667 v3 @ 3.2GHz Haswells (8 cosocket), 2 hardware
threads/core, 64 KB of L1 and 256 KB of L2 (per core), and 20 MB of L3 and 700
GB of RAM (per node).
It represents machines with substantial amounts of memory.
\end{itemize}

\subsubsection{Evaluation Methodology}

We use the arithmetic mean for data summaries. We treat the first 1\% of any
performance data as warmup and we exclude it from the results. We gather enough
data to compute the median and the nonparametric 95\% confidence intervals.



\subsection{Logarithmizing Fine Elements}
\label{sec:eval_log_fine}

%

We first illustrate that logarithmizing fine graph elements, especially vertex
IDs, reduces the size of graphs compared to the traditional adjacency arrays
and incurs negligible performance overheads (in the worst case) or offers
speedups (in the best case). The former is due to overheads from bitwise
manipulations over the input data. Simultaneously, smaller pressure on the memory
subsystem due to less data transferred to and from the CPU results in
performance improvements.
\emph{This class of schemes should be used in order to maintain highest
performance of graph algorithms while enabling moderate reductions in storage
space for the processed graphs.}

\subsubsection{Log(Graph) Variants and Comparison Targets}

We consider four variants of Log(Graph): \texttt{LG-g} (the global approach),
\texttt{LG-g-gap} (the global approach with fixed-size gap encoding), \texttt{LG-l}
(the local approach), and \texttt{LG-l-gap} (the local approach with fixed-size
gap encoding). We also incorporate the ILP heuristic for relabeling
from~\cref{sec:relabeling_scheme} that enhances the local approach.
We compare Log(Graph) to the tuned GAPBS code that uses a traditional adjacency array (Trad).


%
%
\subsubsection{Performance and Size on Single Nodes}

The results can be found in Figure~\ref{fig:lg-fine}.
The collected data confirms our predictions. In many cases Log(Graph) offers
performance comparable or better than that of the default adjacency array, for
example for PR and SSSP. Simultaneously, it reduces $|\mathcal{A}|$ compared to
\texttt{Trad}; the highest advantages are due to \texttt{LG-l-gap}
($\approx$35\% over \texttt{LG-l}); \texttt{LG-g-gap} does not improve much
upon \texttt{LG-g}.

\subsubsection{Scalability}

Log(Graph) advantages directly extend to distributed memories.  Here,
we measure the amount of communicated data and compare it to \texttt{Trad}.
For example, in a distributed BFS and for 1024 compute nodes, this amount is
consistently reduced by $\approx$37\% across the considered graphs.
We also conducted scalability analyses; the performance pattern is not
surprising and all the Log(Graph) variants finish faster as $T$ increases.
Full results are in the Appendix (\cref{sec:A_more_anal_scalability},
\cref{sec:A_more_anal_data}).
%


\subsubsection{ILP Heuristic}

We also investigate the impact from the {ILP} heuristic from~\cref{sec:relabeling_scheme}. It reduces the size of
graphs and we obtain consistent improvements or 1--4\%, for example from 0.614
GB to 0.604 GB for the \texttt{ork} graph. 
%
%
The ILP heuristic could be used on top of gap-encoding
when the user requires the highest compression ratio and still prefers
to logarithmize fine elements (instead of logarithmizing $\mathcal{A}$)
to ensure highest performance of processing graphs.


%

\subsubsection{Key Insights and Answers}

The most important insights are as follows.  First, logarithmizing fine
elements does reduce storage for graphs while ensuring high-performance and
scalability; both on shared- and distributed-memory machines.  Second, both ILP
(\cref{sec:relabeling_scheme}) and fixed-size (\cref{sec:fine_gap_encoding})
gap encoding reduce $|\mathcal{A}|$, with the latter being a definite winner.

\subsection{Logarithmizing Offset Structure $\mathcal{O}$}
\label{sec:eval_log_O}

%
We show that logarithmizing $\mathcal{O}$ with succinct bit vectors
(1) brings large storage reductions over simple bit vectors and offset arrays and
(2) enables low-overhead decompression,
matching performance of adjacency arrays in parallel
settings.


\subsubsection{Log(Graph) Variants and Comparison Targets}

We investigate all the described $\mathcal{O}$ variants of offset arrays and
bit vectors presented in Table~\ref{tab:theory_bit_vectors}. We also
incorporate a variant, denoted as \textsf{ptrLogn}, where we logarithmize each
offset treated as a fine element, as described
in~\cref{sec:fine_offsets}.
We also consider compression with \textsf{zlib}~\cite{Deutsch:1996:ZCD:RFC1950}.

\subsubsection{Size: Which Bit Vector is the Smallest?}

\begin{figure}[h!]
\centering
\includegraphics[width=0.45\textwidth]{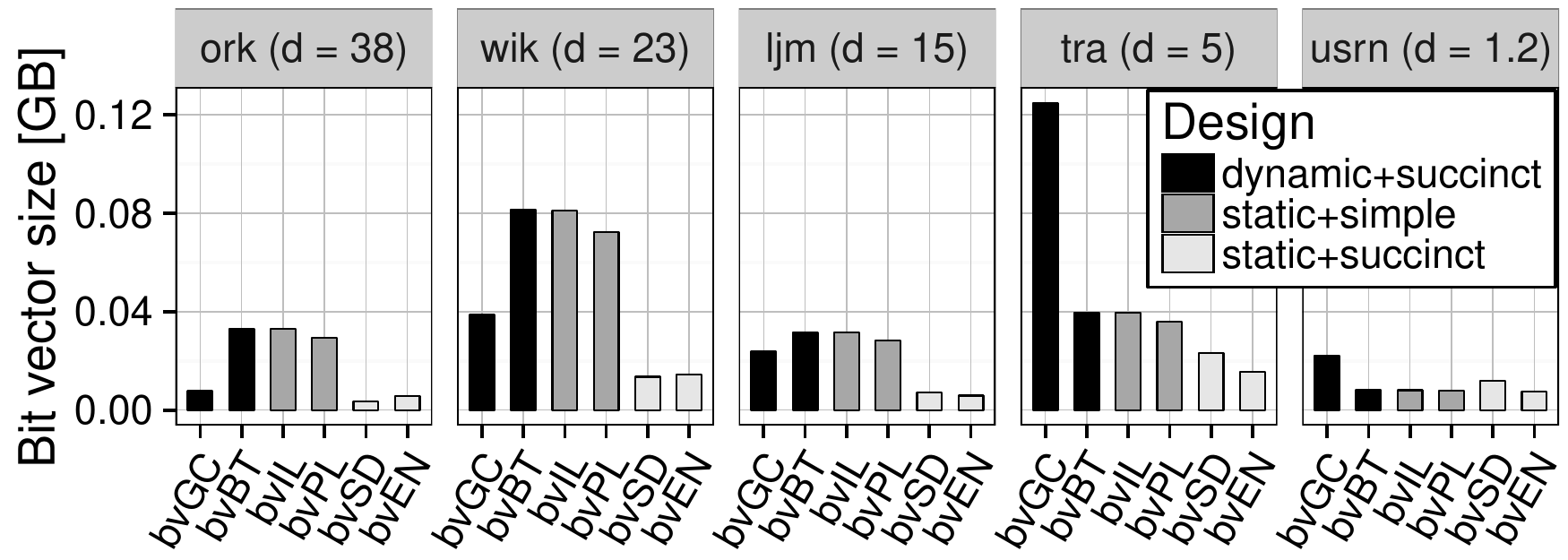}
\caption{Analysis of the sizes of bit vectors (plots
sorted by the average degree $\overline{d}$).}
\label{fig:bitvector_size_all_analysis}
\end{figure}

We first compare the size of all bit vectors for graphs of various sparsities
$\overline{d}$; see Figure~\ref{fig:bitvector_size_all_analysis}. Static
succinct bit vectors consistently use the least space.
Interestingly, \textsf{bvSD} uses more space than \textsf{bvEN} for graphs with
lower $\overline{d} \leq 15$.  This is because the term {\footnotesize$\log \dbinom{\frac{2Wm}{B}}{n}$}
grows faster with the number of edges than that of \textsf{bvEN}.

\subsubsection{Size: Bit Vectors or Offset Arrays?}

We next compare the size of offset arrays ($W \in \{32, 64, \lceil \log n
\rceil\}$) and selected bit vectors in
Figure~\ref{fig:size-analysis-no-changes}.  As expected, offset arrays are the
largest except for graphs with very high $\overline{d}$. The sparser a graph,
the bigger the advantage of bit vectors.  Although $|\mathcal{O}|$ grows
linearly with $m$ for bit vectors, the rate of growth is low (bit vector take
only one bit for a single edge, assuming $B$ is one word size).  Again,
\textsf{bvEN} is the smallest in most cases.

\subsubsection{Size: When To Condense $|\mathcal{O}|$?}

\begin{figure}[!h]
\centering
 \begin{subfigure}[t]{0.23 \textwidth}
  \includegraphics[width=\textwidth]{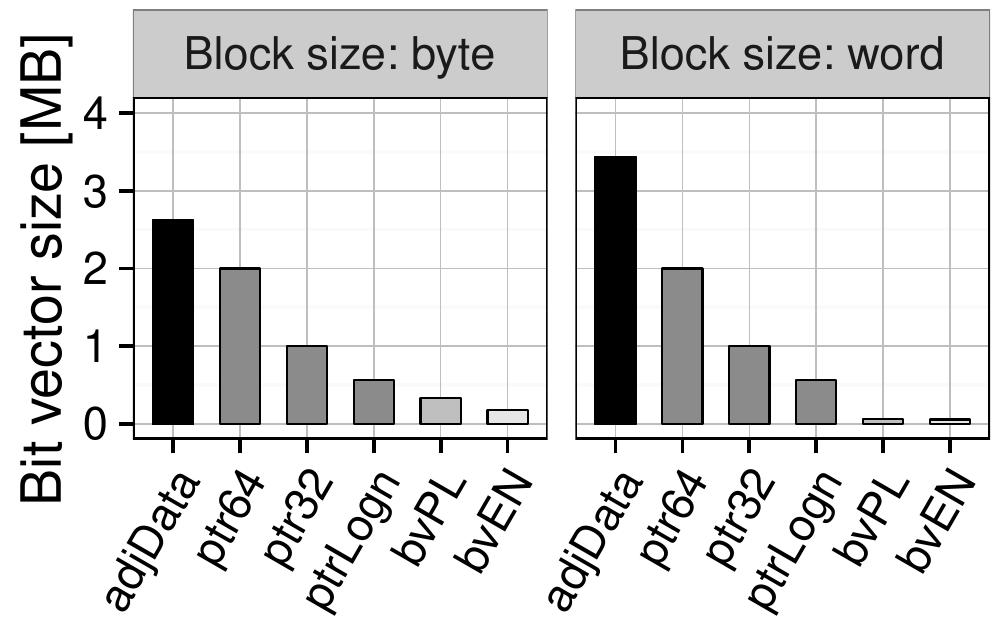}
  \caption{Graph \textsf{am1}; $\bar{d} \approx 5$.}
  \label{fig:offsets_blocks_am1}
 \end{subfigure}
 \begin{subfigure}[t]{0.23 \textwidth}
  \includegraphics[width=\textwidth]{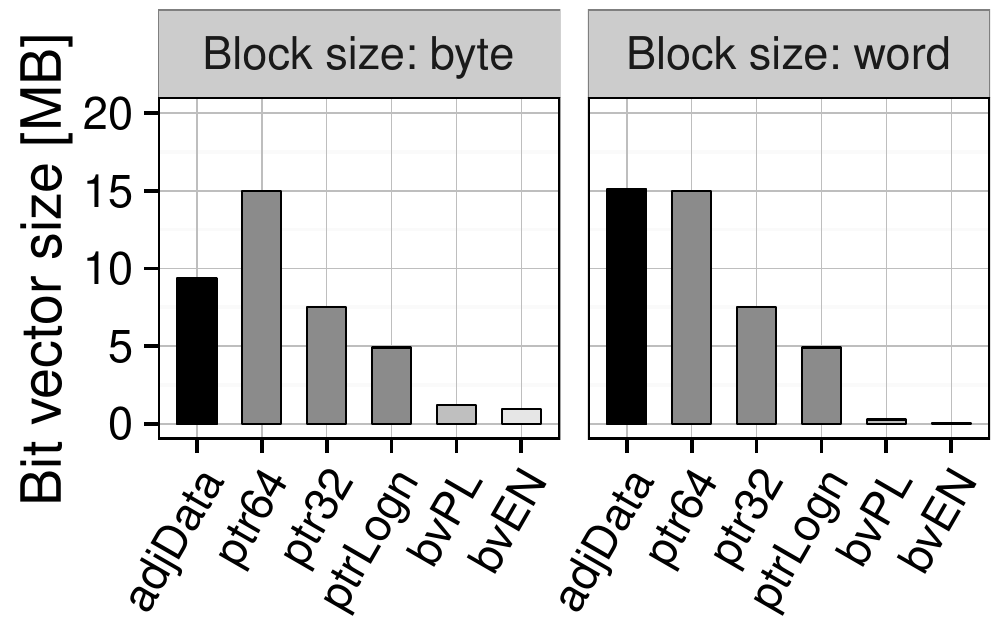}
  \caption{Graph \textsf{rca}; $\bar{d} \approx 1.5$.}
  \label{fig:offsets_blocks_rca}
 \end{subfigure}
 %
%
\caption{Illustration of storage required
for both $\mathcal{A}$ and $\mathcal{O}$ for graphs with various average
degrees $\bar{d}$ when varying the block size $B$.}
\label{fig:offsets_blocks_analysis}
\end{figure}

For many graphs, $|\mathcal{A}| \gg |\mathcal{O}|$ and condensing
$|\mathcal{O}|$ brings only little improvement.  Thus, we also investigate when
to condense $|\mathcal{O}|$.  We analyze the listed graphs and conclude that
$|\mathcal{O}| \approx |\mathcal{A}|$ if the graph is sparse enough:
$\overline{d} \leq 5$.  This is illustrated in
Figure~\ref{fig:offsets_blocks_analysis}.  In such cases, condensing
$\mathcal{O}$ is at least as crucial as $\mathcal{A}$.

\subsubsection{Size: Succinctness or Compression?}

\begin{figure}[h!]
\centering
 \begin{subfigure}[t]{0.112 \textwidth}
  \includegraphics[width=\textwidth]{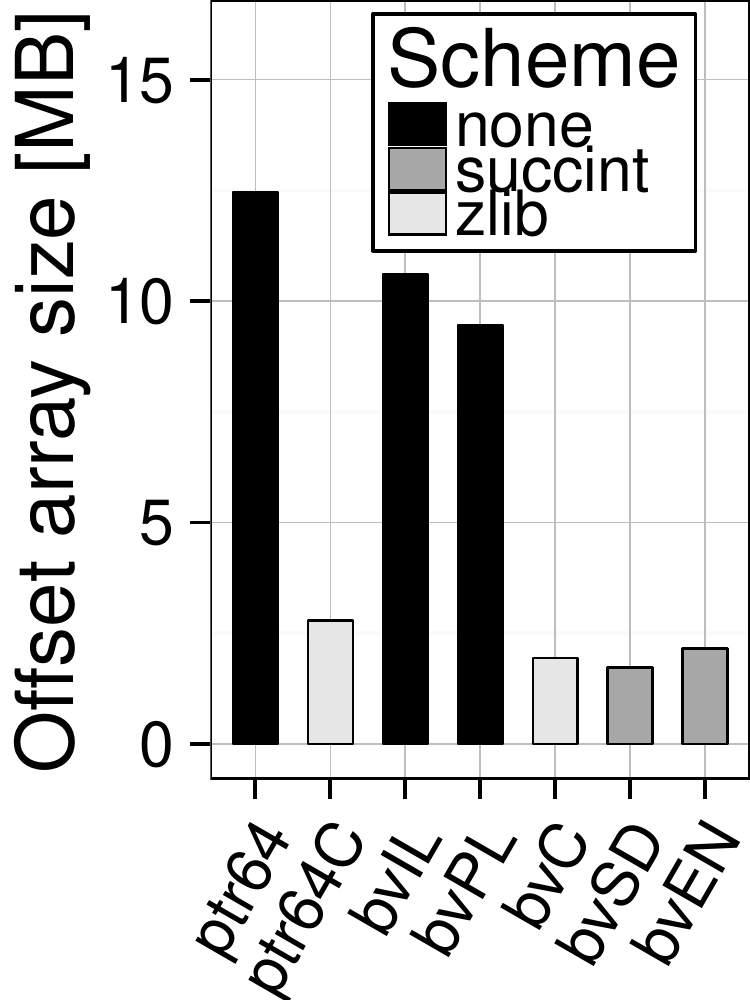}
  \caption{pok; $\bar{d} \approx 20$.}
  \label{fig:offsets_compression_pok}
 \end{subfigure}
 \begin{subfigure}[t]{0.112 \textwidth}
  \includegraphics[width=\textwidth]{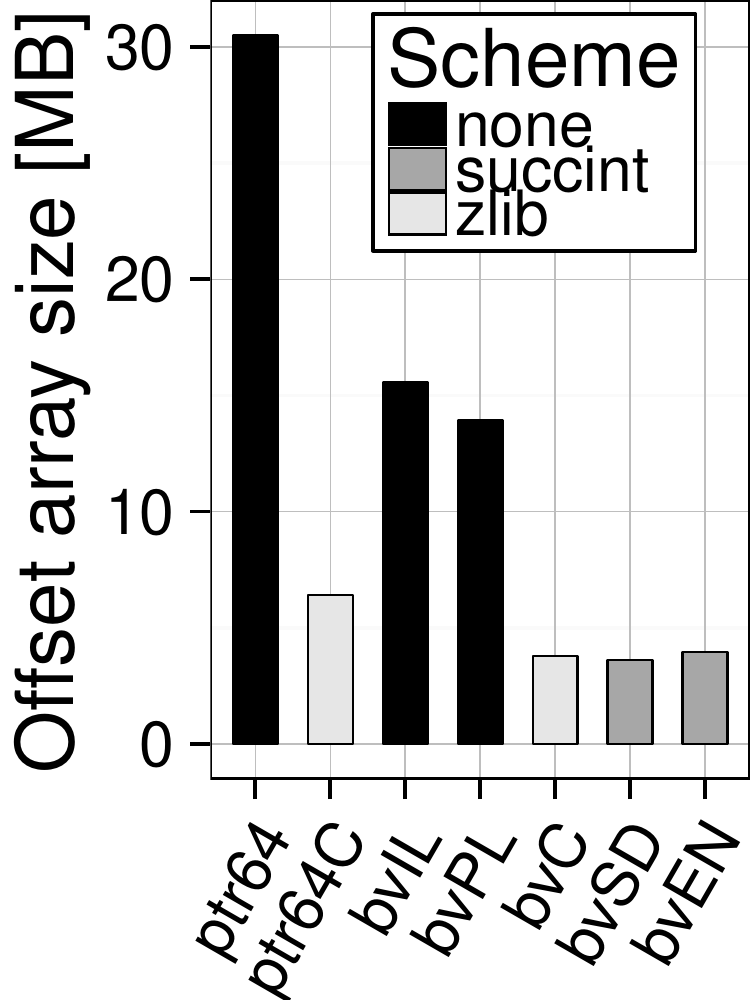}
  \caption{ljn; $\bar{d} \approx 10$.}
  \label{fig:offsets_compression_lj2}
 \end{subfigure}
 \begin{subfigure}[t]{0.112 \textwidth}
  \includegraphics[width=\textwidth]{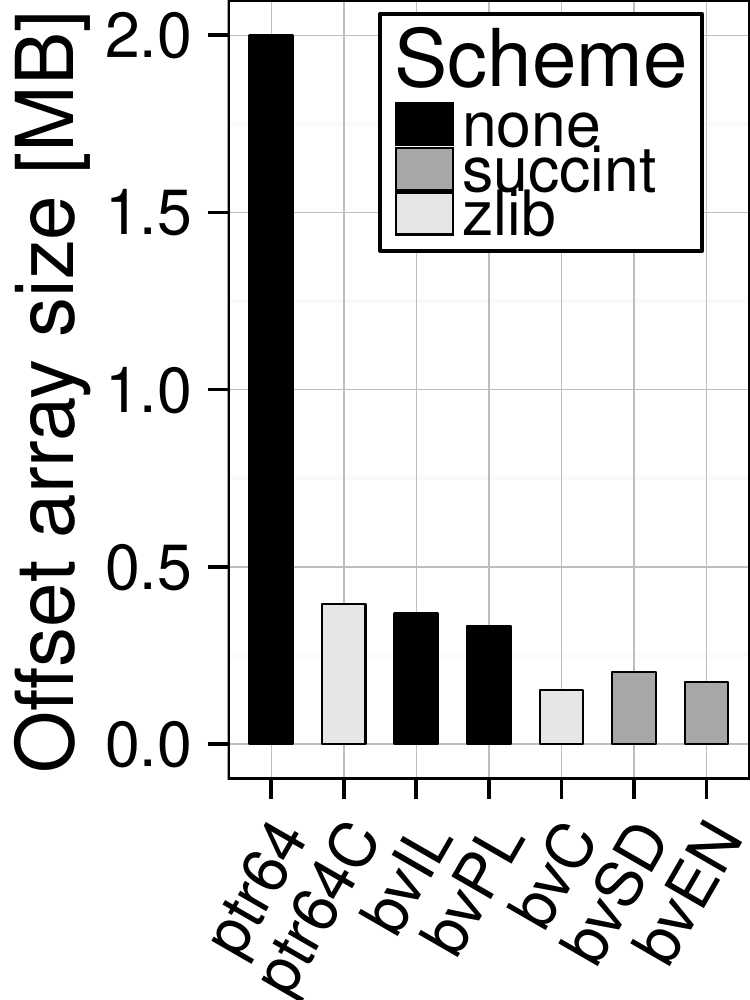}
  \caption{am1; $\bar{d} \approx 5$.}
  \label{fig:offsets_compression_am1}
 \end{subfigure}
 \begin{subfigure}[t]{0.112 \textwidth}
  \includegraphics[width=\textwidth]{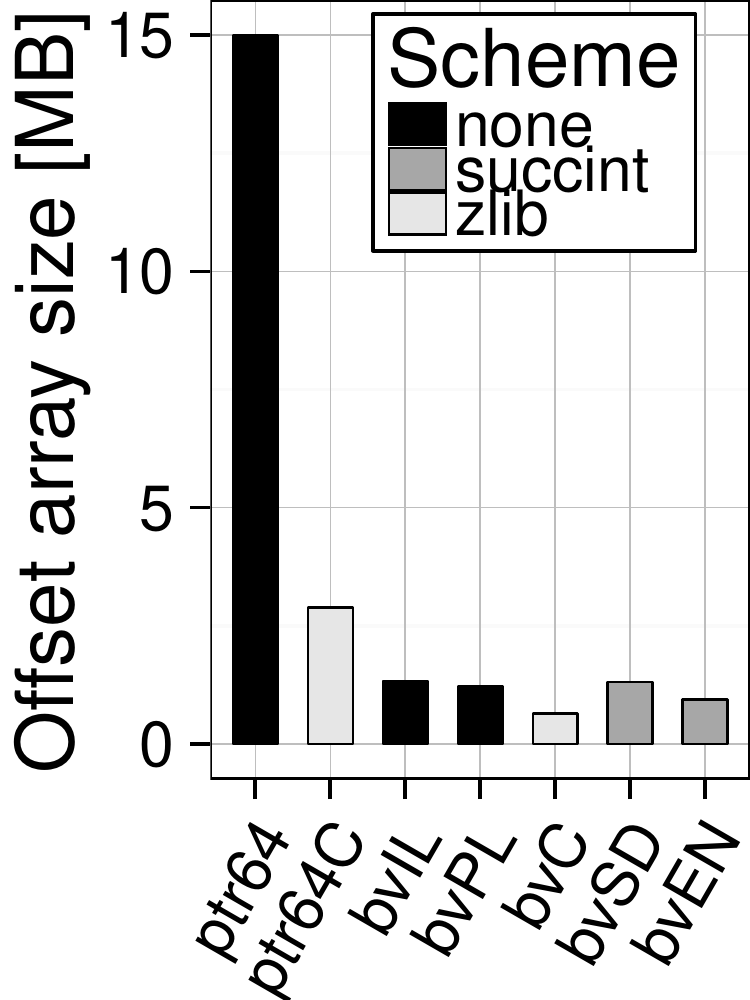}
  \caption{rca; $\bar{d} \approx 1.5$.}
  \label{fig:offsets_compression_rca}
 \end{subfigure}
%
%
\caption{$|\mathcal{O}|$ for various $\overline{d}$ when varying the
compression scheme.}
\label{fig:offsets_compression_analysis}
\end{figure}

We finally analyze the effect of traditional compression included in
$C[\mathcal{O}]$ on the example of the well-known
\textsf{zlib}~\cite{Deutsch:1996:ZCD:RFC1950}.  We present the results in
Figure~\ref{fig:offsets_compression_analysis}; (64-bit offset arrays represent
all pointer schemes that followed similar results). Compressed variants of
\textsf{ptr64} and \textsf{bvPL} are \textsf{ptr64C} and \textsf{bvPLC}.  Once
more, the results heavily depend on $\bar{d}$. For bit vectors, the sparser the
graph, the more compressible $\mathcal{O}$ is; for $\bar{d} < 10$,
\textsf{bvPLC} is smaller than both \textsf{bvSD} and \textsf{bvEN}.
Interestingly, \textsf{ptr64C} is smaller than \textsf{ptr64} by an identical
ratio regardless of $\bar{d}$. We conclude that \textsf{zlib} offers comparable
or slightly (by up to $\approx$20\%) better $|\mathcal{O}|$ reductions than
succinct designs. We now proceed to show that these advantages are annihilated
by performance penalties.

\subsubsection{Performance}

\begin{figure}[h!]
\centering
 \begin{subfigure}[t]{0.23 \textwidth}
  \includegraphics[width=\textwidth]{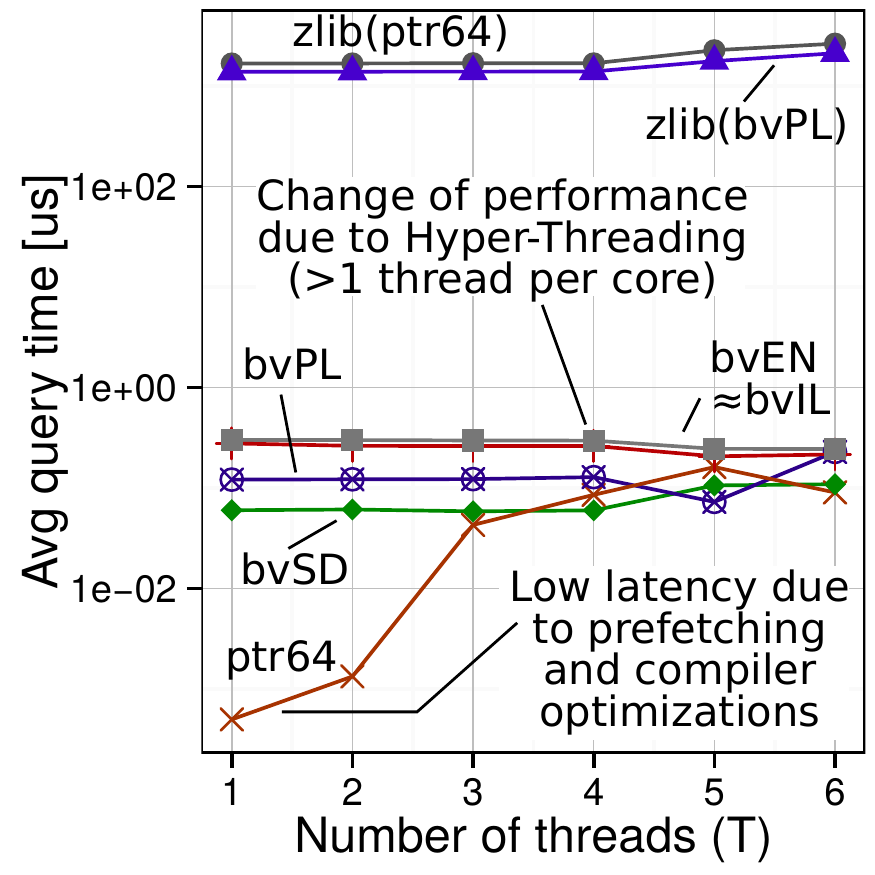}
  \caption{Twitter graph \textsf{tw}.}
  \label{fig:offsets_perf_twt}
 \end{subfigure}
 \begin{subfigure}[t]{0.23 \textwidth}
  \includegraphics[width=\textwidth]{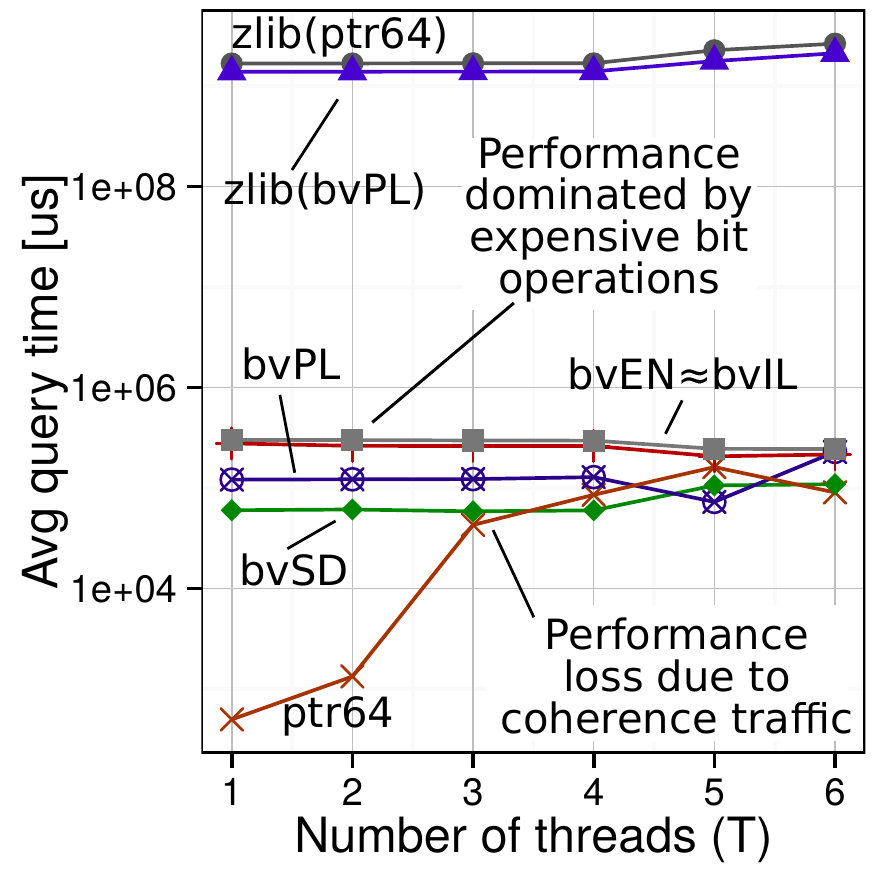}
  \caption{California road graph \textsf{rca}.}
  \label{fig:offsets_perf_rca}
 \end{subfigure}
 %
%
\caption{(\cref{sec:eval_log_O}) Performance analysis of various types of $\mathcal{O}$.}
%
\label{fig:offsets_perf_analysis}
\end{figure}

Finally, we analyze the performance of various $\mathcal{O}$ designs queried by
$T$ threads in parallel. Each thread fetches offsets of 1,000 random vertices.
The results for \textsf{twt} and \textsf{rca} (representing graphs with high
and low $\bar{d}$) are in Figure~\ref{fig:offsets_perf_analysis}.
First,\textsf{bvEN} is consistently slowest due to its complex design; this
dominates any advantages from its small size and better cache reuse.
Surprisingly, \textsf{bvEN} is followed by \textsf{bvIL} that has the biggest
$|\mathcal{O}|$ (cf.  Figure~\ref{fig:size-analysis-no-changes}); its
time/space tradeoff is thus not appealing for graph processing. Finally,
\textsf{bvPL} and \textsf{bvSD} offer highest performance, with \textsf{bvSD}
being the fastest for $T \leq 4$ (the difference becomes diluted for $T>4$ due
to more frequent cache line evictions). \emph{The results confirm the theory}:
\textsf{bvPL} offers $O(1)$ time accesses (while paying a high price in
storage, cf.  Figure~\ref{fig:size-analysis-no-changes}) and \textsf{bvSD} uses
little storage and fits well in cache.
Next, we study offset arrays.  \textsf{ptr64} is the fastest for $T \leq 4$
%
%
due to least memory operations. Interestingly, the smaller $T$, the lower the
latency of \textsf{ptr64}. We conjecture this is because fewer threads cause
less traffic caused by the coherence protocol.
%
%
As for \textsf{zlib}, it entails costly performance overheads as it requires
decompression. We tried a modified blocked \textsf{zlib} variant without
significant improvements.


%

%

\subsubsection{Further Analyses}
\label{sec:further_O_anal}

We vary the block size $B$ that controls the granularity of $\mathcal{A}$ to be
an 8-bit byte or a 64-bit word; see Figure~\ref{fig:offsets_blocks_analysis}.
First, larger $B$ reduces each $|\mathcal{O}|$ (as $|\mathcal{O}|$ is
proportional to $B$). Next, $|\mathcal{A}|$ grows with $B$. This phenomenon is
similar to the internal fragmentation in memory allocation. Here, each
$\mathcal{A}_v$ is aligned with respect to $B$. The larger $B$, the more space
may be wasted at the end of each array.

Other analyses are included in the Appendix (\cref{sec:O_more_anal}).

\subsubsection{Key Insights and Answers}

We conclude that succinct bit vectors are a good match for $\mathcal{O}$.
First, they reduce $|\mathcal{O}|$ more than any offset array and are
comparable to traditional compression methods such as \textsf{zlib}. Next, they
closely match the performance of offset arrays for higher thread counts and are
orders of magnitude faster than \textsf{zlib}. Finally, they consistently
retain their advantages when varying the multitude of parameters, both related
to input graphs ($\overline{d}$) and to the utilized \textsf{AA} ($B$ and
$\mathcal{A}$).  They can enhance any system for condensing static or slowly
changing graphs that uses $\mathcal{O}$.

\begin{figure*}[t!]
  \centering
  \begin{minipage}{0.25\textwidth}
  \centering
  \includegraphics[width=1\textwidth]{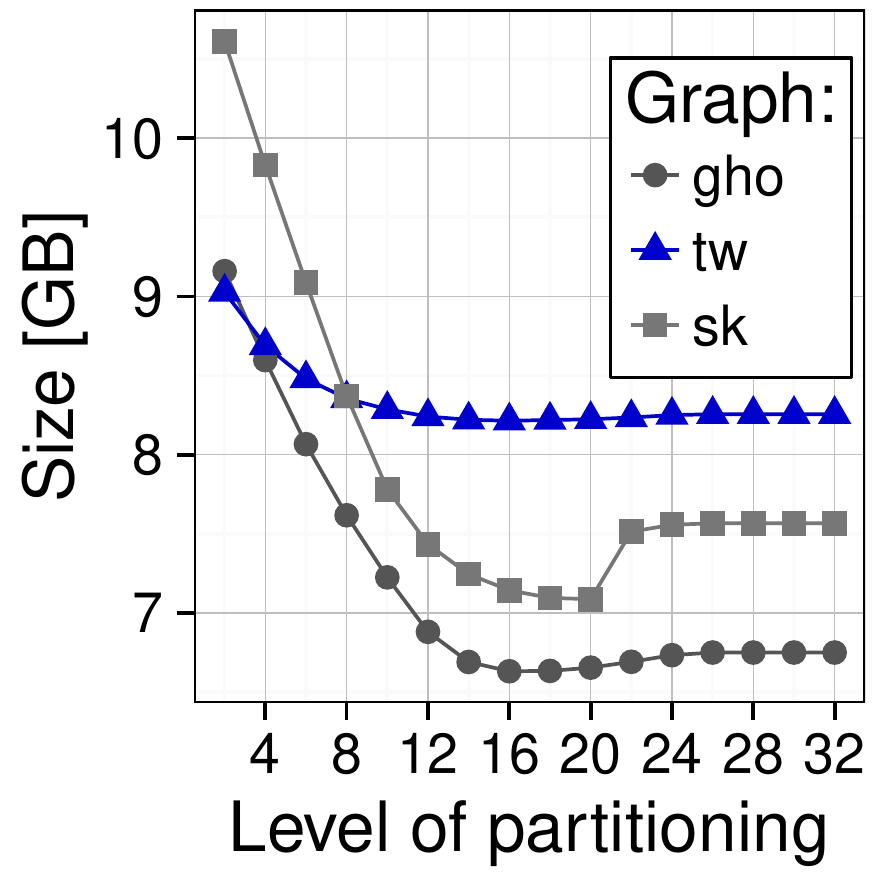} 
  \caption{(\cref{sec:eval_log_A}) BRB analysis.}
  \label{fig:brb-size}
  \end{minipage}
  \begin{minipage}{0.28\textwidth}
  \centering
  \includegraphics[width=1\textwidth]{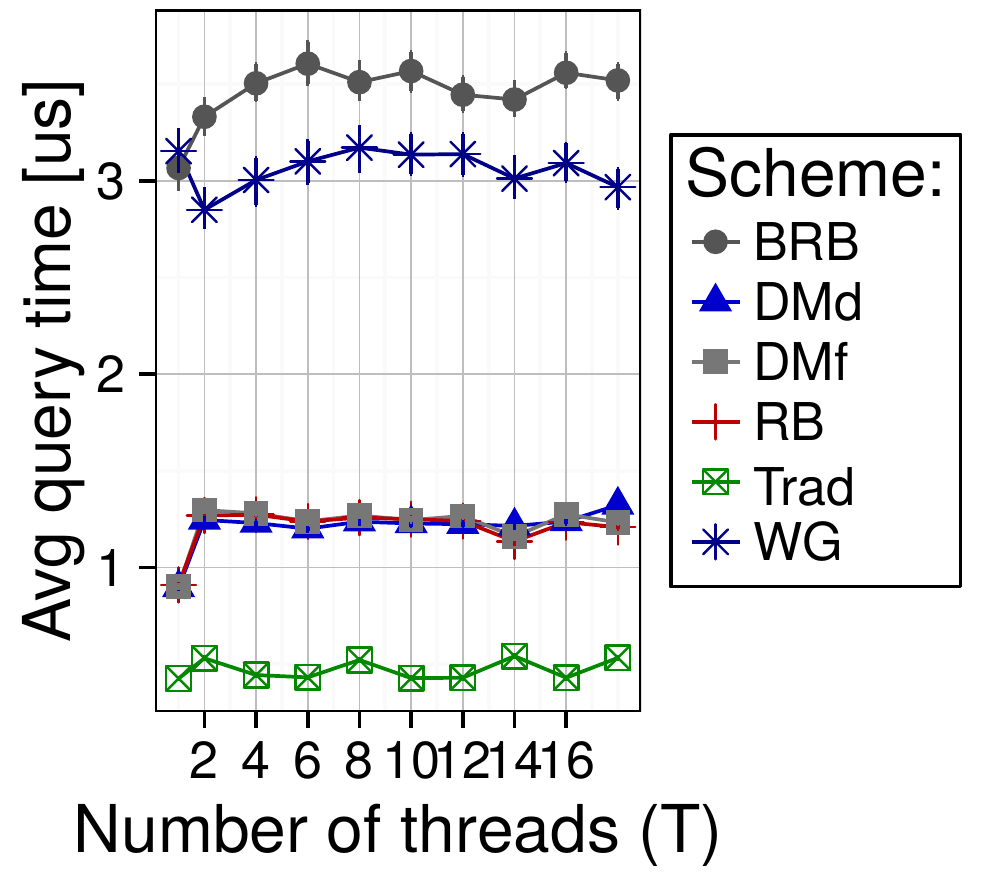} 
  \caption{(\cref{sec:eval_log_A}) The performance of $N_v$ for the orm graph.}
  \label{fig:queries-nv}
  \end{minipage}
\\
  \begin{minipage}{0.68\textwidth}
  \centering
  \includegraphics[width=1\textwidth]{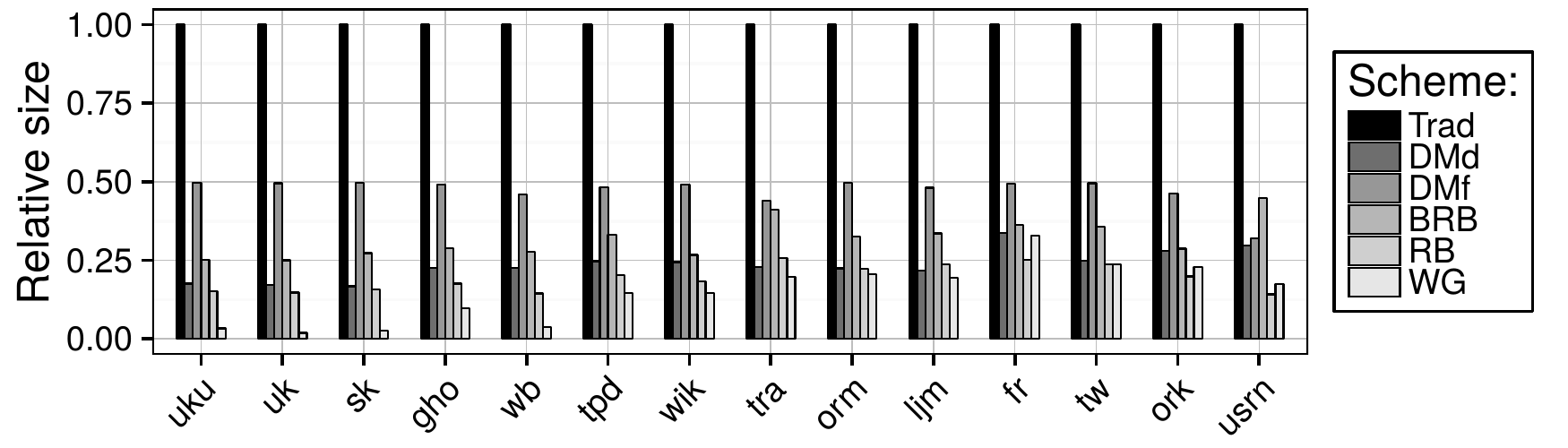} 
  \caption{(\cref{sec:eval_log_A}) Illustration of the storage overhead of different types of $\mathcal{A}$.}
  \label{fig:size-ad-analysis}
  \end{minipage}
%
\end{figure*}

%

\subsection{Logarithmizing Adjacency Structure $\mathcal{A}$}
\label{sec:eval_log_A}

Finally, we evaluate the logarithmization of $\mathcal{A}$
and show that it offers storage reductions that are in many cases comparable to that of modern
graph compression schemes while providing
significant speedups due to low-overhead decompression.
\emph{This class of schemes should be used in order to maintain highest
reductions in storage space for graphs while enabling large speedups over
the existing graph compression schemes.}
%

%

\subsubsection{Log(Graph) Variants and Comparison Targets}

We evaluate all the discussed schemes: RB (\cref{sec:bf_scheme}), BRB
(\cref{sec:br_scheme}), DMd as well as DMf (\cref{sec:dm_scheme}), the
traditional adjacency array ({Trad}), the state-of-the-art WebGraph
(WG)~\cite{boldi2004webgraph} compression system, POD (\cref{sec:pod_scheme}),
and the combination of these two (\cref{sec:hybridization}).  We use the
WebGraph original tuned Java implementation for gathering the data on
compression ratios but, as Log(Graph) is developed in C++, we use a
proof-of-a-concept C++ implementation of WG schemes for a fair C++ based
performance analysis.
%


\begin{figure*}[t]
\centering
\begin{subfigure}[t]{0.65 \textwidth}
  \includegraphics[width=1.0\textwidth]{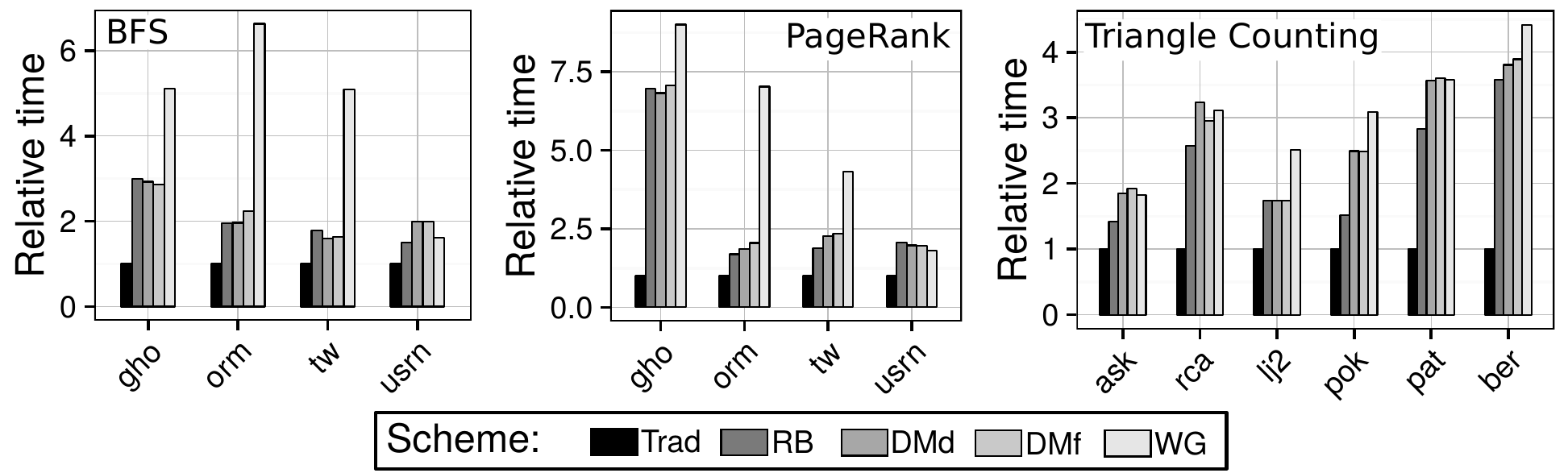}
\caption{BFS, PageRank, and Triangle Counting.}
  \label{fig:my_bfs_wg}
  \end{subfigure}
  \begin{subfigure}[t]{0.65 \textwidth}
  \includegraphics[width=1.0\textwidth]{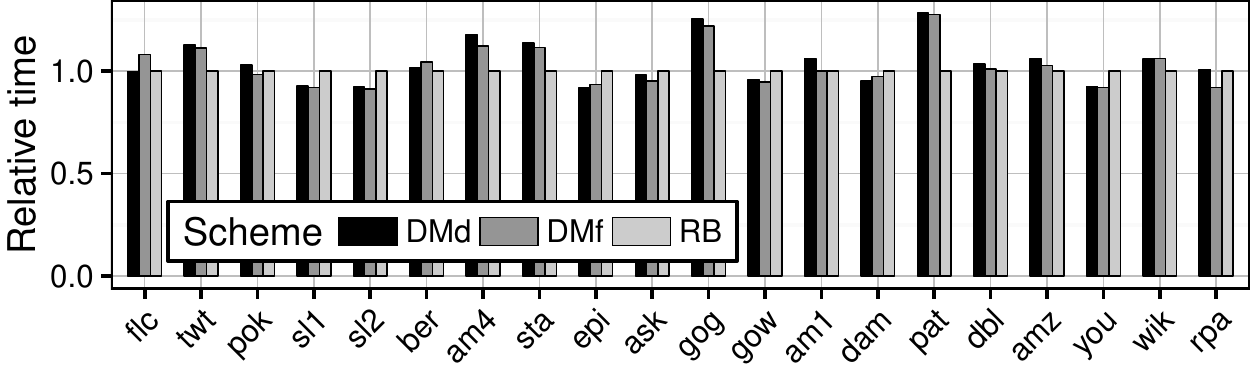}
  \caption{Comparison of DM and RB (for BFS) on SNAP graphs.}
  \label{fig:my_bfs}
\end{subfigure}
%
%
\caption{(\cref{sec:eval_log_A}) The analysis of the performance impact of
Log(Graph) on parallel graph algorithms and comparison to the C++
implementation of WebGraph schemes. The geometric mean of ratios of DMd and WG
is $\approx$0.5 (BFS), $\approx$0.6 (PageRank), and $\approx$0.9 (TC).}
%
\label{fig:alg_analysis}
\end{figure*}

\subsubsection{BRB: Alleviating RB's Preprocessing}

We start with illustrating that BRB alleviates preprocessing overhead inherent
to RB. Table~\ref{tab:rb-sucks} shows the overhead from RB compared to a
simple AA. Now, BRB's preprocessing takes equally long if we build the full
separator tree.  The idea is to build a given limited number of the separator
tree levels. We illustrate this analysis in Figure~\ref{fig:brb-size}.  Using
fewer partitioning levels increases $|\mathcal{A}|$ but also reduces the
preprocessing time (it approximately doubles for each new level).
Interestingly, the storage overhead from preserving the recursive graph
structure begins to dominate at a certain level, annihilating further
$|\mathcal{A}|$ reductions.

Yet, BRB comes with overheads while resolving $N_v$ because one must
construct vertex IDs from bit strings.  This results in a 2-2.5x slowdown of
obtaining $N_v$, depending on the graph.  We
conclude that whether to use RB or BRB should depend on the targeted workload:
for frequent accesses to $N_v$ one should use RB while to handle
large or evolving graphs that require continual preprocessing one should use
BRB.

\begin{table}[!h]
\centering
\footnotesize
\sf
\begin{tabular}{@{}l|lllllll@{}}
\toprule
\textbf{Graph} & uku & gho & orm & tw & usrn & ema & am1 \\ \midrule
\textbf{Generation of \textsf{RB}} & 981.5 & 458.9 & 101.6 & 572.3 & 47.7 & 0.33 & 0.41 \\
\textbf{Generation of \textsf{AA}} & 19.5 & 5.9 & 1.1 & 5.8  & 0.3 & 0.02 & 0.02 \\ \bottomrule
\end{tabular}
\caption{(\cref{sec:eval_log_A}) Illustration of preprocessing overheads [seconds].}
%
\label{tab:rb-sucks}
\end{table}

\subsubsection{DMd: Approaching the Time/Space Sweetspot}

Next, we illustrate that DMd significantly reduces $|\mathcal{A}|$, resolves
$N_v$ fast, outperforms WG, uses less storage than DMf, and can be generated fast.  The size analysis is
shown in Figure~\ref{fig:size-ad-analysis}.  We use relative sizes for clarity
of presentation; the largest graphs use over 60 GB in size (in \texttt{Trad}).
DMf and DMd generate much smaller $\mathcal{A}$ than \texttt{Trad}, with DMd
outperforming DMf, being comparable or in many cases better than either RB or BRB (e.g., for ljm).
Now, in various cases DMd closely matches WebGraph, for example for tw, fr,
ljm. For others, it gives slightly larger $\mathcal{A}$ (e.g., for wik).
Next, we also derive time to obtain $N_v$; WG is consistently slower ($>$2x)
than DMd; more results are in Figure~\ref{fig:queries-nv} and the Appendix (\cref{sec:A_more_anal_queries}).
We conclude that DMd offers the storage/performance sweetspot: it ensures high
level of condensing, trades a little storage for fast $N_v$, and finally takes
significantly (>10x for RB) less time to generate than any other scheme
$\mathcal{A}$.

\subsubsection{Preprocessing}

Log(Graph) preprocessing time is negligible, except for BRB.
WebGraph is consistently slower.

\subsubsection{Further Analyses}
\label{sec:further_A_anal}

Other analyses include: investigating the ILP schemes, using various types
of cuts while building the separator tree, and varying the maximum allowed
difference in the sizes of subgraphs derived while partitioning. These
analyses are included in the Appendix (\cref{sec:A_more_anal}). Here, we conclude that ILP does
improve upon RB and DM by reducing sums of differences between consecutive
IDs.

\subsubsection{Key Insights and Answers}

We conclude that BRB alleviates RB's preprocessing overheads while DMd offers the best
space/performance tradeoff.

\subsection{The Log(Graph) Library}
\label{sec:eval_log_LG}

We finally evaluate the Log(Graph) library and show that it ensures high
performance.

\subsubsection{Performance: Graph Algorithms}

We use the Log(Graph) library to implement graph algorithms. We present the
results for BFS, PR, and TC.
%
%
We use succinct bit vectors (\textsf{bvSD}) as $\mathcal{O}$ and various
schemes for $\mathcal{A}$. Our modular design based on the established model
enables quick and easy implementation of $\mathcal{A}$; each
variant requires at most 20 lines of code.
The results are shown in Figure~\ref{fig:alg_analysis}. The BFS and PR analyses
for large graphs (gho, orm, tw, usrn) illustrate that DMd is comparable to RB and DMf,
  merely up to 2x slower than the uncompressed Trad, and significantly faster
  (e.g., $>$3x for orm) than WebGraph.
The relative differences for TC are smaller because the high computational
complexity of TC makes decompression overheads less severe.
Finally, we also study the differences between DMd and RB as well as DMf in more detail in
Figure~\ref{fig:my_bfs}) for a broader set of SNAP graphs.  We conclude that
{DMd} offers performance comparable to the state-of-the-art {RB} as well as DMf,
while avoiding costly overheads from recursive partitioning.

\subsubsection{Performance: Graph Accesses}

We also evaluate obtaining $d_v$ and $N_v$. This also enables understanding the
performance of succinct structures in a parallel setting, which is of
independent interest. Full results are in the Appendix
(\cref{sec:A_more_anal_queries}). Trad is the fastest (no decoding). The
difference is especially visible for bvSD and $f_v$ due to the complex
$\mathcal{O}$ design. DMd, DMf, and RB differ only marginally (1-3\%) due to
decoding.

\subsection{Discussion of Results}

Our evaluation confirms the characteristics of three logarithmization
families of schemes.

First,
\textbf{logarithmizing fine elements}
does deliver storage reductions (20-35\%)
compared to the traditional adjacency array and it enables very high
performance close to or even exceeding that of tuned graph processing codes. It 
enables its merits on both shared- and distributed-memory machines. 

Next,
\textbf{logarithmizing adjacency data}
is somewhat an opposite to the logarithmization of fine elements: it aggressively
reduces storage, in some cases by up to $\approx$80\% compared
to the adjacency array, approaching the compression ratios
of modern graph compression schemes and simultaneously
offering speedups of around 3$\times$ over these schemes.
Specifically, the BRB scheme alleviates RB's preprocessing overheads while the DMd scheme offers the best
space/performance tradeoff.
Yet, this family of schemes leads to higher overheads in performance
than the logarithmization of fine elements.
Thus, it should be used when reducing storage outweighs achieving highest performance.


Finally,
\textbf{logarithmizing offset structures}
can enhance \emph{any} parallel graph processing
computation because it does not incur performance overheads
in parallel settings (for $T \ge 4$ in our tests) while it does reduce storage required for
offsets (a part of the adjacency array) even by >90\%. 
We conclude that succinct bit vectors are a good match for $\mathcal{O}$.
First, they reduce $|\mathcal{O}|$ more than any offset array and are
comparable to traditional compression methods such as \textsf{zlib}. Next, they
closely match the performance of offset arrays for higher thread counts and are
orders of magnitude faster than \textsf{zlib}. Finally, they consistently
retain their advantages when varying the multitude of parameters.

%

%


\section{RELATED WORK}
We now discuss how Log(Graph) differs from or complements various aspects of graph
processing and compression.
As we illustrated, \emph{Log(Graph) is a tool that can enhance any graph processing engine, benchmark, or algorithm
that stores graphs as adjacency arrays}, such as GAPBS~\cite{beamer2015gap},
Pregel~\cite{Malewicz:2010:PSL:1807167.1807184},
HAMA~\cite{Seo:2010:HEM:1931470.1931872},
GraphLab~\cite{low2010graphlab},
Spark~\cite{Zaharia:2012:RDD:2228298.2228301},
Galois~\cite{Kulkarni:2007:OPR:1250734.1250759},
PBGL~\cite{Gregor05theparallel},
GAPS~\cite{beamer2015gap},
Ligra~\cite{shun2013ligra},
Gemini~\cite{zhu2016gemini},
Tux$^2$~\cite{xiao2017tux2},
Green-Marl~\cite{Hong:2012:GDE:2150976.2151013},
and others~\cite{besta2017push, gianinazzi2018communication, besta2015accelerating}.
It could also be used to enhance systems and schemes where graphs are modeled
with their adjacency matrix~\cite{besta2017slimsell, mattson2014standards,
bulucc2011combinatorial, solomonik2017scaling}.
%
%
For example, one could use logarithmized vertex IDs to
accelerate graph processing and reduce the pressure on the memory subsystem~\cite{besta2018slim} or network
in distributed-memory environments~\cite{besta2015active, besta2014fault, schmid2016high, fompi-paper, besta2014slim}.
%


%

\subsection{Log(Graph) and Compact Schemes}

A graph representation based on recursive partitioning, proposed by
Blandford et al.~\cite{Blandford:2003:CRS:644108.644219}, was proved
to be \emph{compact}: it takes $O(n)$ bits for an input
graph with $n$ vertices.
It reduces
$|\mathcal{A}|$ for several real-world graphs. Yet, its preprocessing is
costly. \emph{Log(Graph) alleviates it with the BRB scheme.}
%
%
%
%



\subsection{Log(Graph) and Succinct Schemes}

\emph{Log(Graph) uses and puts in practice succinct designs to enhance
graph storage and processing.}
There are various succinct graph representations~\cite{succinct_bound, labeled,
Kannan92implicitrepresentation__, succ-category, DBLP:journals/corr/abs-0705-0552,
Jacobson:1988:SSD:915547, Jacobson:1989:SST:1398514.1398646,
gonzalez2005practical, vigna2008broadword, gog2014optimized, Munro:2002:SRB:586840.586885,
Jansson:2007:URO:1283383.1283445, Claude08practicalrankselect,
alvarez2017succinct,
Okanohara07practicalentropycompressed}
but they are mostly theoretical structures with large
hidden constants, negligible asymptotic enhancements over the respective storage lower bound, or no practical codes.
Succinct~\cite{agarwal2015succinct} is a data store that uses succinct data structures; yet, it does not
specifically target graphs or graph processing.
%
%
%
Some works~\cite{shun2015parallel} construct succinct structures in parallel, but they do \emph{not}
process them in parallel.
Finally, there are several libraries of succinct data
structures~\cite{grossi2013design, sux, libcds, rsdic, succinct, gbmp2014sea}.
\emph{Contrarily to our work, none of these designs enhances graph processing
and they do not address parallel processing of a succinct data structure}.



\subsection{Log(Graph) and Compression Schemes}

A mature compression system for graphs is WebGraph~\cite{boldi2004webgraph}. 
There are also other works~\cite{besta2018survey, suel2001compressing, adler2001towards, buehrer2008scalable,
raghavan2003representing, claude2007fast, navarro2007compressing, claude2010extended, brisaboa2009k2, ladra2011algorithms, brisaboa2014compact, claude2011practical, asano2008efficient, hernandez2012compressed, randall2002link, stanley2017compressing,
navlakha2008graph, khan2017summarizing, maneth2017grammar, asadi2017compressing, tian2008efficient, diaz2002survey, chierichetti2009compressing}.
%
%
Some mention encoding some vertex
IDs with the logarithmic number of bits~\cite{adler2001towards, suel2001compressing}; \emph{Log(Graph) extends them with
schemes such as local logarithmization~\cref{sec:local_approach_log}}.
%
%
Several works use ILP to relabel vertices to reduce
$|\mathcal{A}|$~\cite{diaz2002survey, chierichetti2009compressing}.
Others collapse specified subgraphs into {supervertices} and merge edges
between them into a {superedge}~\cite{buehrer2008scalable, stanley2017compressing, raghavan2003representing}. 
%
%
%
These systems come with \emph{complex compression} and \emph{costly
decompression}.
Next, Ligra+~\cite{shun2015smaller} compresses graphs while ensuring high
performance of graph algorithms.  It is orthogonal to Log(Graph) as it uses
parallel compute power to provide fast decoding while Log(Graph) relies on
simplicity and \emph{it can can be used to enhance Ligra+} (e.g., with local
vertex ID or $\mathcal{O}$ logarithmization) and ensure \emph{even more
performance}.
Moreover, G-Store~\cite{kumar2016g} is a storage system for graphs that, among
others, removes most signigicant bit (MSB) zeros of vertex IDs within one tile
of 2D partitioning. In general, removing MSB zeros was proposed even before
G-Store~\cite{suel2001compressing, adler2001towards}.  We enhance this
technique and apply it holistically to all the considered fine graph elements. 
%
%
The technique in G-Store is orthogonal to ours and can be combined with the
hierarchical logarithmization to reduce space even further. 
There are also several works that reorder vertex IDs for more performance.  For
example, Wei et al.~\cite{wei2016speedup} reduce cache miss rate. As their most
important goal is to accelerate graph processing without storage reductions, we
exclude this work from a more detailed discussion as less related.
%
%
%
We conclude that Log(Graph), on one hand, \emph{enables simple and generic
logarithmization of fine elements for inexpensive storage reductions and
possible performance improvements}. Simultaneously, it comes with \emph{more
sophisticated schemes for graphs with more specific properties such as
separability.}


\section{CONCLUSION}

Reducing graph storage overheads is important in large-scale
computations. Yet, established schemes such as
WebGraph~\cite{boldi2004webgraph}
negatively impact performance.
To address this, 
we propose Log(Graph): a graph representation that
applies logarithmic storage lower
bounds to (aka ``logarithmizes'') various graph elements. 

First, logarithmizing fine elements offers simplicity and negligible
performance overheads or even speedups from reducing data transfers. It can
enhance virtually any graph processing engine in shared- and distributed-memory
settings. For example, we accelerate SSSP in the GAP
Benchmark~\cite{beamer2015gap} by $\approx$20\% while reducing the required
storage by 20-35\%.

To logarithmize offset or adjacency data, we use \emph{succinct data
structures}~\cite{Jacobson:1988:SSD:915547,
Okanohara07practicalentropycompressed} and ILP. We investigate the associated
tradeoffs and identify as well as tackle the related issues, enhancing the
processing and storing of both specific and general graphs. For example,
Log(Graph) outperforms WebGraph schemes
%
%
while nearly
matching its compression ratio with various schemes. We provide a carefully crafted and
extensible, high-performance implementation.


Finally, to the best of our knowledge, our work is the first performance
analysis of accessing succinct data structures in a parallel environment. It illustrates
surprising differences between succinct bit vectors and offset arrays when varying
the amount of parallelism.
Our insights can be used by both theoreticians and practitioners
to develop more efficient succinct schemes for 
parallel settings.

{\vspace{0em}\section*{Acknowledgements} We thank Juraj
Hromkovič for inspiring discussions and useful comments that helped
us improve the quality of the paper. We thank Guy Blelloch for providing us
with the source code of some succinct and compact designs, and Simon Gog for
help and useful information. We thank the CSCS
and ALCF teams granting access to the Piz Dora and Vesta machines, and for
their excellent technical support.}

\begin{figure*}
\centering
  \begin{subfigure}[t]{0.18 \textwidth}
    \centering
    \includegraphics[width=\textwidth]{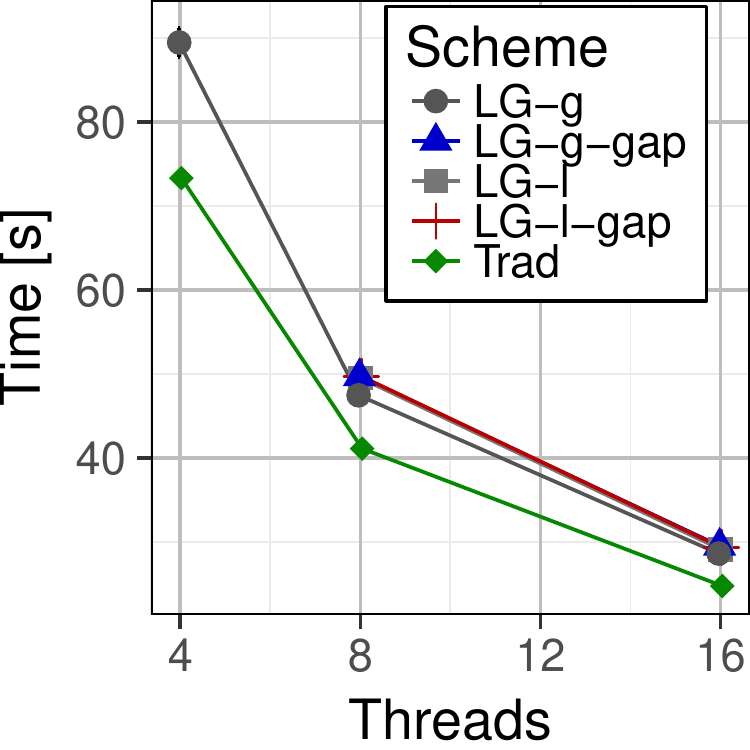}
    \caption{CC, fr.}
    \label{fig:x}
  \end{subfigure}
  \begin{subfigure}[t]{0.18 \textwidth}
    \centering
    \includegraphics[width=\textwidth]{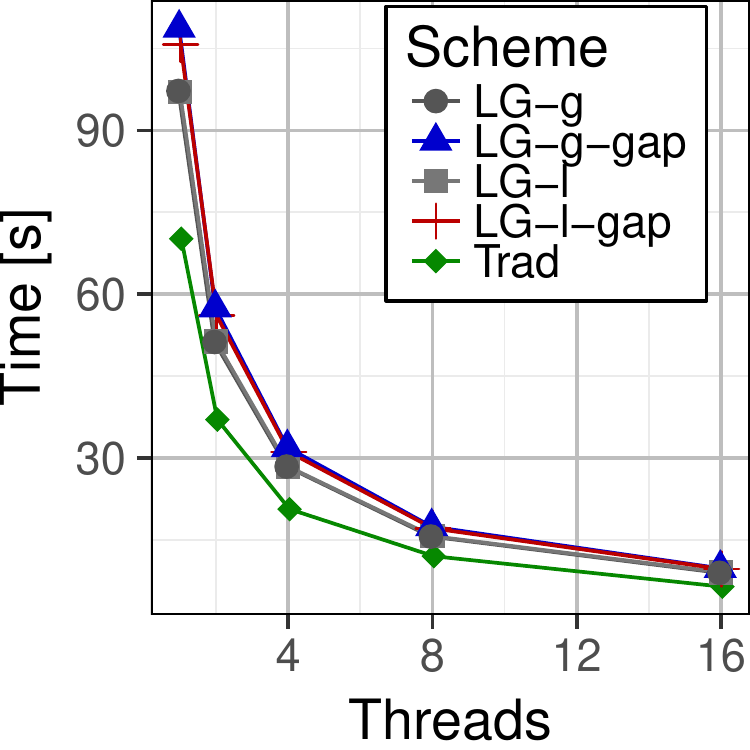}
    \caption{CC, uku.}
    \label{fig:x}
  \end{subfigure}
  \begin{subfigure}[t]{0.18 \textwidth}
    \centering
    \includegraphics[width=\textwidth]{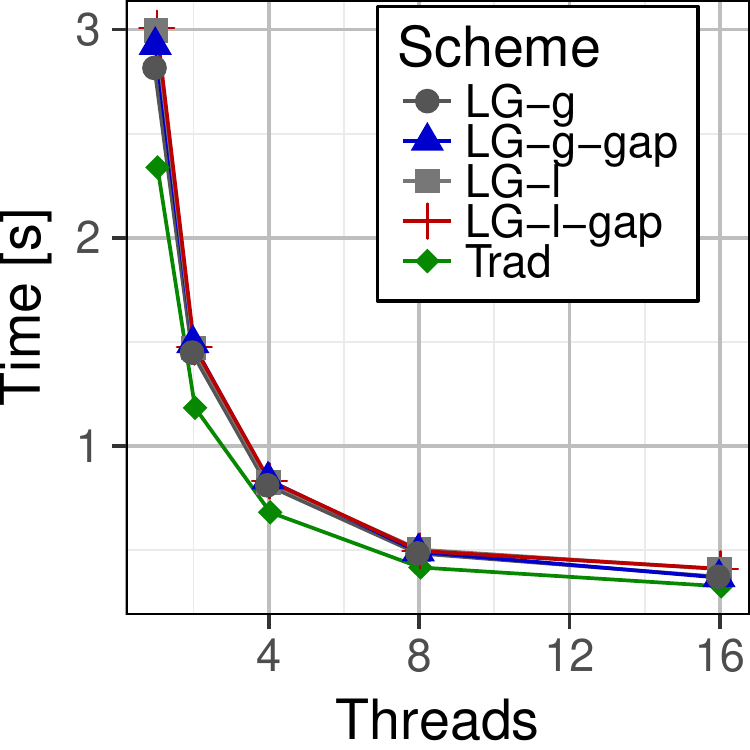}
    \caption{CC, usrn.}
    \label{fig:x}
  \end{subfigure}
  \begin{subfigure}[t]{0.18 \textwidth}
    \centering
    \includegraphics[width=\textwidth]{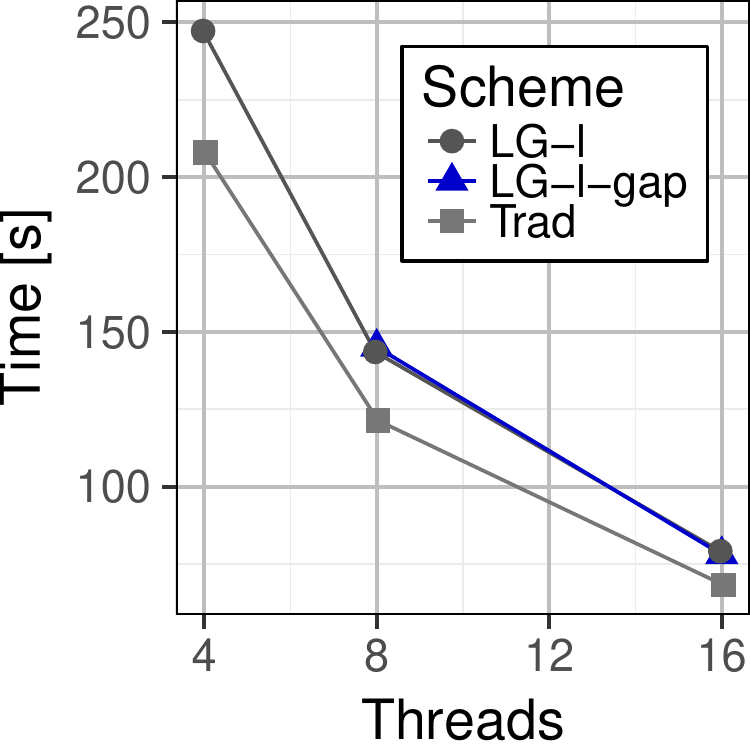}
    \caption{PR, fr.}
    \label{fig:x}
  \end{subfigure}
  \begin{subfigure}[t]{0.18 \textwidth}
    \centering
    \includegraphics[width=\textwidth]{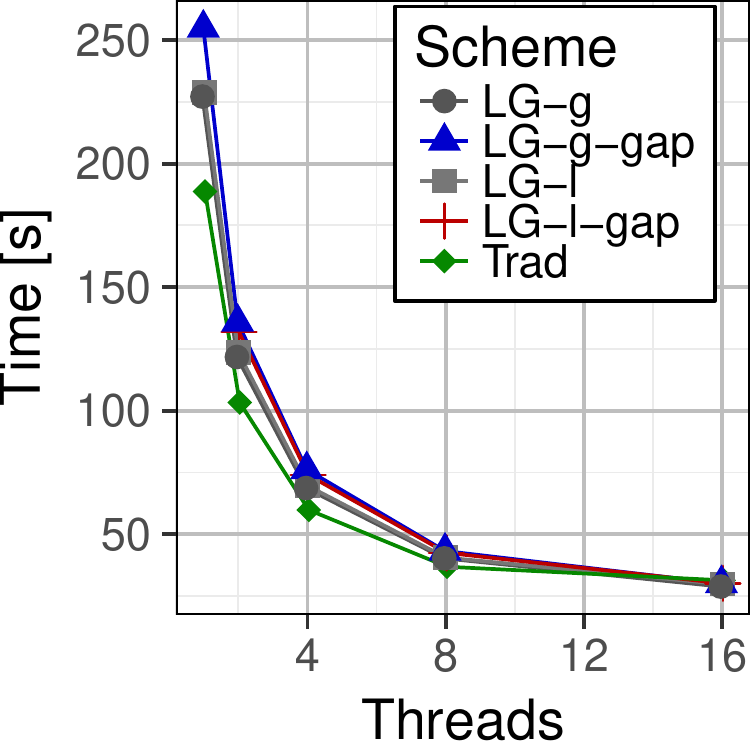}
    \caption{PR, uku.}
    \label{fig:x}
  \end{subfigure}
\\
  \begin{subfigure}[t]{0.18 \textwidth}
    \centering
    \includegraphics[width=\textwidth]{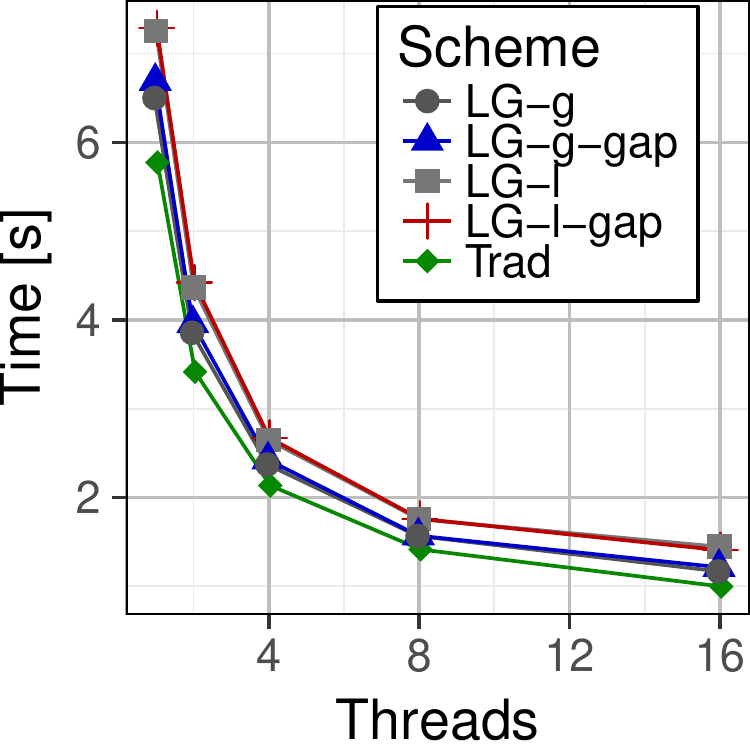}
    \caption{PR, usrn.}
    \label{fig:x}
  \end{subfigure}
  \begin{subfigure}[t]{0.18 \textwidth}
    \centering
    \includegraphics[width=\textwidth]{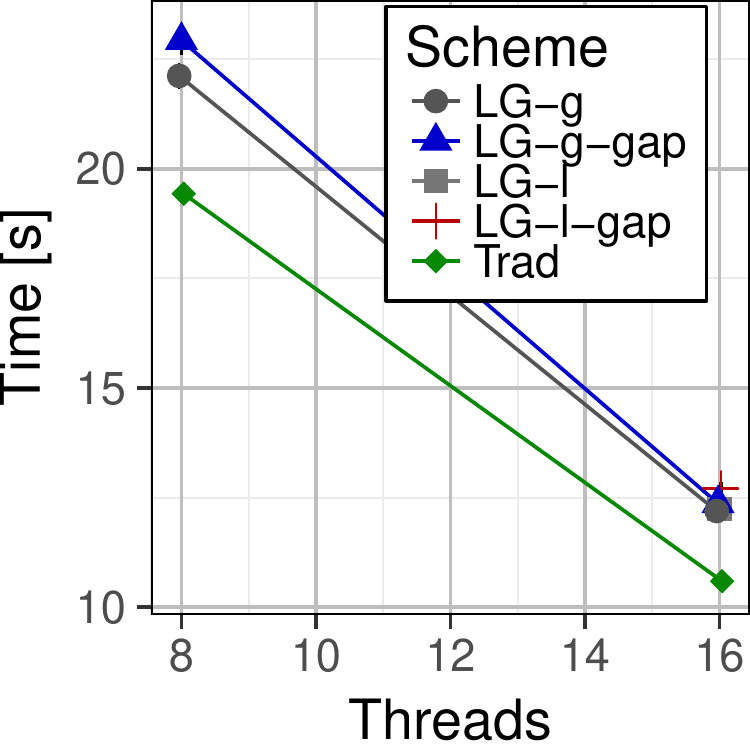}
    \caption{BC, fr.}
    \label{fig:x}
  \end{subfigure}
  \begin{subfigure}[t]{0.18 \textwidth}
    \centering
    \includegraphics[width=\textwidth]{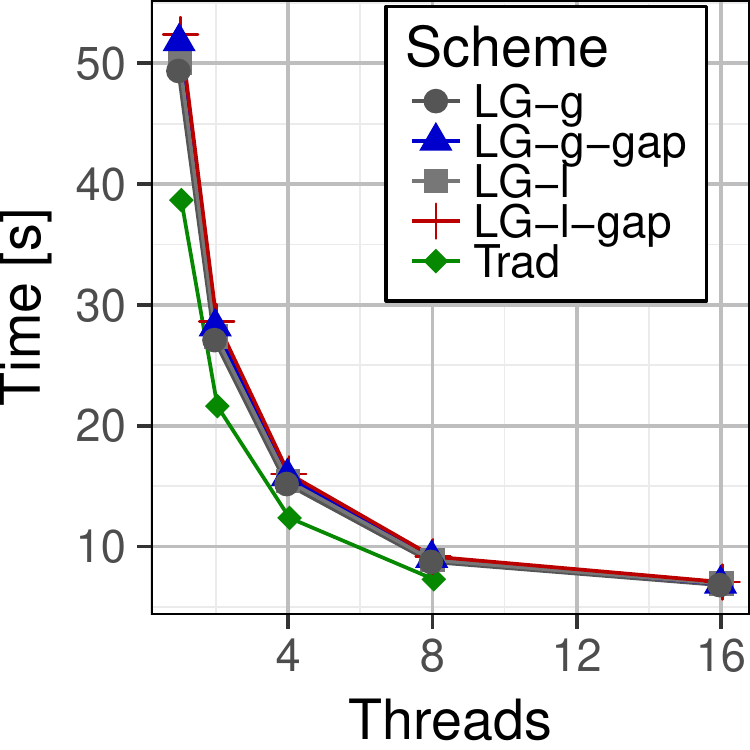}
    \caption{BC, uku.}
    \label{fig:x}
  \end{subfigure}
  \begin{subfigure}[t]{0.18 \textwidth}
    \centering
    \includegraphics[width=\textwidth]{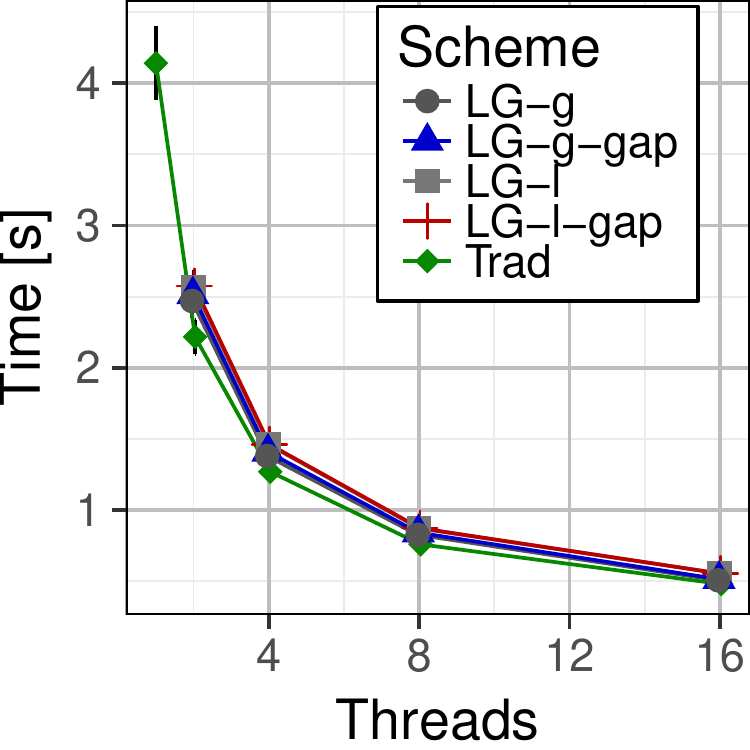}
    \caption{BFS, fr.}
    \label{fig:x}
  \end{subfigure}
  \begin{subfigure}[t]{0.18 \textwidth}
    \centering
    \includegraphics[width=\textwidth]{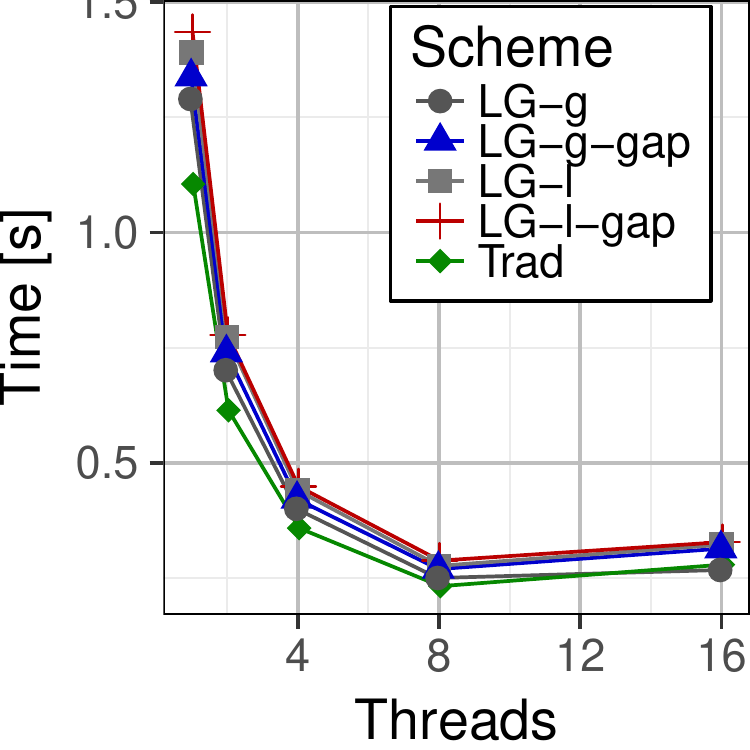}
    \caption{BFS, usrn.}
    \label{fig:x}
  \end{subfigure}
\\
  \begin{subfigure}[t]{0.18 \textwidth}
    \centering
    \includegraphics[width=\textwidth]{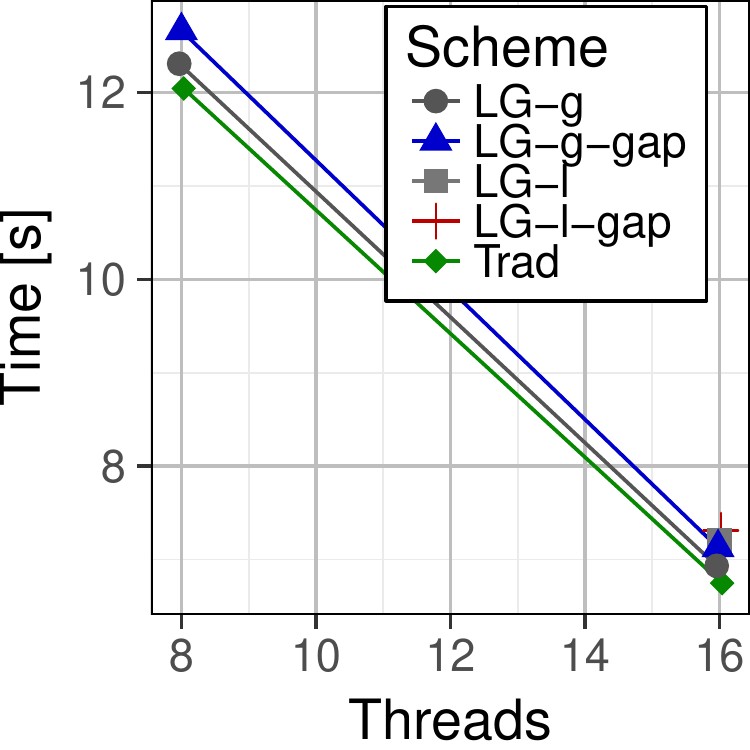}
    \caption{SSSP, fr.}
    \label{fig:x}
  \end{subfigure}
  \begin{subfigure}[t]{0.18 \textwidth}
    \centering
    \includegraphics[width=\textwidth]{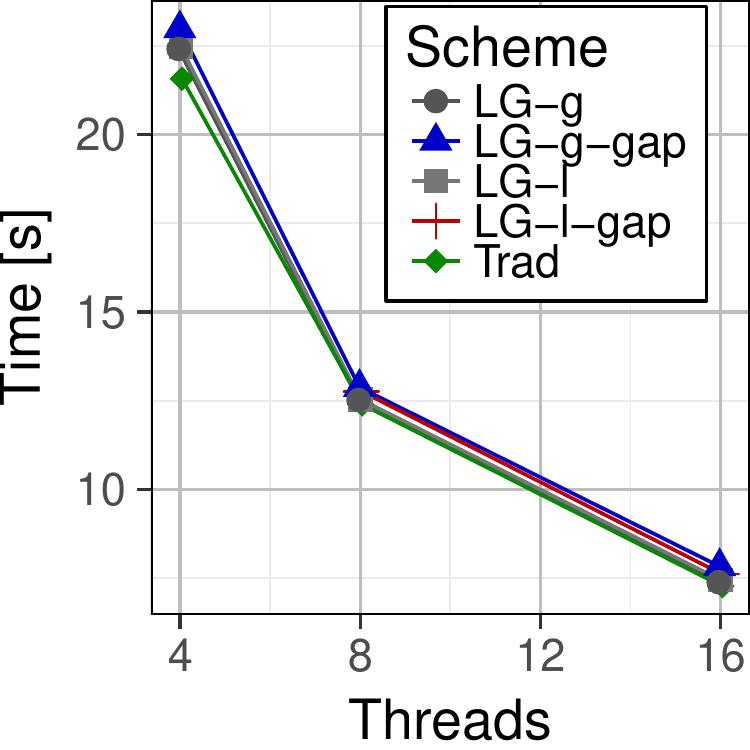}
    \caption{SSSP, uku.}
    \label{fig:x}
  \end{subfigure}
  \begin{subfigure}[t]{0.18 \textwidth}
    \centering
    \includegraphics[width=\textwidth]{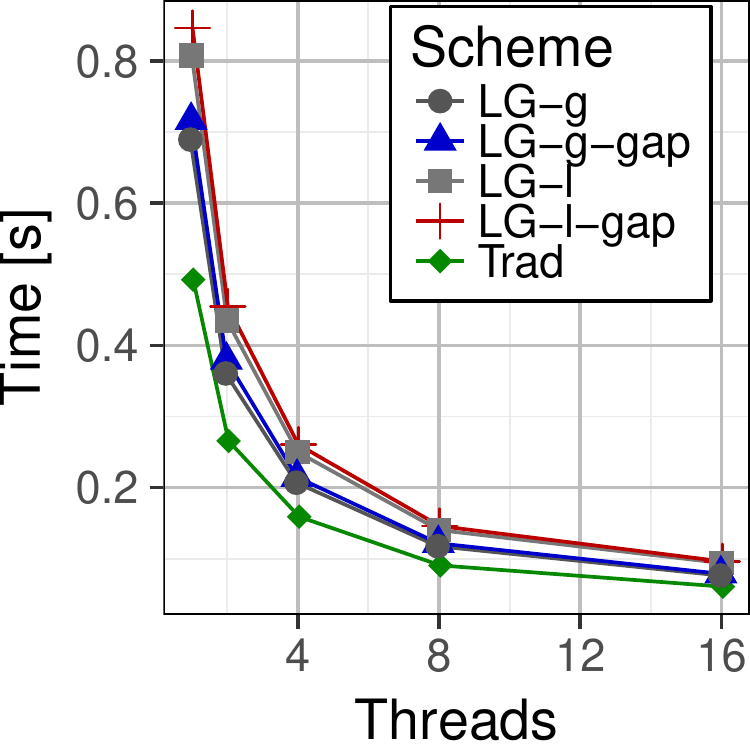}
    \caption{TC, usrn.}
    \label{fig:x}
  \end{subfigure}
%
\caption{(\cref{sec:A_more_anal_scalability}) The scalability analysis of Log(Graph), real-world graphs.}
\label{fig:rw-perf}
\end{figure*}


\section{APPENDIX}

\subsection{Theory: Additional Analyses}
\label{sec:more_theory}

%
Here, we first provide the derivation of the expressions for $\mathscr{O}$
and $\mathscr{A}$ for power-law graphs.
We assume that the minimum degree is 1 and also use the recent result that
bounds the maximum degree in a power-law graph with high $\left(1-\frac{1}{\log
n}\right)$ probability~\cite{besta2017slimsell}: $\hat{d} \le \left(\frac{\alpha n \log
n}{\beta - 1}\right)^{\frac{1}{\beta - 1}}$ . We now aim to derive an
expression for $m$ as a function of $\alpha$ and $\beta$.  Using the degree
distribution we have

\begin{gather}
m = \frac{1}{2} \sum_{x = 1}^{\hat{d}} x f(x) = \frac{1}{2} \sum_{x = 1}^{\hat{d}} \alpha x^{1-\beta}
\end{gather}
\normalsize

This can be approximated with an integral

\begin{gather}
m \approx \frac{1}{2} \int_{x = 1}^{\hat{d}} \alpha x^{1-\beta} dx = \frac{\alpha}{2} \frac{1}{(2 - \beta)} \left(\hat{d} ^{2 - \beta} - 1\right)
\end{gather}
\normalsize

Plugging this into the storage expression, we obtain

\begin{gather}
E[|\mathscr{A}|] \approx \frac{\alpha}{2-\beta} \left(\left(\frac{\alpha n \log n}{\beta - 1}\right)^{\frac{2-\beta}{\beta - 1}} - 1\right) \left(\left\lceil \log n \right\rceil + \left\lceil \log \rwh{\mathcal{W}} \right\rceil\right) \\
E[|\mathscr{O}|] \approx n \left\lceil \log \left(\frac{\alpha}{2-\beta} \left( \left(\frac{\alpha n \log n}{\beta - 1}\right)^{\frac{2-\beta}{\beta - 1}} - 1 \right) \right) \right\rceil
\end{gather}
\normalsize

\begin{figure*}
\centering
\includegraphics[width=0.8\textwidth]{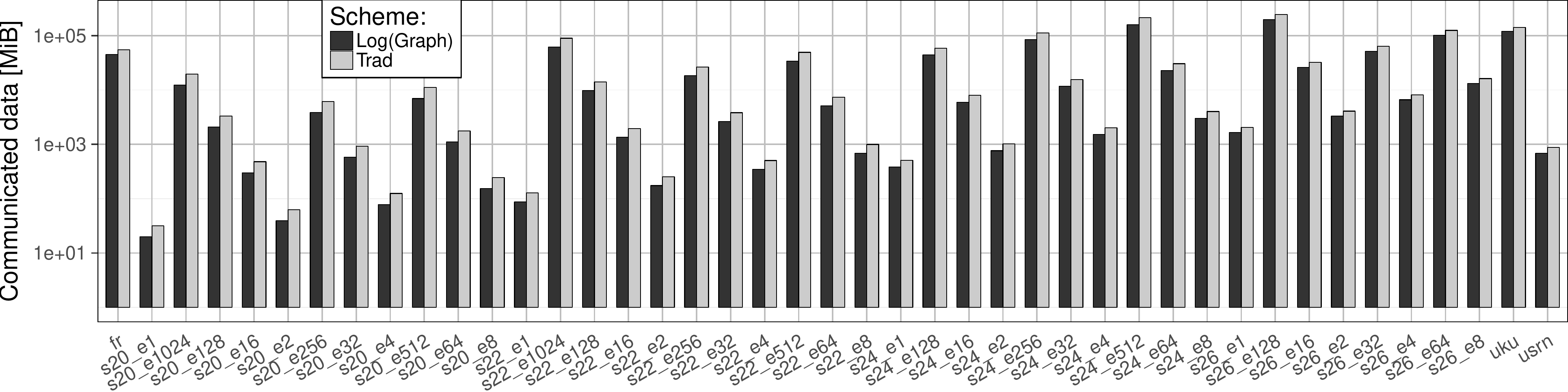}
\caption{(\cref{sec:A_more_anal_data}) The amount of data communicated in a
distributed-memory BFS for 1,024 compute nodes when logarithmizing fine-grained
graph elements.}
\label{fig:lossless-dist}
%
\end{figure*}

\begin{figure*}[t!]
\centering
\includegraphics[width=1.0\textwidth]{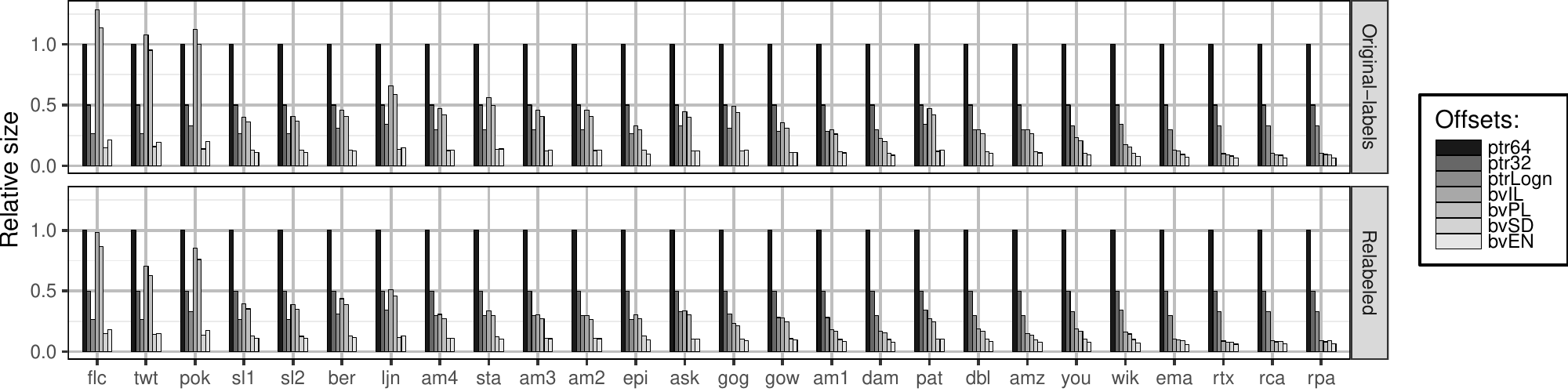}
 \caption{(\cref{sec:O_more_anal}) Illustration of the size of various $\mathscr{O}$ with and without relabeling vertex IDs (with a traditional adjacency data structure $\mathscr{A}$).} 
\label{fig:size-analysis-no-changes}
\end{figure*}


\begin{figure*}[!t]
\centering
\begin{subfigure}[t]{0.22 \textwidth}
  \includegraphics[width=\textwidth]{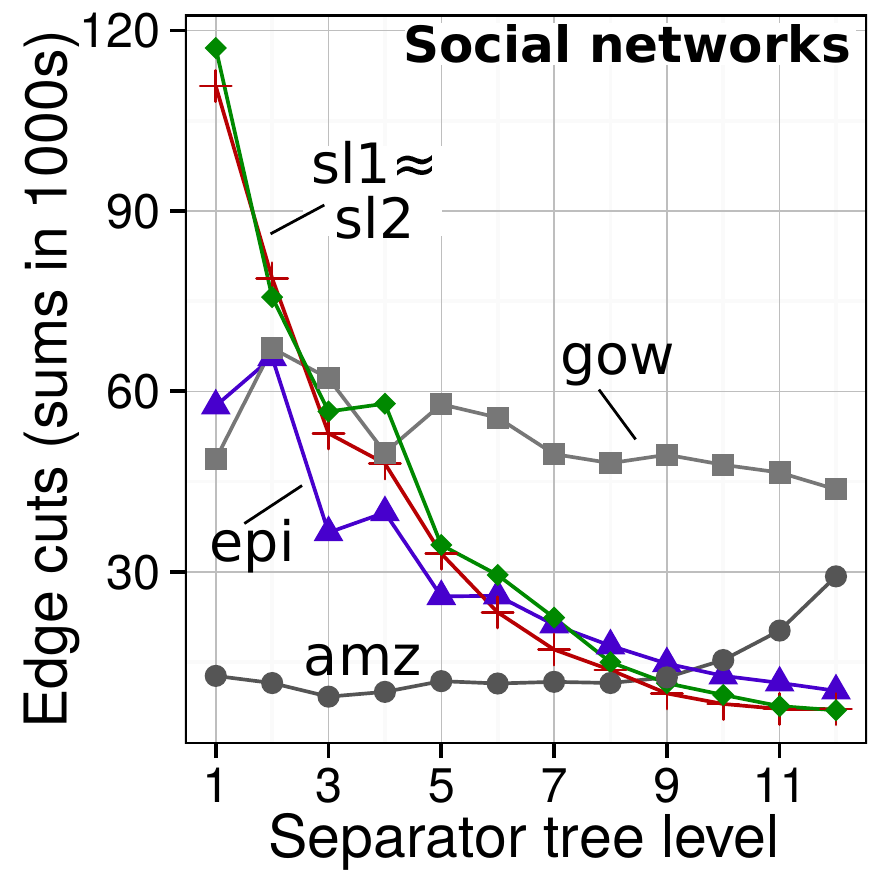}
 \caption{Edge cuts.}
  \label{fig:cut-lvl-analysis-sn-e-1}
 \end{subfigure}
\begin{subfigure}[t]{0.22 \textwidth}
  \includegraphics[width=\textwidth]{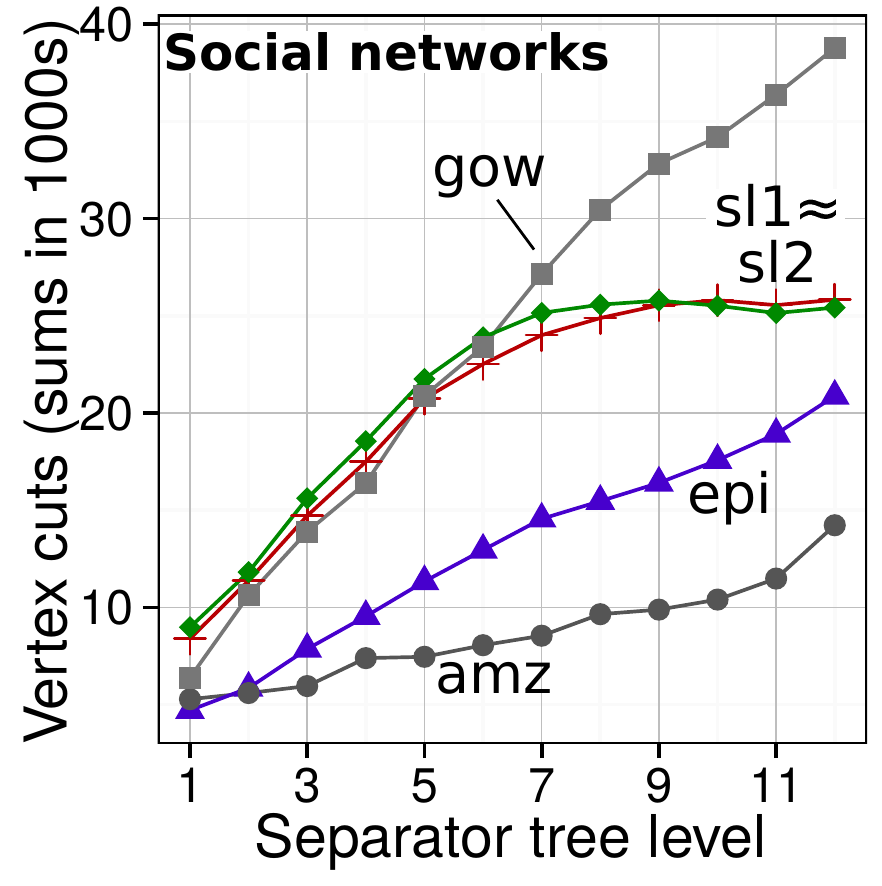}
 \caption{Vertex cuts.}
  \label{fig:cut-lvl-analysis-sn-v-1}
 \end{subfigure}
\begin{subfigure}[t]{0.22 \textwidth}
  \includegraphics[width=\textwidth]{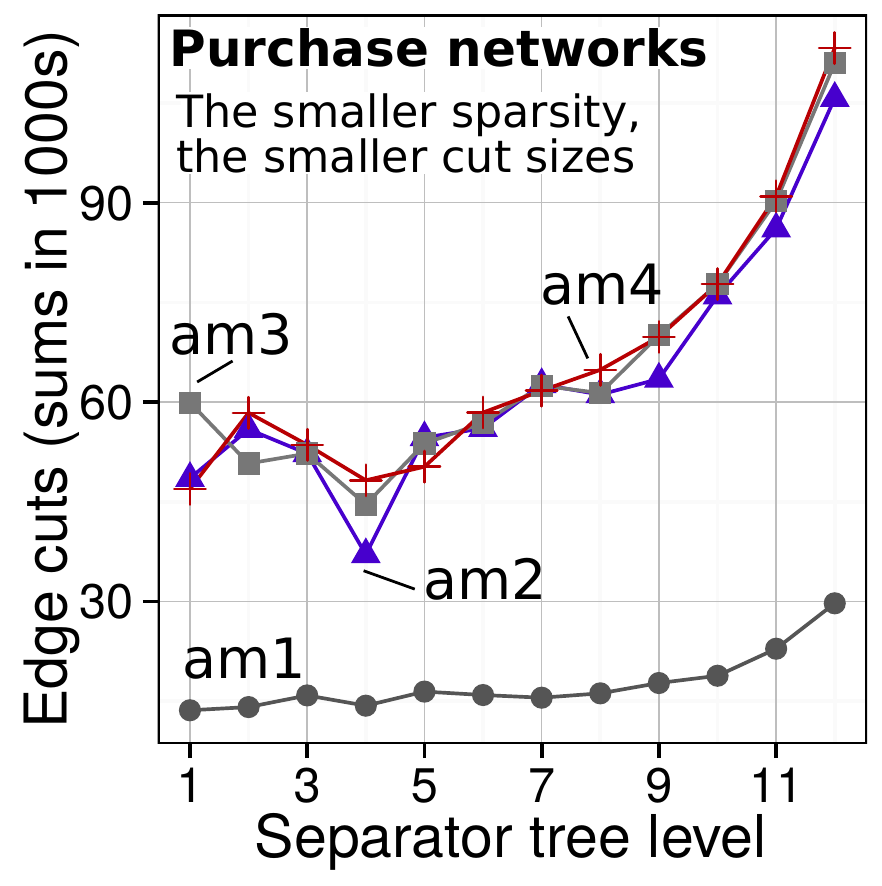}
 \caption{Edge cuts.}
  \label{fig:cut-lvl-analysis-pn-e-1}
 \end{subfigure}
\\
\begin{subfigure}[t]{0.22 \textwidth}
  \includegraphics[width=\textwidth]{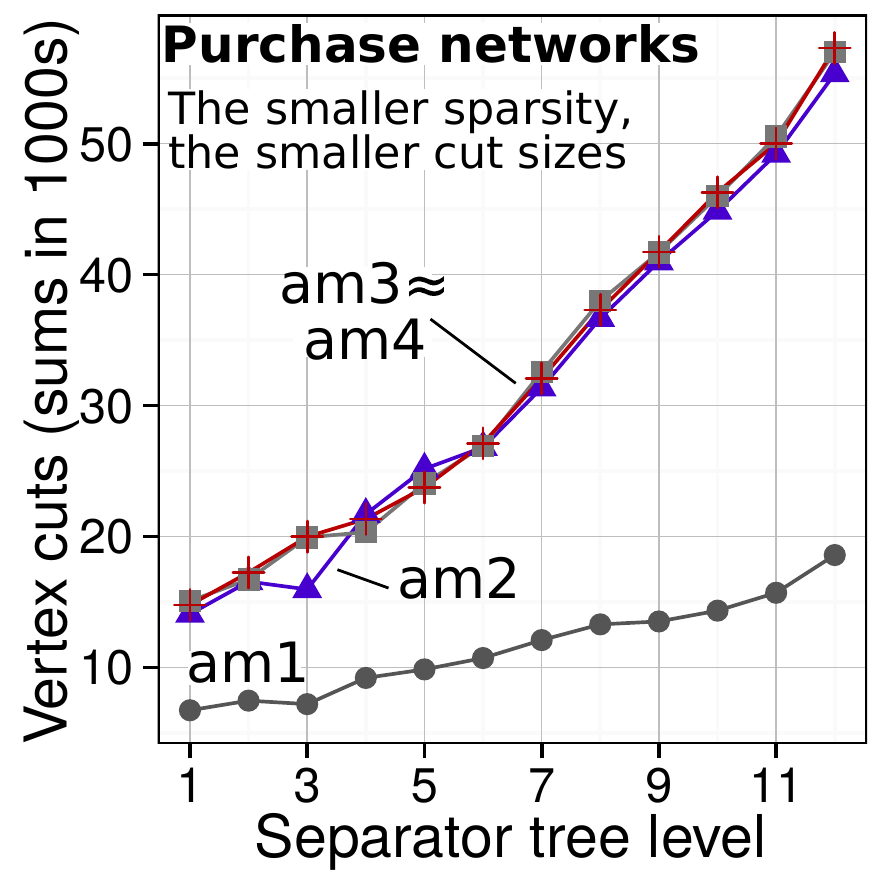}
 \caption{Vertex cuts.}
  \label{fig:cut-lvl-analysis-pn-v-1}
 \end{subfigure}
\begin{subfigure}[t]{0.22 \textwidth}
  \includegraphics[width=\textwidth]{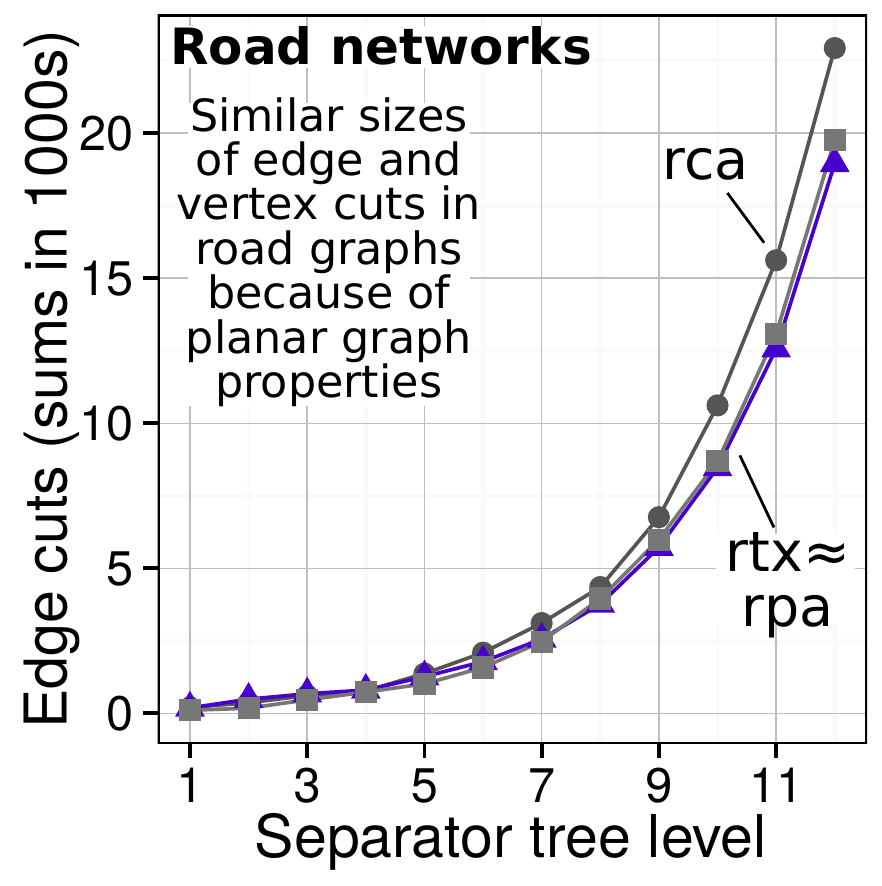}
 \caption{Edge cuts.}
  \label{fig:cut-lvl-analysis-rn-e-1}
 \end{subfigure}
\begin{subfigure}[t]{0.22 \textwidth}
  \includegraphics[width=\textwidth]{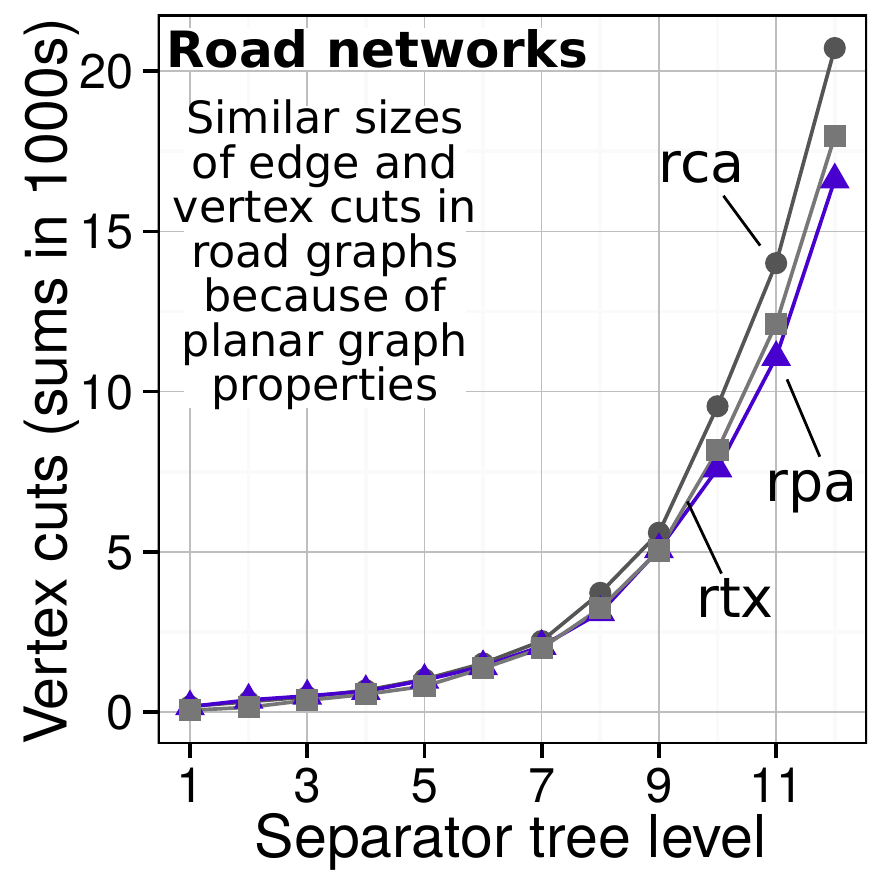}
 \caption{Vertex cuts.}
  \label{fig:cut-lvl-analysis-rn-v-1}
 \end{subfigure}
\caption{(\cref{sec:A_more_anal_cuts}) An illustration of sums of edge and vertex cuts at various levels of
separator trees in different types of graphs.} 
\label{fig:holistic_analysis_cuts_recursive}
\end{figure*}

\begin{figure*}[!t]
\begin{subfigure}[t]{0.35 \textwidth}
  \includegraphics[width=\textwidth]{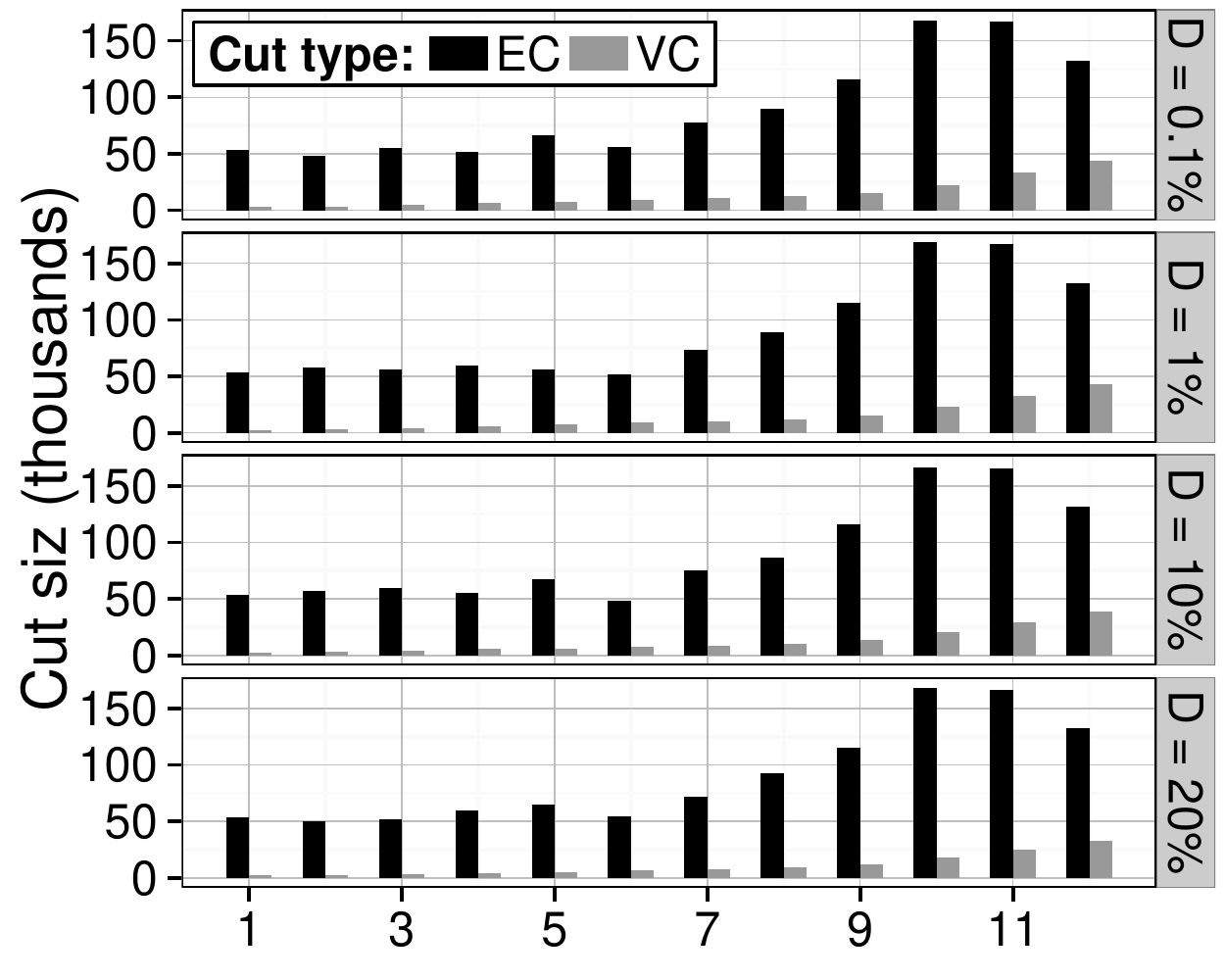}
 \caption{Graph \textsf{twt}; $\bar{d} \approx 20$.}
  \label{fig:offsets_bal_twt}
  \end{subfigure}
\begin{subfigure}[t]{0.35 \textwidth}
  \includegraphics[width=\textwidth]{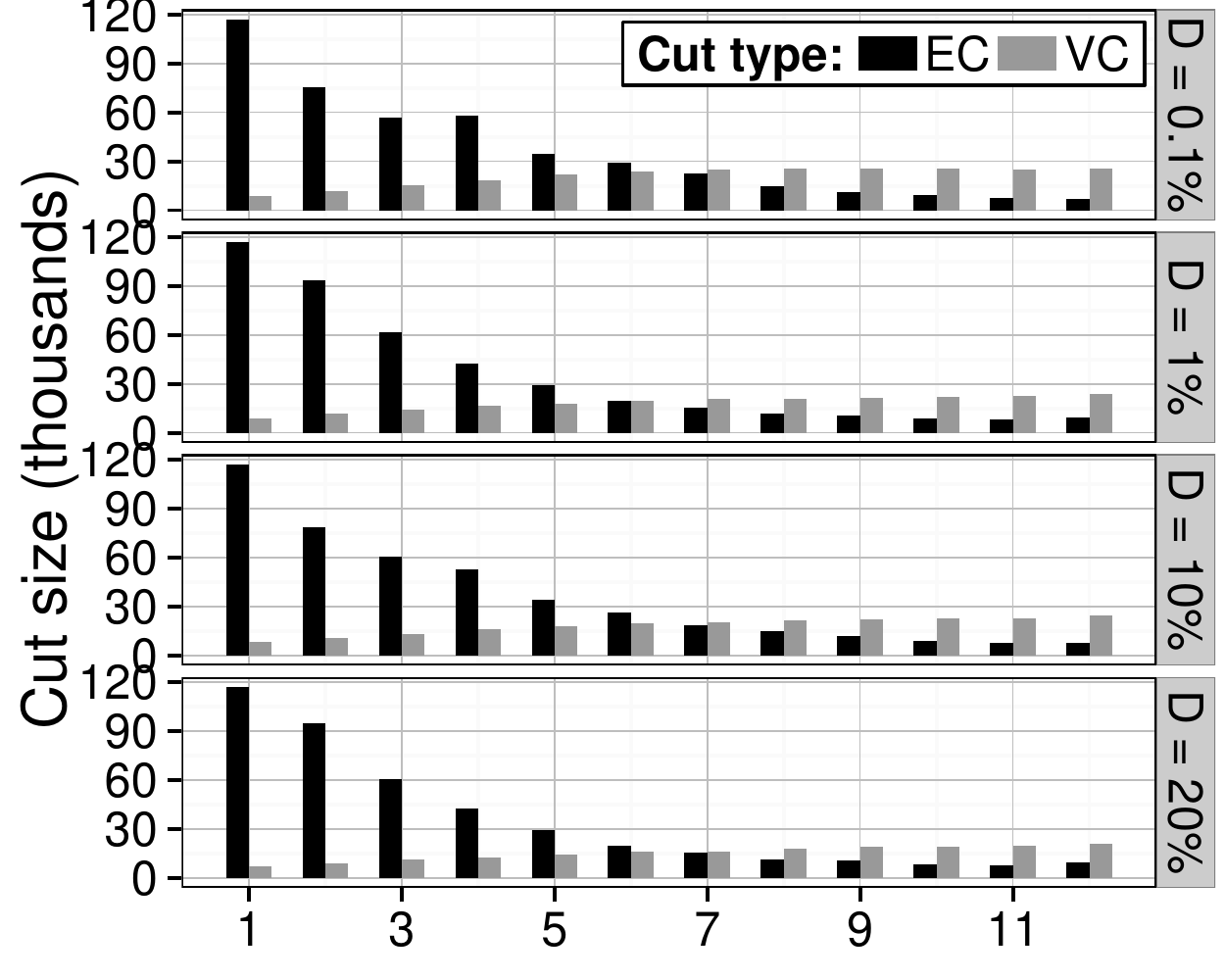}
 \caption{Graph \textsf{sl2}; $\bar{d} \approx 10$.}
  \label{fig:offsets_bal_sl2}
 \end{subfigure}
\\
\begin{subfigure}[t]{0.35 \textwidth}
  \includegraphics[width=\textwidth]{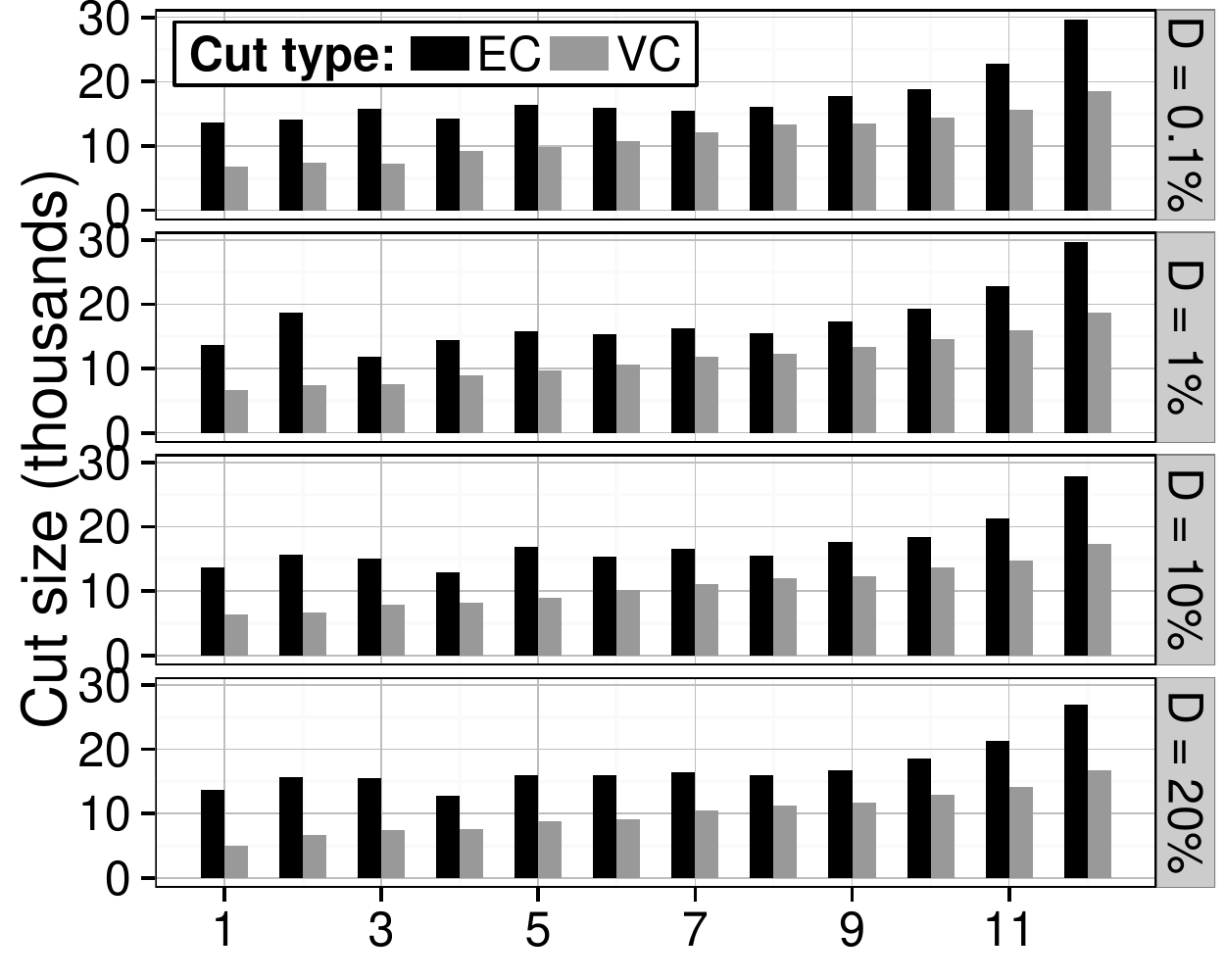}
 \caption{Graph \textsf{am1}; $\bar{d} \approx 5$.}
  \label{fig:offsets_bal_am1}
 \end{subfigure}
\begin{subfigure}[t]{0.35 \textwidth}
  \includegraphics[width=\textwidth]{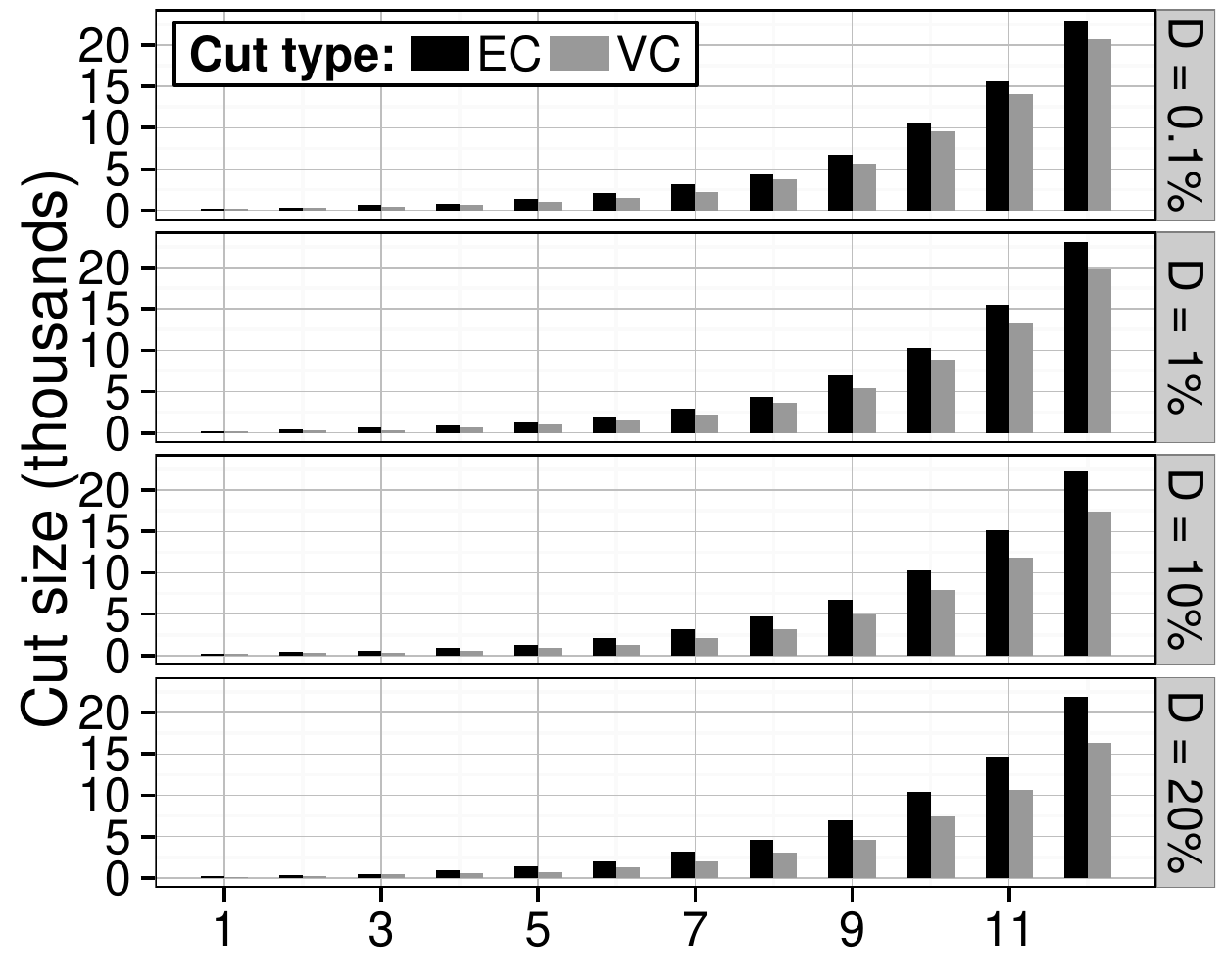}
 \caption{Graph \textsf{rca}; $\bar{d} \approx 1.5$.}
  \label{fig:offsets_bal_rca}
 \end{subfigure}
\caption{(\cref{sec:A_more_anal_relaxing}) Illustration of vertex/edge cut sizes 
when varying
the balancedness ratio $\mathcal{D}$ and levels of respective separator trees.}
%
\label{fig:balancedness_analysis}
\end{figure*}

\begin{figure*}[t]
\centering
 \begin{subfigure}[t]{0.48 \textwidth}
  \includegraphics[width=\textwidth]{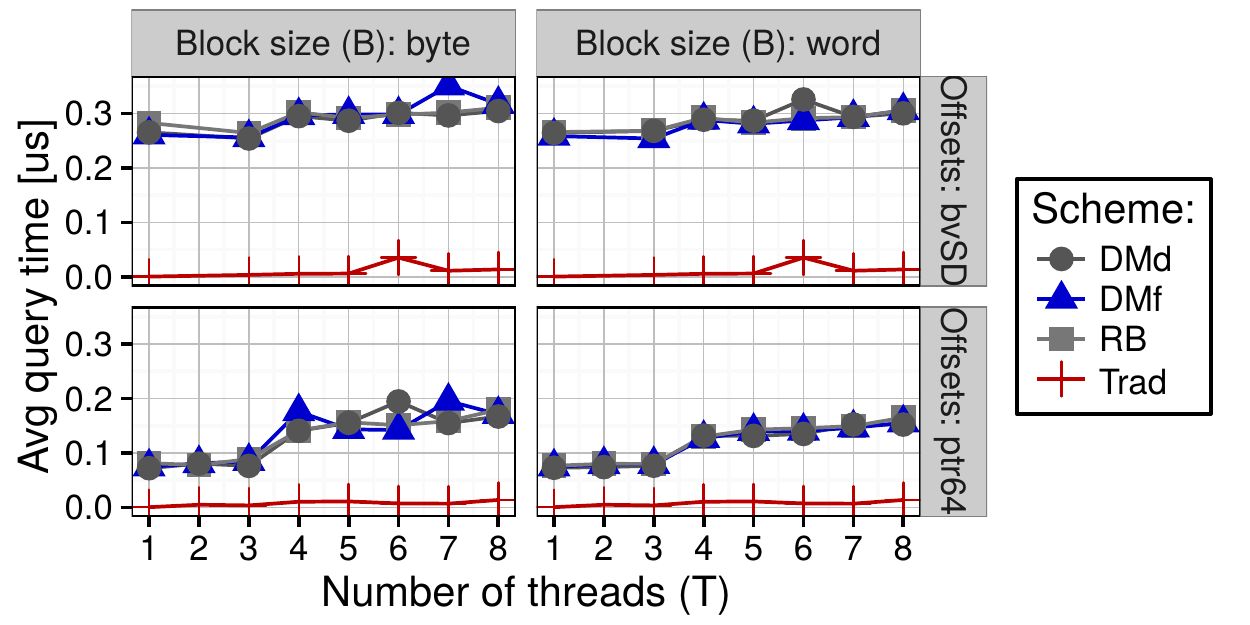}
 \caption{The analysis of obtaining $d_v$; \textsf{pok} graph.}
  \label{fig:perf_A_deg}
 \end{subfigure} 
 \begin{subfigure}[t]{0.48 \textwidth}
   \includegraphics[width=\textwidth]{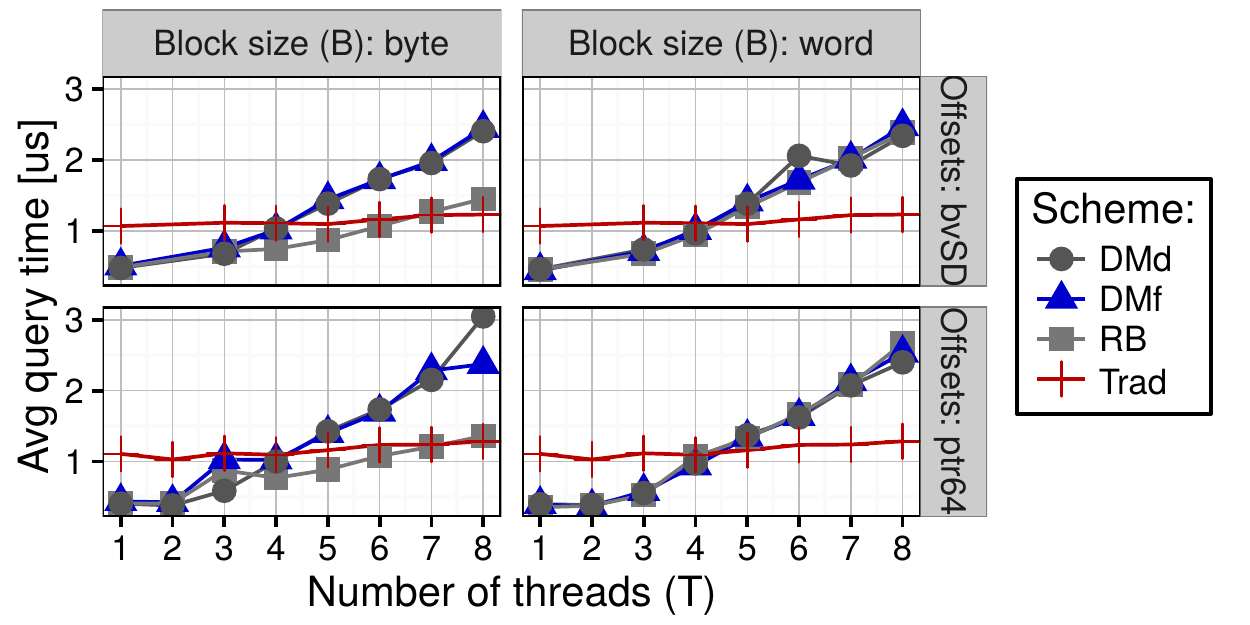}
  \caption{The analysis of obtaining $N_v$; \textsf{pok} graph.}
   \label{fig:perf_A_N}
  \end{subfigure}
%
%
%
\caption{(\cref{sec:A_more_anal_queries}) Performance analysis of accessing a graph in a parallel setting.}
%
\end{figure*}

\subsection{Logarithmizing Fine Elements: Additional Analyses}
\label{sec:FINE_more_anal}

We now analyze in more detail how logarithmizing fine elements impacts
aspects such as scalability or communicated data.

\subsubsection{Investigating Scalability}
\label{sec:A_more_anal_scalability}

We also provide the results of scalability analyses. We vary the number of threads $T$
for various real-world graphs, see in Figure~\ref{fig:rw-perf}.

\subsubsection{Investigating Distributed-Memory Settings}
\label{sec:A_more_anal_data}

Finally, we also present results that show how Log(Graph) reduces the amount of
communicated data in a distributed-memory environment when logarithmizing fine-grained
graph elements. This approach benefits from explicitly considering
the locality of data~\cite{tate2014programming}. The results are illustrated in Figure~\ref{fig:lossless-dist}.

\subsection{Logarithmizing $\mathscr{O}$: Additional Analyses}
\label{sec:O_more_anal}

In the main body of the work, we have only analyzed the influence of
$\mathscr{O}$'s properties on $|\mathscr{O}|$. Yet, the scope of the interplay
between \textsf{AA}'s parameters is much broader: $|\mathscr{O}|$ is also
impacted by $\mathscr{A}$'s design. Specifically, if $\mathscr{A}$ is encoded
using a scheme that shrinks adjacency arrays, such as Blandford's RB scheme,
then the size of $\mathscr{O}$ based on pointer arrays should remain the same
(as $W$ is fixed) while \textsf{bvPL} and \textsf{bvIL} should shrink.
Succinct designs are harder to predict due to more complex dependencies; for
example \textsf{bvSD}'s length also gets smaller but as the ratio of ones to
zeros gets higher, $|\mathscr{O}|$ may as well increase; a similar argument
applies to \textsf{bvEN}.
We now analyze these effects by comparing the original $|\mathscr{O}|$ to
$|\mathscr{O}|$ after relabeling of vertices according to the inorder traversal
of the separator tree as performed in the Blandford's scheme; see
Figure~\ref{fig:size-analysis-no-changes}. As expected, all offset arrays
remain identical because they only depend on fixed parameters. 
Contrarily, all the bit vectors shrink (e.g., \textsf{bvPL} and \textsf{bvSD}
are on average $\approx$25\% and $\approx$12\% smaller). This is because the
relabelled $\mathscr{A}$ uses less storage, requiring shorter bit vectors.
%

\subsection{Logarithmizing $\mathscr{A}$: Additional Analyses}
\label{sec:A_more_anal}

We also illustrate more analyses related to logarithmizing $\mathscr{A}$.

\subsubsection{Using Vertex Cuts Instead of Edge Cuts}
\label{sec:A_more_anal_cuts}

So far, we have only considered edge cuts (ECs) in the considered recursive partitioning schemes (RB and BRB). Yet, as explained
in~\cref{sec:bf_scheme},
vertex cuts (VCs) can also be incorporated to enhance RB. They
seem especially attractive as it can be proven that they are always smaller or
equal than the corresponding ECs~\cite{west2001introduction}. In our setting,
this relationship is more complicated as we partition graphs recursively and
the correspondence between ECs and VCs is lost.
Thus, we first compute the total sums of sizes of ECs and VCs at various levels
of respective separator trees, see
Figure~\ref{fig:holistic_analysis_cuts_recursive}. 
%
%
We ensure that the respective partitions are of almost equal sizes. We did not
find strong correlation between cut sizes and sparsities $\overline{d}$.  We
group selected representative graphs into social networks (SNs), purchase
networks (PNs), and road graphs (RNs).
For most SNs, ECs start from numbers much larger than in the corresponding VCs,
and then steadily decrease. This is due to vertices with very high
degrees whose edges are cut early during recursive partitioning. Contrarily,
VCs in SNs start from very small values and then grow. This is because it is
easy to partition initial input SNs using VCs as they are rich in
communities~\cite{yang2015defining}, but later on, as communities
become harder to find, cut sizes grow.
Then, ECs and VCs in PNs follow similar patterns; they increase
together with levels of separator trees. This suggests that in both cases it
becomes more difficult to find clusters after several initial partitioning
rounds. Finally, ECs and VCs in RNs differ marginally because these
graphs are almost planar.

We conclude that, in most cases, VCs are significantly smaller than ECs, being
potentially a more appealing tool in reducing $|\mathscr{A}|$ because less
information crosses partitions and has to be encoded as large differences in
RB. Yet, $\mathscr{A}$ based on VCs introduces redundancy as now some vertices
are present in graph separators as well as in graph partitions. It requires
additional lookup structures with \emph{shadow
pointers}~\cite{Blandford:2003:CRS:644108.644219} for mapping between such
vertex clones (i.e., \emph{shadow vertex
trees}~\cite{Blandford:2003:CRS:644108.644219}). We calculated the number of
shadow pointers in such structures and the resulting storage overheads in
$|\mathscr{A}|$ based on VCs.
%
%
All the graphs follow similar trends. For example, \textsf{twt} requires 3.53M
additional pointers, giving 4.81MB for $|\mathscr{A}|$ (VCs, \textsf{ptrLogn}),
as opposed to 3.1MB (ECs).
%
%
Thus, the additional complexity from shadow vertex trees removes advantages
from smaller cuts, motivating us to focus on ECs.

\subsubsection{Relaxing Balancedness of Partitions}
\label{sec:A_more_anal_relaxing}

While bisecting $G$, we now let the maximum relative imbalance
between the partition sizes be at most $\mathcal{D}$. We first analyze
how changing $\mathcal{D}$ influences sizes of ECs and VCs
for each level of separator trees. We plot the findings for representative graphs in
  Figure~\ref{fig:balancedness_analysis}. 
As expected, cuts become smaller with 
growing $\mathcal{D}$: more imbalance more often prevents partitioning clusters.
Yet, these differences in sizes of both ECs and VCs are surprisingly small.
For example, the difference between ECs for \textsf{twt} (level 1) is only
around $\approx$2\%; other cases follow similar patterns. They do not impact
the final $|\mathscr{A}|$, resulting in minor ($\approx$1\%) differences.

\subsubsection{Approaching the Optimal Labeling with POD/CMB}
\label{sec:A_more_anal_opt}

\begin{figure}
  \centering
  \includegraphics[width=0.3\textwidth]{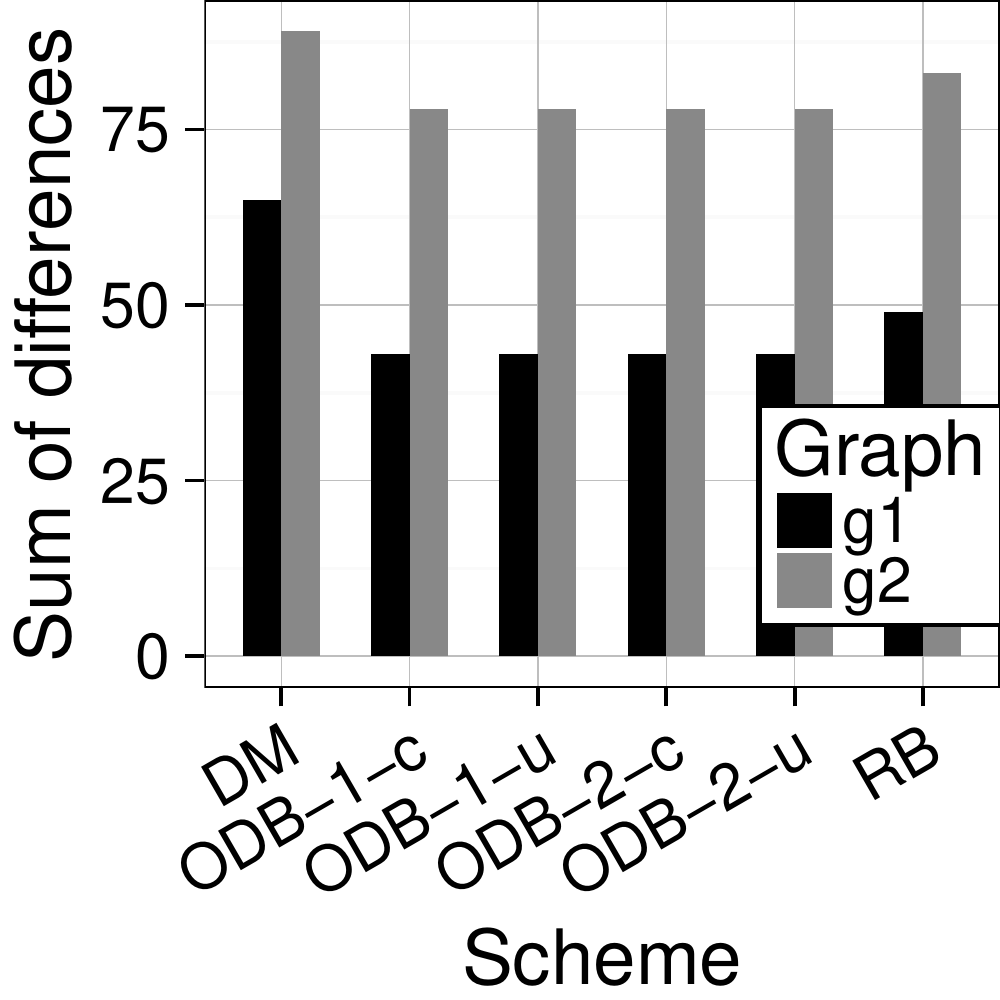} 
\caption{(\cref{sec:A_more_anal_opt}) Analysis of POD/CMB.}
  \label{fig:ilp}
\end{figure}

We now use POD and HYB to approach optimal labeling and outperform RB and DM.
%
We use IBM CPLEX~\cite{cplex2009v12} to solve the ILP problems formulated
in~\cref{sec:odb_scheme} and~\cref{sec:pod_scheme}.
We use two graphs
\textsf{g1} and \textsf{g2},
both consisting of two communities with few (<$0.2m$) edges in-between.  
%
%
Here, we illustrate (Figure~\ref{fig:ilp}) that POD/HYB do improve upon RB and DM
by reducing sums of differences between consecutive labels. Each scheme is
denoted as \texttt{ODB-y-z}: 
%
%
\texttt{y} indicates the variant of the objective function (\texttt{y=1} for
Eq.~(\ref{eq:obj_fun_1}) and \texttt{y=2} for Eq.~(\ref{eq:obj_fun_2})) and
\texttt{z} determines if we force the obtained labels to be contiguous
(\texttt{z=c}) or not (\texttt{z=u}).
Each proposed scheme finds a better labeling than RB (by 5-10\%) and
DM (by 30-40\%).

\subsubsection{Investigating Performance of Graph Accesses}
\label{sec:A_more_anal_queries}

Here, we present the full results of the performance of obtaining
$d_v$ and $N_v$ that we discussed briefly in~\cref{sec:eval_log_A}.

\begin{figure*}
\centering
  \begin{subfigure}[t]{0.2 \textwidth}
    \centering
    \includegraphics[width=\textwidth]{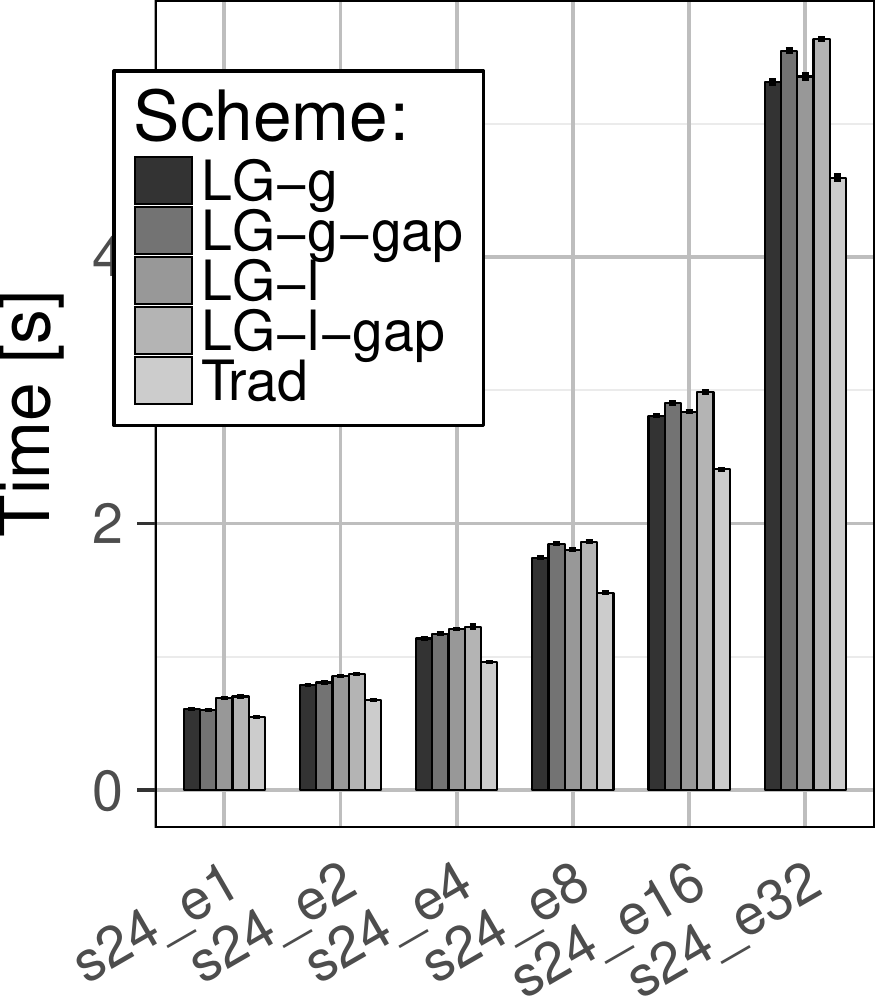}
    \caption{PR, sparse graphs.}
    \label{fig:xx}
  \end{subfigure}
  \begin{subfigure}[t]{0.2 \textwidth}
    \centering
    \includegraphics[width=\textwidth]{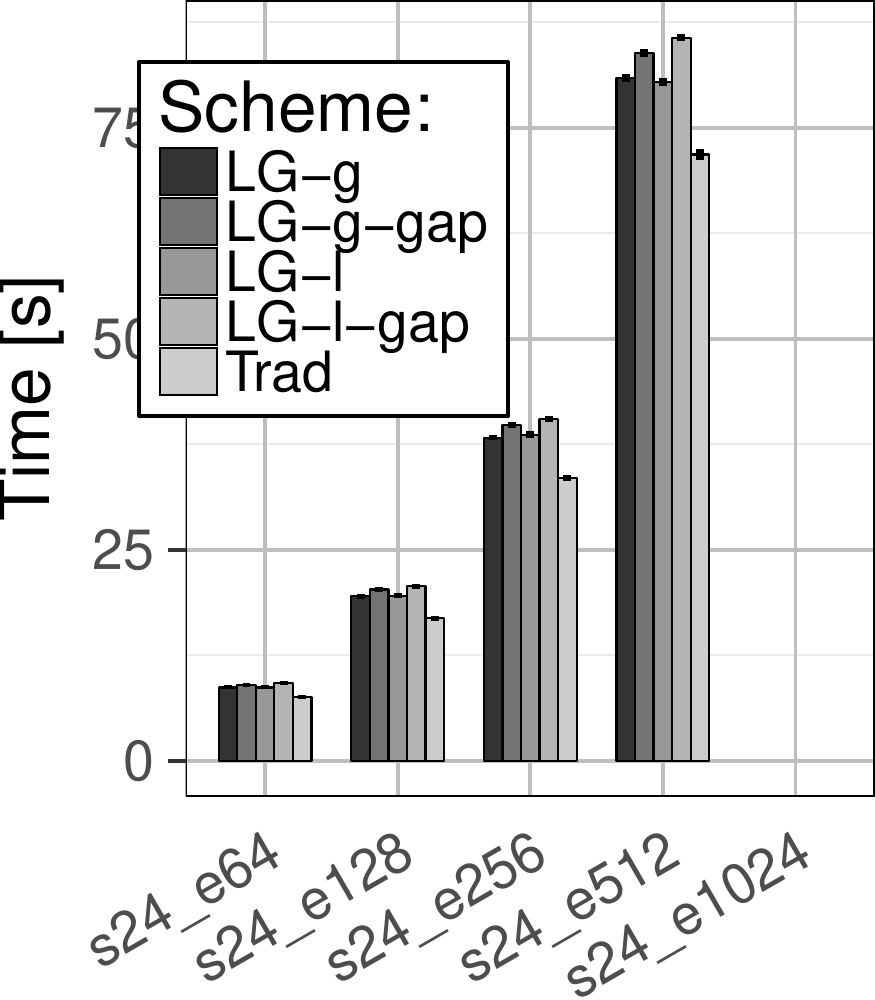}
    \caption{PR, dense graphs.}
    \label{fig:xx}
  \end{subfigure}
  \begin{subfigure}[t]{0.2 \textwidth}
    \centering
    \includegraphics[width=\textwidth]{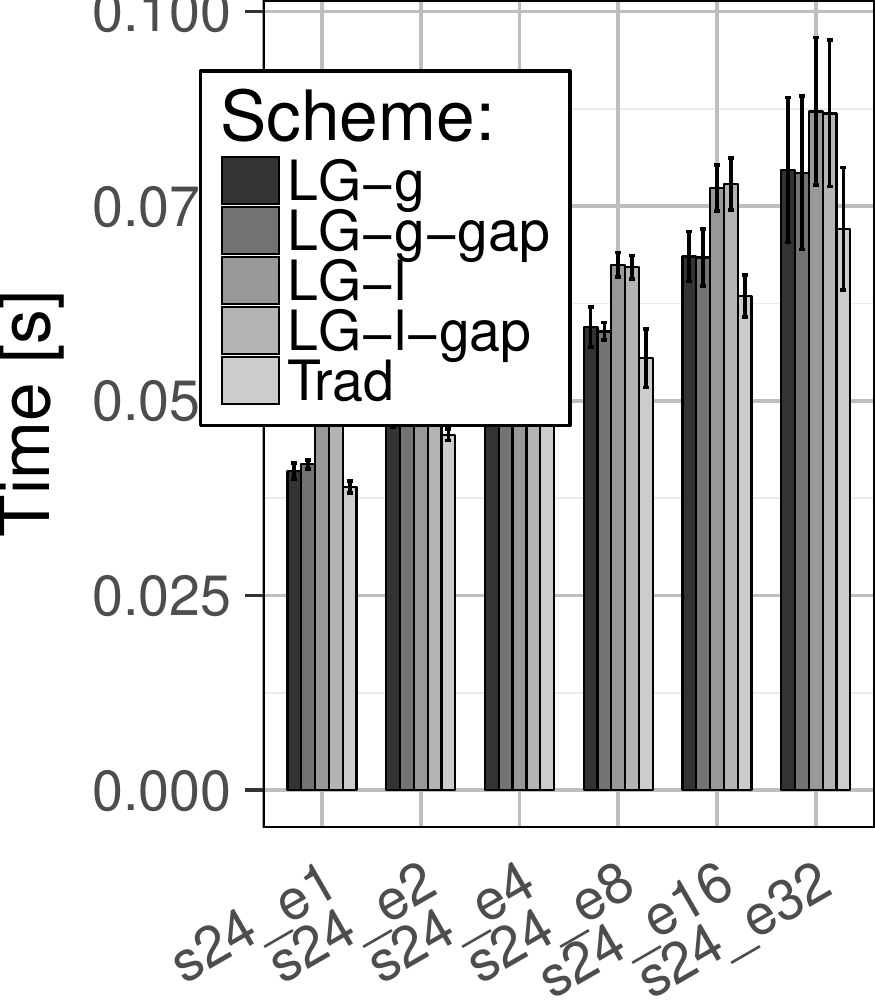}
    \caption{BFS, sparse graphs.}
    \label{fig:xx}
  \end{subfigure}
\\
  \begin{subfigure}[t]{0.2 \textwidth}
    \centering
    \includegraphics[width=\textwidth]{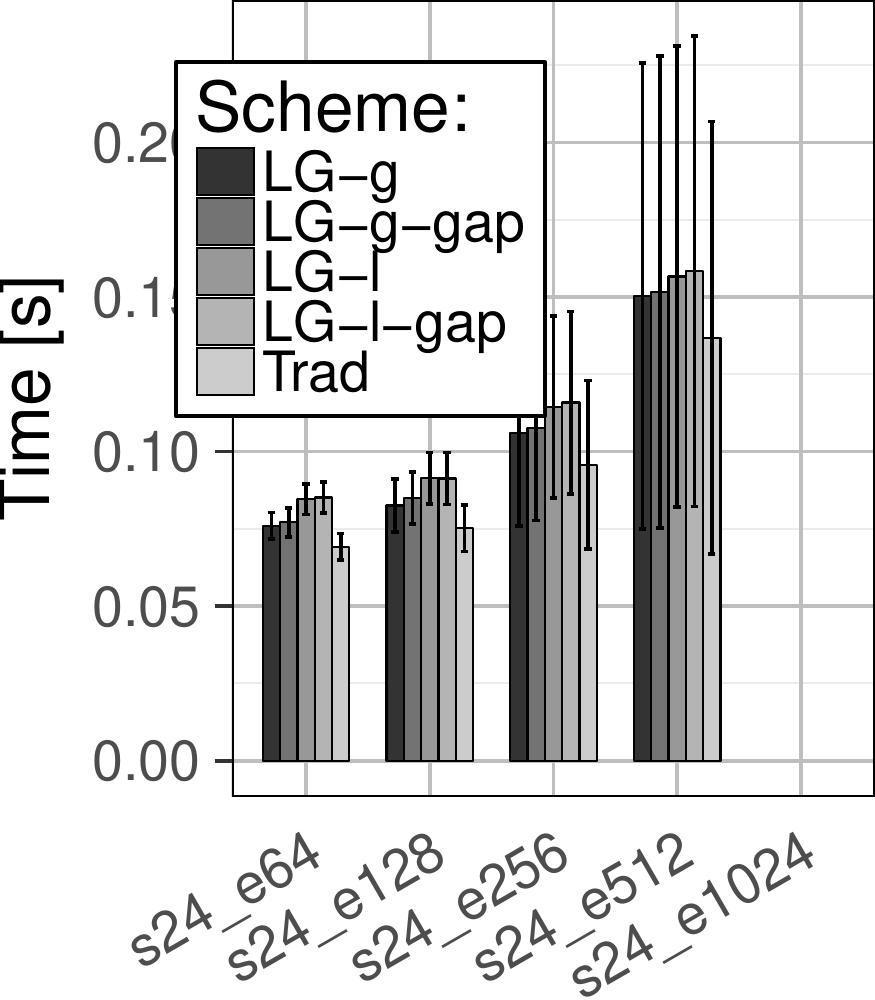}
    \caption{BFS, dense graphs.}
    \label{fig:xx}
  \end{subfigure}
  \begin{subfigure}[t]{0.2 \textwidth}
    \centering
    \includegraphics[width=\textwidth]{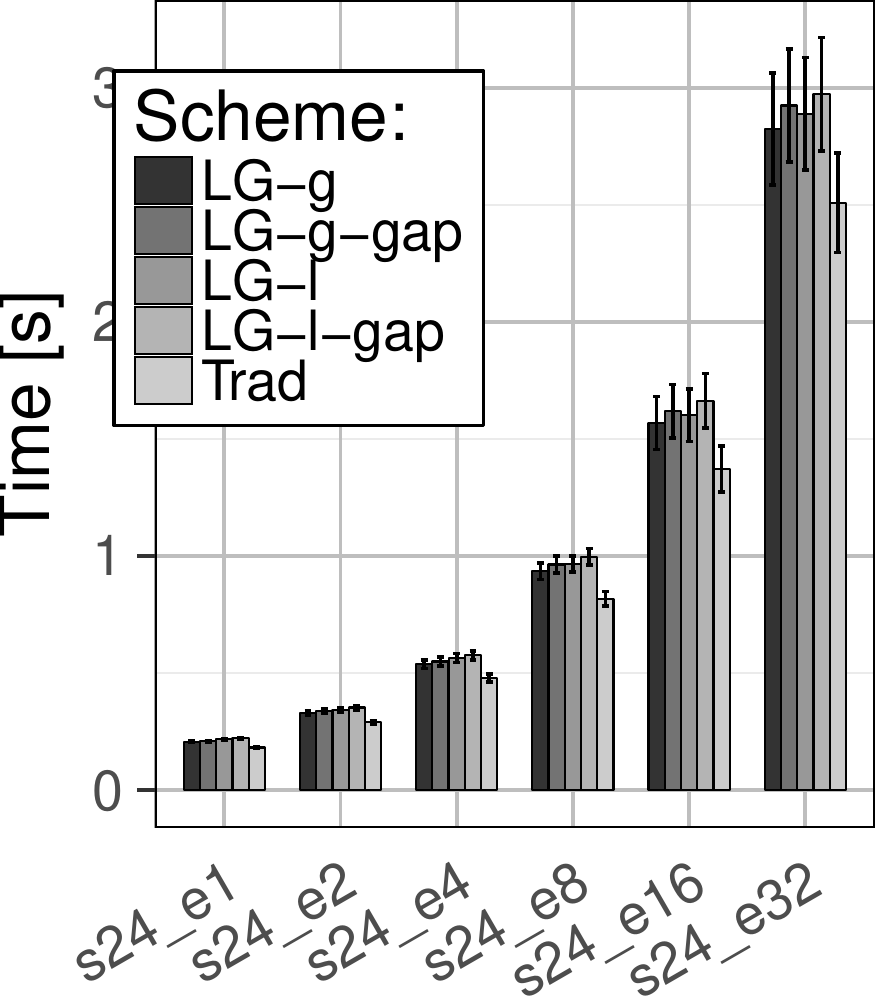}
    \caption{BC, sparse graphs.}
    \label{fig:xx}
  \end{subfigure}
  \begin{subfigure}[t]{0.2 \textwidth}
    \centering
    \includegraphics[width=\textwidth]{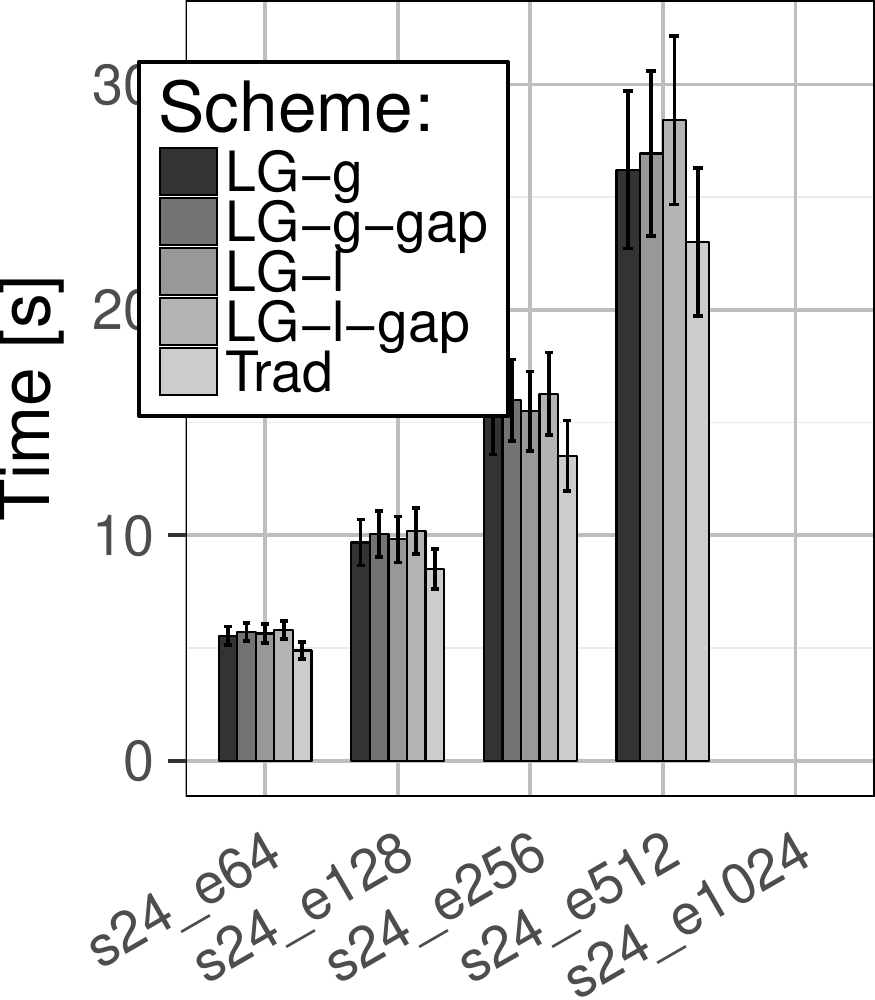}
    \caption{BC, dense graphs.}
    \label{fig:xx}
  \end{subfigure}
\\
  \begin{subfigure}[t]{0.2 \textwidth}
    \centering
    \includegraphics[width=\textwidth]{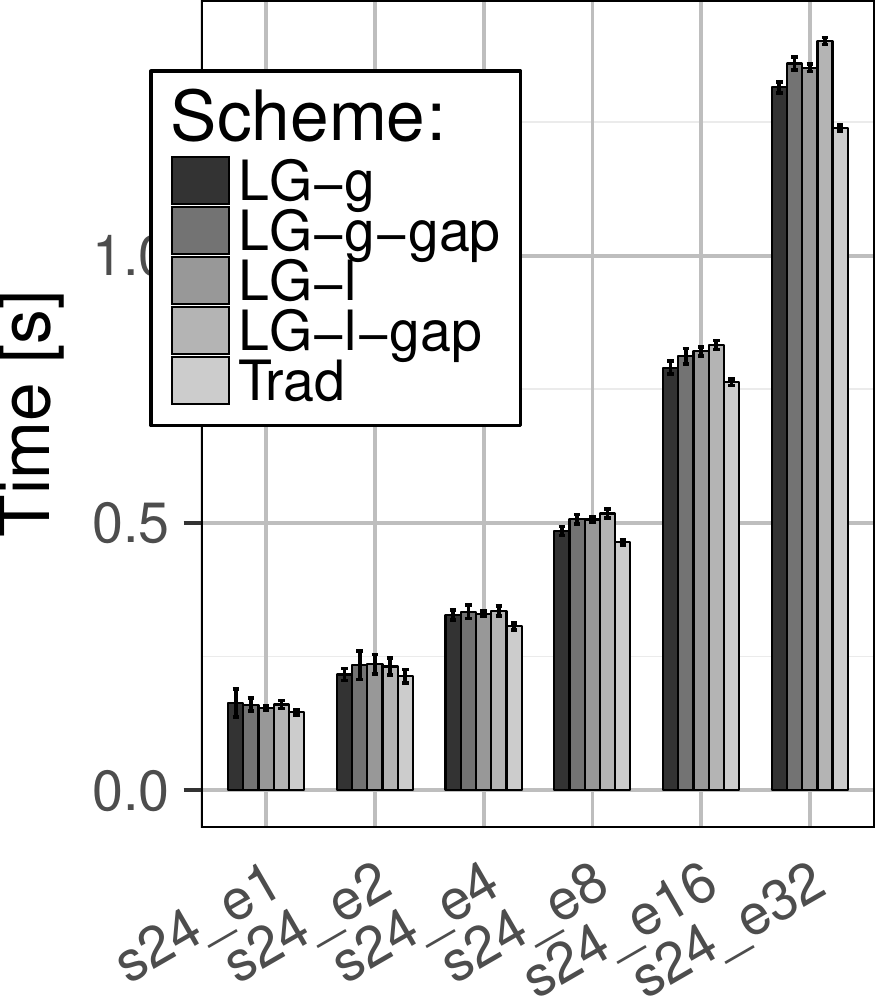}
    \caption{SSSP, sparse graphs.}
    \label{fig:xx}
  \end{subfigure}
  \begin{subfigure}[t]{0.2 \textwidth}
    \centering
    \includegraphics[width=\textwidth]{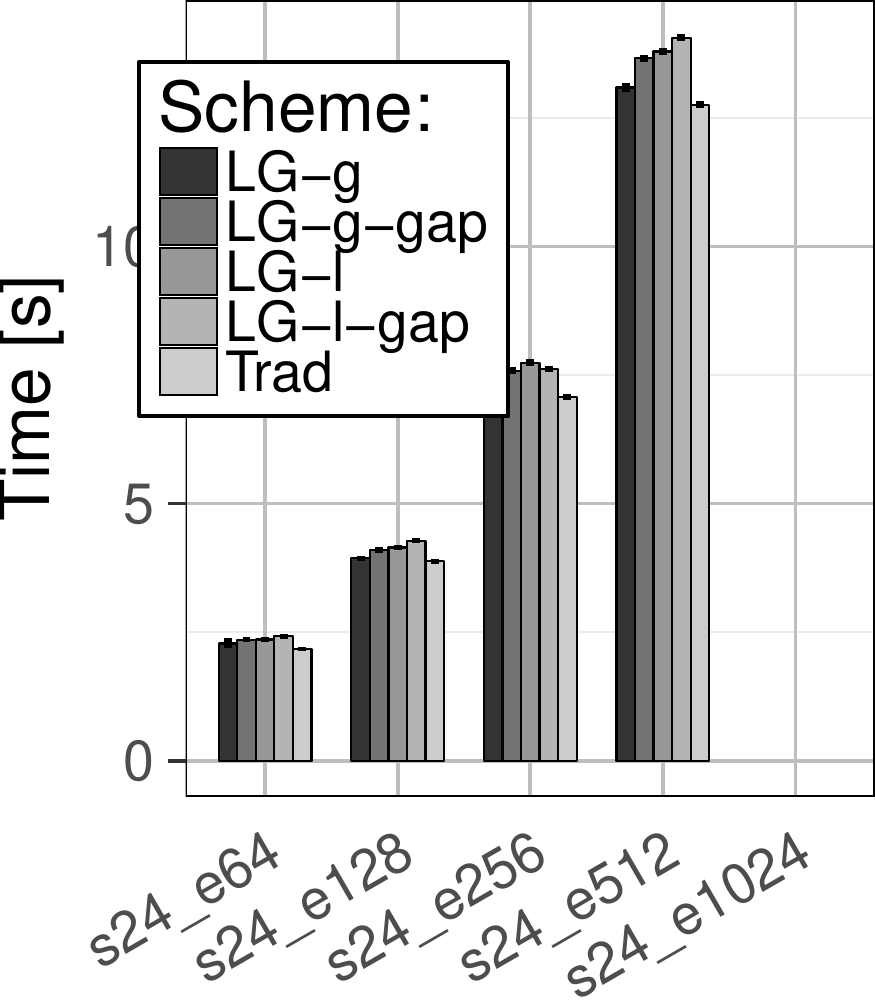}
    \caption{SSSP, dense graphs.}
    \label{fig:xx}
  \end{subfigure}
  \begin{subfigure}[t]{0.2 \textwidth}
    \centering
    \includegraphics[width=\textwidth]{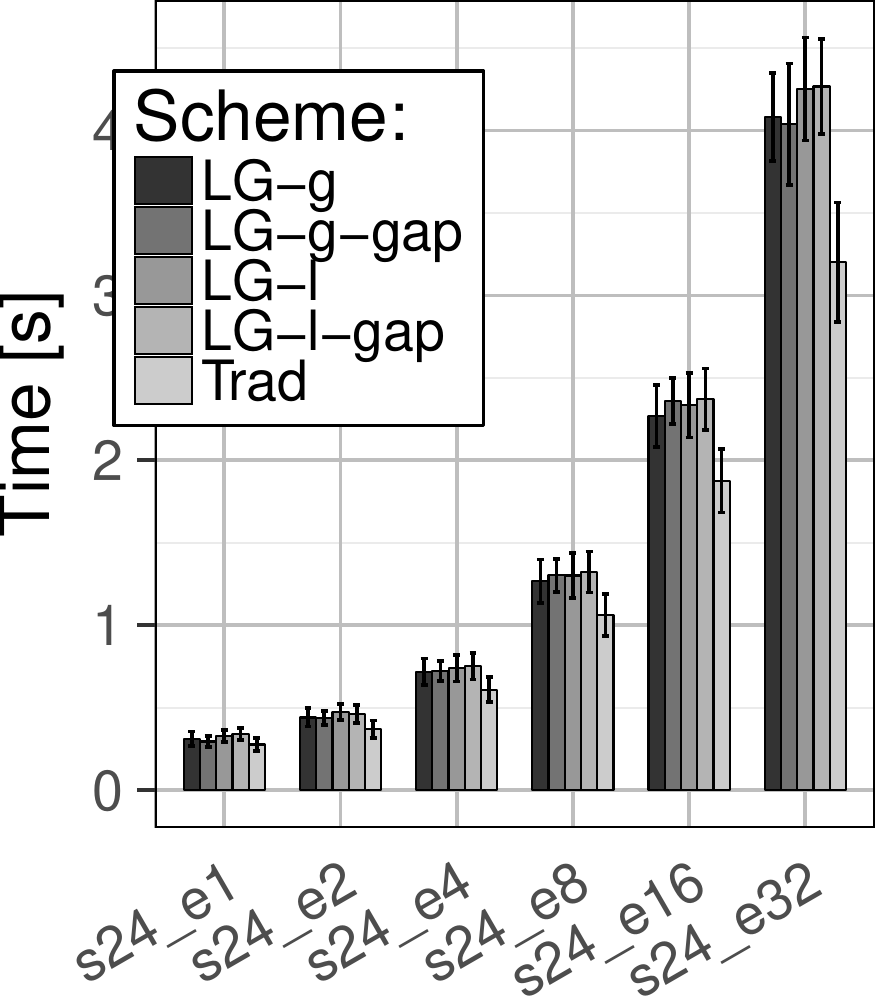}
    \caption{CC, sparse graphs.}
    \label{fig:xx}
  \end{subfigure}
\\
  \begin{subfigure}[t]{0.2 \textwidth}
    \centering
    \includegraphics[width=\textwidth]{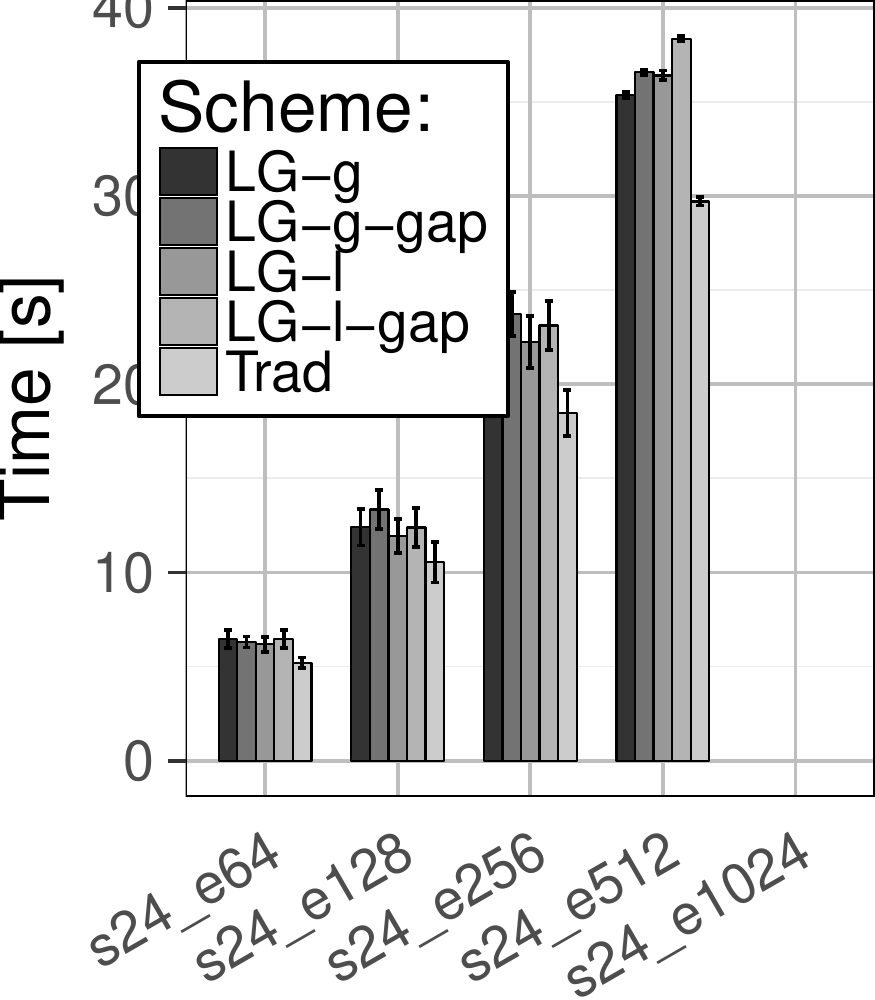}
    \caption{CC, dense graphs.}
    \label{fig:xx}
  \end{subfigure}
  \begin{subfigure}[t]{0.2 \textwidth}
    \centering
    \includegraphics[width=\textwidth]{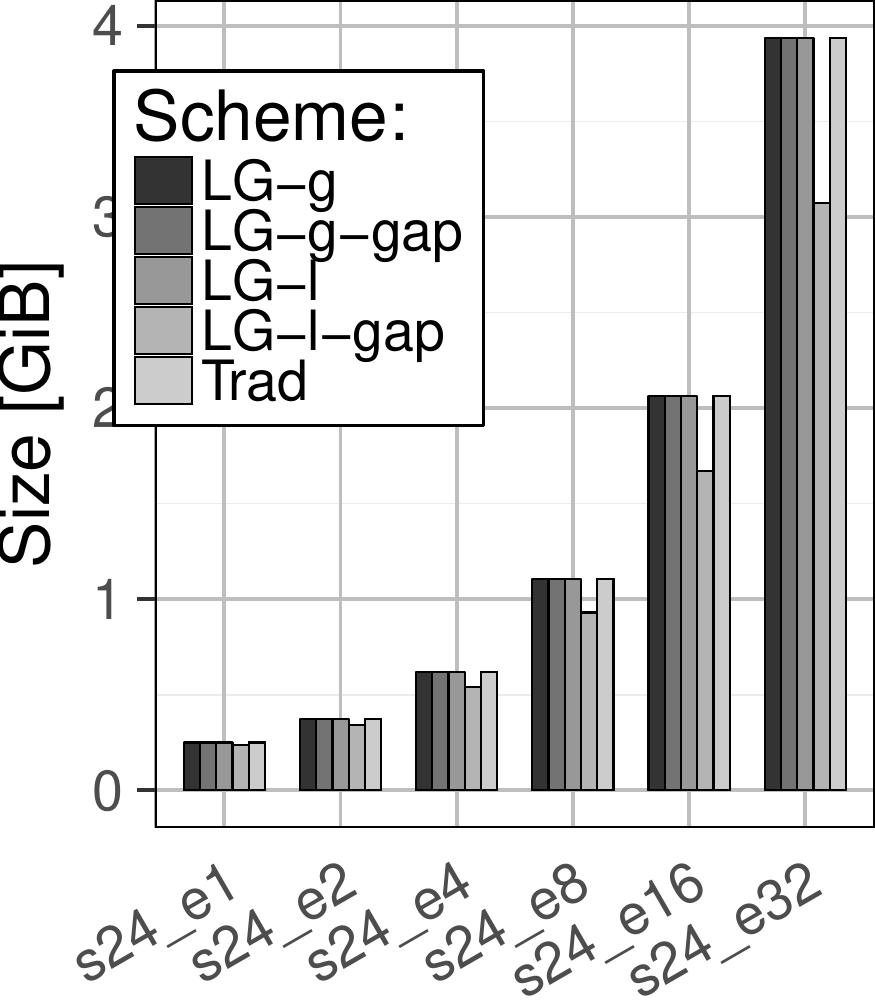}
    \caption{Size, sparse graphs.}
    \label{fig:xx}
  \end{subfigure}
  \begin{subfigure}[t]{0.2 \textwidth}
    \centering
    \includegraphics[width=\textwidth]{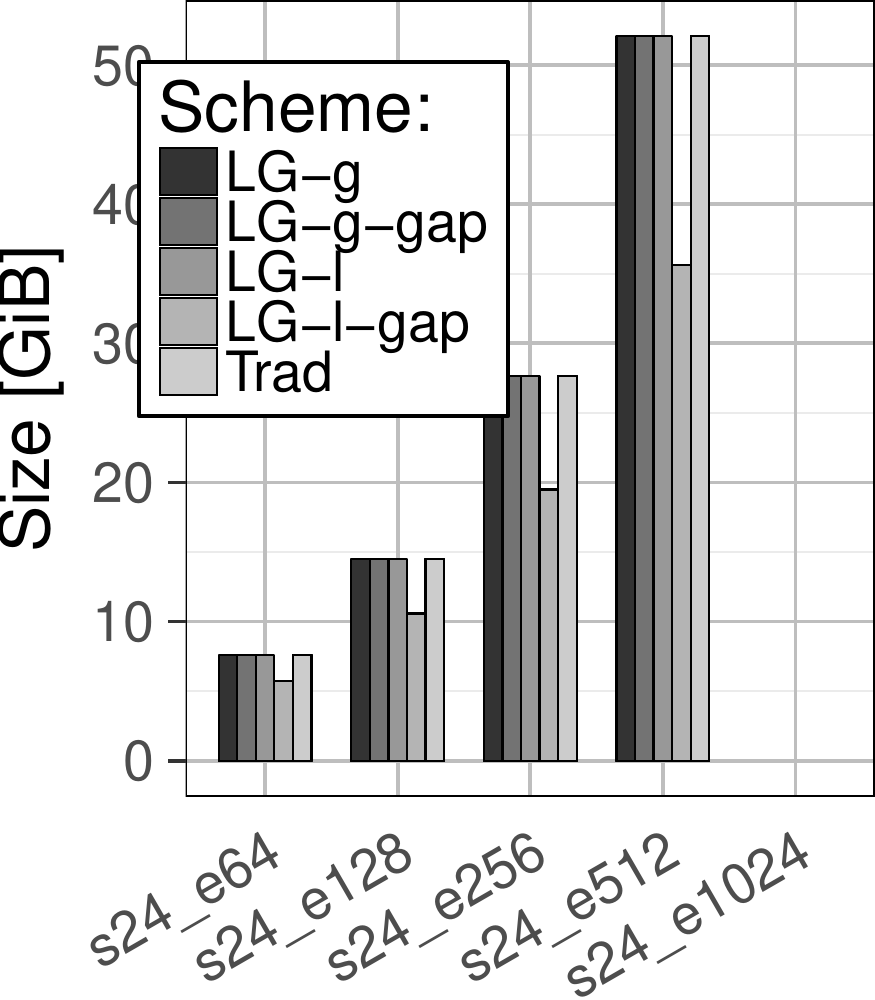}
    \caption{Size, dense graphs.}
    \label{fig:xx}
  \end{subfigure}
\caption{Log(Graph) performance analysis, logarithmizing fine elements, $n=2^{24}$, $T=16$ (full parallelism), Kronecker graphs.}
\label{fig:lg-fine-24}
\end{figure*}

{
\bibliographystyle{abbrv}
\bibliography{references}
}

\end{document}